\documentclass[12pt]{article}
\usepackage{amsfonts,amssymb,epsfig,amsmath}
\usepackage[centertableaux]{ytableau}

\usepackage{graphicx}
\usepackage{subcaption}
\renewcommand{\baselinestretch}{1.2}

\newcommand{\be}{\begin{eqnarray}}
\newcommand{\ee}{\end{eqnarray}}

\newcommand{\bn}{\begin{enumerate}}
\newcommand{\en}{\end{enumerate}}

\begin{document}

\makeatletter \@addtoreset{equation}{section} \makeatother
\renewcommand{\theequation}{\thesection.\arabic{equation}}
\renewcommand{\thefootnote}{\alph{footnote}}

\begin{titlepage}

\begin{center}
\hfill {\tt SNUTP18-003}\\
%\hfill{\tt KIAS-P18076}\\

\vspace{2cm}

{\Large\bf Quantum vortices, M2-branes and black holes}

\vspace{2cm}

\renewcommand{\thefootnote}{\alph{footnote}}

{\large Sunjin Choi$^1$, Chiung Hwang$^2$ and Seok Kim$^1$}

\vspace{0.7cm}

\textit{$^1$Department of Physics and Astronomy \& Center for
Theoretical Physics,\\
Seoul National University, Seoul 08826, Korea.}\\

\vspace{0.2cm}

\textit{$^2$Dipartimento di Fisica, Universit\`{a} di Milano-Bicocca
\& INFN,\\
Sezione di Milano-Bicocca, I-20126 Milano, Italy.}\\

\vspace{0.7cm}

E-mails: {\tt csj37100@snu.ac.kr, chiung.hwang@unimib.it,
skim@phya.snu.ac.kr}

\end{center}

\vspace{1cm}

\begin{abstract}

We study the partition functions of BPS vortices and magnetic monopole operators,
in gauge theories describing $N$ M2-branes. In particular, we explore two closely
related methods to study the Cardy limit of the index on $S^2\times\mathbb{R}$.
The first method uses the factorization of this index to vortex partition functions,
while the second one uses a continuum approximation for the monopole charge sums.
Monopole condensation confines most of the $N^2$ degrees of freedom except
$N^{\frac{3}{2}}$ of them, even in the high temperature
deconfined phase. The resulting large $N$ free energy statistically accounts for the
Bekenstein-Hawking entropy of large BPS black holes in $AdS_4\times S^7$.
Our Cardy free energy also suggests a finite $N$ version of the $N^{\frac{3}{2}}$ degrees
of freedom.

\end{abstract}

\end{titlepage}

\renewcommand{\thefootnote}{\arabic{footnote}}

\setcounter{footnote}{0}

\renewcommand{\baselinestretch}{1}

\tableofcontents

\renewcommand{\baselinestretch}{1.2}

\section{Introduction}

M2/M5-branes provide valuable insights to quantum field theories at strong coupling.
An intriguing feature is that $N$ M2/M5-branes exhibit $N^{\frac{3}{2}}$ and $N^3$
degrees of freedom, respectively. These behaviors were first discovered from their
black brane solutions \cite{Klebanov:1996un}. Recent studies from field theory
shed more lights on it, e.g. from the partition function on $S^3$
\cite{Drukker:2010nc,Herzog:2010hf} or $S^5$ \cite{Kim:2012ava}.
However, these studies on $N^{\frac{3}{2}}$, $N^3$ have
been on vacuum properties, such as vacuum entanglement entropy or
vacuum energy. For M5-branes, more interesting quantities could be
studied using anomalies \cite{Harvey:1998bx}, which see $N^3$.
For instance, certain higher derivative terms proportional to $N^3$ are
studied in \cite{Maxfield:2012aw}, and the $N^3$ scaling of the D0-D4 system
at high temperature was studied in \cite{Kim:2017zyo}, which are all related to 6d
anomalies. More recently, these anomalies are used to count the microstates
of BPS black holes in AdS$_7$ \cite{Choi:2018hmj,nahmgoong}.
For M2-branes, 3d QFTs deformed by topological twisting were studied,
in which one finds a macroscopic number of
ground states \cite{Benini:2015eyy}. The entropy of these
ground states scales like $N^{\frac{3}{2}}$, which accounts for the
magnetic/dyonic black holes
in the AdS$_4$ dual \cite{Benini:2015eyy,Benini:2016rke}.

In this paper, we study $N^{\frac{3}{2}}$ degrees of freedom of the radially
quantized SCFT on M2-branes. We shall find the $N^{\frac{3}{2}}$ scaling of an
entropic free energy, by counting excited states of this CFT.
This free energy will account for the thermodynamic properties of
the electrically charged rotating BPS black holes
in $AdS_4\times S^7$ \cite{Cvetic:2005zi,Hristov:2019mqp}. From the field theory side,
we find the deconfined $N^{\frac{3}{2}}$ degrees of freedom at
high `temperature' (meaning a suitable inverse chemical potential).
The physics of magnetic monopoles
or vortices makes the structures much richer and subtler than 4d deconfinement,
whose details we explore in this paper.

As an intermediate observable, we first study an index for vortices
in the M2-brane QFT deformed by massive parameters. Our QFT
lives on $N$ D2-branes and $1$ D6-brane. This is a 3d
$\mathcal{N}=4$ Yang-Mills theory with one adjoint and
one fundamental hypermultiplet, which flows in IR to the $\mathcal{N}=8$ SCFT
on M2-branes. It has been a useful setting to study M2-branes
\cite{Kapustin:2010xq,Gang:2011xp}. We shall study its vortices in the Higgs
branch, after a deformation by the Fayet-Iliopoulos (FI) parameter.
This index is related to our main observable, the index on $S^2\times\mathbb{R}$
\cite{Bhattacharya:2008zy,Bhattacharya:2008bja,Kim:2009wb}, in two closely related
ways. One is by the factorization of the latter into various vortex partition
functions. Another relation is obtained by taking the large angular momentum limit
on $S^2$, which we call the Cardy limit. In this limit, we make a continuum
approximation of the magnetic monopole's charge sum, finding another asymptotic
factorization to vortex partition functions. Using these relations, we compute the asymptotic free energy of the index on $S^2\times\mathbb{R}$ at large
temperature-like parameter, also in the large $N$ limit. This free energy
is proportional to $N^{\frac{3}{2}}$, and precisely accounts for the
Bekenstein-Hawking entropies of large BPS black holes in $AdS_4\times S^7$
\cite{Cvetic:2005zi,Hristov:2019mqp}. A crucial role is played by the so-called entropy
function of BPS AdS$_4$ black holes, recently discovered in \cite{Choi:2018fdc}.

Recently, supersymmetric AdS$_D$ black holes in $D>3$ were studied rather
explicitly from the radially quantized dual SCFTs in $D-1$ dimensions,
which have been somewhat
enigmatic for more than a decade. Studies are made for black holes in:
$AdS_5$ \cite{Cabo-Bizet:2018ehj,Choi:2018hmj,Choi:2018vbz,Benini:2018ywd,
Honda:2019cio,ArabiArdehali:2019tdm,Kim:2019yrz,Cabo-Bizet:2019osg,Larsen:2019oll},
$AdS_6$ \cite{Choi:2019miv},
and $AdS_7$ \cite{Choi:2018vbz,nahmgoong}. In particular, it has been shown
in $D=5,6,7$ that black holes with
large angular momenta in $AdS_D$ can be studied using the Cardy formulae
of the dual SCFTs \cite{Choi:2018hmj,Choi:2019miv,nahmgoong}. (See also
\cite{Honda:2019cio,ArabiArdehali:2019tdm,Kim:2019yrz,Cabo-Bizet:2019osg}.) In
this paper, we study the case with $D=4$, establishing the microscopic studies
of black holes in all higher dimensional AdS/CFT with SUSY.

The structures of our Cardy and large $N$ saddle points are intriguing.
In 4d Cardy formulae studied recently, the Cardy saddle point (or high temperature
saddle point) is `maximally deconfining' in that the gauge symmetry is unbroken by the
Polyakov loop operator. This makes the $N^2$ degrees of freedom fully visible.
In 3d gauge theories, one also has to
sum over the GNO charges of magnetic monopoles.
We argue that this GNO charge sum will forbid the analogous maximally deconfining
saddle point for the M2-brane system, following the ideas of \cite{Aharony:2012ns,Jain:2013py} for the vector-Chern-Simons model.
On the other hand, magnetic monopole operators condense
at the physical saddle point. The condensation effectively breaks the gauge symmetry
of the QFTs, confining most of the $N^2$ degrees of freedom even at high temperature.
The number of the remaining light degrees of freedom scales like $N^{\frac{3}{2}}$.

Our Cardy approximation is applicable to the Chern-Simons-matter theories \cite{Bagger:2006sk,Gustavsson:2007vu,Aharony:2008ug} such as the ABJM theory,
which we explore.
Also, one can study the Cardy asymptotic free energy at finite $N$.
We find a finite $N$ version of $N^{\frac{3}{2}}$ in this set-up.

The rest of this paper is organized as follows. In section 2, we study
semi-classical vortices in the Higgs branch, and
study their index. We also explain how the index on
$S^2\times\mathbb{R}$ factorizes into vortex partition functions.
In section 3, we explain a Cardy approximation of the index on $S^2\times\mathbb{R}$,
based on approximating the GNO charge sum by an integral. We compare it with
the vortex factorization formula of section 2.
In section 4, we study the large $N$ and Cardy limit of the index on
$S^2\times\mathbb{R}$, which accounts for the entropies of the dual AdS$_4$
black holes. We also comment on the monopole condensation, partial confinement
and the behaviors of the Wilson-Polyakov loops.
We then study the Cardy limit at finite $N$,
suggesting a finite $N$ version of $N^{\frac{3}{2}}$.
Section 5 concludes with remarks.

\section{Vortices on M2-branes and their indices}

We first explain the 3d QFTs that describes M2-branes. Among others,
there are Chern-Simons-matter type theories at level $1$
\cite{Bagger:2006sk,Gustavsson:2007vu,Aharony:2008ug}.
We find this approach somewhat tricky for various reasons.
The subtle aspects will be commented on below, but we shall also use these
QFT approaches in section 4.3.

The gauge theory description that we shall mainly use
is a Yang-Mills-matter theory engineered on $N$ D2-branes on top of one D6-brane.
The UV theory has 3d
$\mathcal{N}=4$ SUSY and $U(N)$ gauge symmetry. It consists of the following
fields:
\begin{eqnarray}
  \textrm{vector multiplet}&:&A_\mu\ ,\ \ \Phi^i\ ,\ \ \textrm{fermions}\\
  \textrm{adjoint hypermultiplet}&:&\phi_A=(\phi,\tilde{\phi}^\dag)\ ,\ \
  \textrm{fermions}\nonumber\\
  \textrm{fundamental hypermultiplet}&:&q_A=(q,\tilde{q}^\dag)\ ,\ \
  \textrm{fermions}\nonumber
\end{eqnarray}
where $i=1,2,3$ is an $SU(2)_r$ triplet index, and $A=1,2$ is an
$SU(2)_R$ doublet index. The $\mathcal{N}=4$ SUSY is associated with
$SO(4)\sim SU(2)_r\times SU(2)_R$ R-symmetry. The adjoint hypermultiplet
can be decomposed to two half-hypermultiplets, $\phi_A\rightarrow\phi_{Aa}$,
with $a=1,2$ being a doublet index of $SU(2)_L$ flavor symmetry.
The $SU(2)_L\times SU(2)_R\sim SO(4)$ acts on $\mathbb{R}^4$ along the D6-brane,
transverse to D2's. $SU(2)_r$ acts on $\mathbb{R}^3$ transverse to the D6-brane.
Finally, there is a topological $U(1)_T$ symmetry coming
from the current $j_\mu\sim{\rm tr}(\star F_\mu)$. In string theory, this
corresponds to the D0-brane charge, or
the momentum charge along the M-theory circle. Here, note that the D6-brane
(with transverse direction $\mathbb{R}^3$ spanned by $\Phi^i$) uplifts to a
single-centered Taub-NUT ($TN$) space in M-theory. So the QFT describes $N$ M2-branes
probing the transverse space $\mathbb{R}^4\times TN$. In the asymptotic
$\mathbb{R}^3\times S^1$ region
of Taub-NUT, $U(1)_T$ acts as the translation along the circle.
The circle is fibered over $\mathbb{R}^3$ to form $\mathbb{R}^4$ near the
Taub-NUT center. Near the center, $U(1)_T\times SU(2)_r$ enhances to
$SO(4)$ rotation symmetry of $\mathbb{R}^4$. In particular, $U(1)_T$ becomes
a Cartan of the rotation symmetry of $SO(8)$ acting on $\mathbb{R}^8$.
The strong-coupling limit of 3d QFT corresponds to the large circle limit
of M-theory, so the Taub-NUT effectively decompactifies to
$\mathbb{R}^4$. So this QFT is expected to flow to the $\mathcal{N}=8$
SCFT describing $N$ M2-branes on flat spacetime.
In particular, $SU(2)_L\times SU(2)_R\times U(1)_T\times SU(2)_r$
symmetry of our gauge theory is expected to enhance to $SO(8)$.

We are interested in the Higgs branch of this system,
and the vortex solitons in this branch. We study the system with
nonzero Fayet-Iliopoulos (FI) parameter. One can turn on three
FI parameters $\zeta^I$, where $I=1,2,3$ is a triplet index of $SU(2)_R$.
We shall only turn on $\zeta\equiv\zeta^3>0$, which breaks $SU(2)_R$ to $U(1)$.
The Higgs branch vacuum condition is given by
the following triplet of D-term conditions:
\begin{equation}
  qq^\dag-\tilde{q}^\dag\tilde{q}+[\phi,\phi^\dag]
  +[{\tilde{\phi},\tilde{\phi}}^\dag]=\zeta
  \ ,\ \ q\tilde{q}+[\phi,\tilde{\phi}]=0\ .
\end{equation}
$q$ is an $N\times 1$ matrix, $\tilde{q}$ is a $1\times N$ matrix, and
$\phi,\tilde{\phi}$ are $N\times N$ matrices. These equations describe
the moduli space of $N$ $U(1)$ instantons, which is
real $4N$ dimensional after modding out by the $U(N)$ gauge orbit.
The instanton moduli space appears since the Higgs branch describes $N$ D2-branes
dissolved into the $\mathbb{R}^4$ part of D6 world-volume.
$\zeta^I$ come from NS-NS B-fields on $\mathbb{R}^4$.

We study the vortex solitons on a subspace of the Higgs branch.
With $\zeta>0$, we shall consider the subspace $\tilde{q}=0$ with nonzero $q$.
The vortex partition functions appearing in the factorization formulae
in section 2.2 will all assume $\tilde{q}=0$.
Adjoint scalars $\phi,\tilde\phi$ may have very rich possibilities
which allow vortices. In most of our discussions in this paper,
we shall consider a simple subspace in which only $q,\phi$ are nonzero,
with $\tilde{q}=0,\tilde\phi=0$. Only in section 2.2, we shall briefly comment
on branches with nonzero $q,\phi,\tilde\phi$, and the vortex partition functions
in these branches.
Setting $\tilde{q}=0$, $\tilde{\phi}=0$, the vacuum condition is
\begin{equation}
  qq^\dag+[\phi,\phi^\dag]=\zeta{\bf 1}_{N\times N}\ .
\end{equation}
$q$ satisfies $q^\dag q=N\zeta$. We can
set $q^\dag=(\sqrt{N\zeta},0,\cdots,0)$ using $U(N)$ rotation.
Then one obtains
\begin{equation}
  [\phi,\phi^\dag]=\zeta~{\rm diag}(-(N-1),1,\cdots,1)\ .
\end{equation}
A particular solution to this equation takes the following form:
\begin{equation}\label{higgs-reference}
  \phi=\sqrt{\zeta}\left(\begin{array}{cccccc}
    0&\cdots\\
    \sqrt{N\!-\!1}\!\!&0&\cdots\\
    0&\!\!\sqrt{N\!-\!2}\!\!&0&\cdots\\
    \vdots&&&\ddots\\
    0&\cdots&&\sqrt{2}~~&0~~&0\\
    0&\cdots&&0~~&1~~&0
  \end{array}\right)\ .
\end{equation}
This vacuum breaks $U(N)$ gauge symmetry.
There are more general solutions labeled by $2N$ real parameters.
Below, we discuss the classical vortex solitons only at the point (\ref{higgs-reference}),
which will provide enough intuitions to understand our partition function.

In the above vacuum, vortex solitons are semi-classically described as follows.
Each $U(1)$ of the spontaneously broken $U(1)^N\subset U(N)$ can host its
own vortex charges, i.e. a $U(1)$ flux. On the other hand, vorticities are given
by space-dependent VEV's of the $N$ nonzero elements of $q$ and $\phi$ above, with
winding numbers at asymptotic infinity of $\mathbb{R}^2$. Consider the following
energy density, involving
$\phi_1\equiv q_1,\phi_i\equiv \phi_{i,i-1}$ ($i=2,\cdots,N$), $A_\mu$,
where $\mu=1,2$:
\begin{eqnarray}\label{vortex-energy}
  \mathcal{E}&=&|(\partial_\mu-iA^1_\mu)\phi_1|^2+
  \sum_{i=2}^N|(\partial_\mu-i(A^i_\mu-A^{i\!-\!1}_\mu))\phi_i|^2
  +\frac{1}{2g_{YM}^2}\sum_{i=1}^N(F_{12}^i)^2\\
  &&+\frac{g_{YM}^2}{2}
  \left[{\rm diag}(|\phi_1|^2-|\phi_2|^2-\zeta,|\phi_2|^2-|\phi_3|^2-\zeta,
  \cdots,|\phi_N|^2-\zeta)\right]^2
  \nonumber\\
  &=&\sum_{i=1}^N\left|(D_1+iD_2)\phi_i\right|^2
  +\sum_{i=1}^{N-1}\frac{1}{2g_{YM}^2}
  \left[F^i_{12}+g_{YM}^2(|\phi_i|^2-|\phi_{i+1}|^2-\zeta)\right]^2
  \nonumber\\
  &&+\frac{1}{2g_{YM}^2}\left[F^N_{12}+g_{YM}^2(|\phi_N|^2-\zeta)\right]^2
  +\zeta\sum_{i=1}^N F_{12}^i
  -i\epsilon^{\mu\nu}\sum_{i=1}^N\partial_\mu\left(\phi_i^\ast D_\nu\phi_i\right)\ .
  \nonumber
\end{eqnarray}
Here $D_\mu$'s are covariantized with $A^1,A^2-A^1,\cdots,A^N-A^{N-1}$ for
$\phi_1,\phi_2,\cdots,\phi_N$, respectively.
The last surface term can be ignored if $D_\nu\phi_i$ falls off sufficiently
fast at infinity. One thus obtains the following BPS equations for vortices
in this Higgs vacuum:
\begin{equation}
  (D_1+iD_2)\phi_i=0\ ,\ \ F_{12}^i=g_{YM}^2(\zeta-|\phi_i|^2+|\phi_{i+1}|^2)
  \ ,\ \ F_{12}^N=g_{YM}^2(\zeta-|\phi_N|^2)\ .
\end{equation}
The vorticities $n_i\geq 0$ for $\phi_i$ are defined by the number of
phase rotations made by $\phi_i$ at spatial infinity. This is related to
the fluxes $k_i$ carried by $A_\mu^i$ by
\begin{equation}
  n_1=k_1\ ,\ \ n_2=k_2-k_1\ ,\ \ \cdots\ ,\ \ n_{N}=k_N-k_{N-1}\ ,
\end{equation}
from the ways in which $A_\mu^i$ appear in the covariant derivatives.
Therefore, from the second term of the last line of (\ref{vortex-energy}),
one finds the multi-vortex mass given by
\begin{equation}
  M=2\pi\zeta\sum_{i=1}^N k_i\ \ ,\ \
  k_1\leq k_2\leq\cdots \leq k_N\ .
\end{equation}
The vortex masses are proportional to $\zeta$.
The masses for elementary particles in the Higgs phase are proportional to
$g_{YM}\cdot({\rm VEV})\sim g_{YM}\zeta^{\frac{1}{2}}$. Therefore, at
`weak coupling' $g_{YM}\ll\zeta^{\frac{1}{2}}$, vortex solitons are non-perturbative
and much heavier than elementary particles. At `strong coupling'
$g_{YM}\gg\zeta^{\frac{1}{2}}$, vortices are lighter than elementary particles.
We stress that the $N$ vortices are constrained as
$k_1\leq k_2\leq \cdots\leq k_N$. This is an important aspect
which will enable the partition function to have a smooth large $N$ limit.
These vorticities are naturally parametrized by Young diagrams
with $N$ or less rows, whose lengths are $k_N,k_{N-1},\cdots,k_1$, respectively.

\subsection{Indices on $D_2\times S^1$ and $\mathbb{R}^2\times S^1$}

We study an index which counts the BPS vortices discussed so far.
This is a partition function on $\mathbb{R}^2\times S^1$, where $S^1$ is for the
Euclidean time, in the Higgs branch.
The index is defined by
\begin{equation}\label{index-def}
  Z(q,t,z,Q)={\rm Tr}\left[(-1)^Fq^{R+r+2j}t^{R-r}z^{2L}Q^T\right]\ ,
\end{equation}
with suitable boundary conditions for fields assumed at infinity of $\mathbb{R}^2$,
to be explained below. $r$, $R$, $L$ are the Cartans of
$SU(2)_r\times SU(2)_R\times SU(2)_L$, $T$ is the $U(1)_T$ charge (the
vorticity), and $j$ is the $SO(2)$ angular momentum on $\mathbb{R}^2$.
The factors in the trace are chosen so that they commute with
a supercharge within the $\mathcal{N}=4$ SUSY. More concretely, the $\mathcal{N}=4$ supercharges take the form of $Q^{\dot{A}B}_{\alpha}$, where $\dot{A}$, $B$ and
$\alpha$ are doublet indices of $SU(2)_r$, $SU(2)_R$, $SO(2,1)$, respectively.
The supercharge $Q^{\dot{+}+}_-$ has charges
$r=R=\frac{1}{2}$, $j=-\frac{1}{2}$, $L=0$, $T=0$, so it commutes
with the whole factor inside the trace. This supercharge and its Hermitian
conjugate $Q^{\dot{-}-}_+$ annihilate the BPS states captured by this index.
The supercharges
$Q^{\dot{+}+}_\alpha$ and their conjugates $Q^{\dot{-}-}_\alpha$
define a 3d $\mathcal{N}=2$ supersymmetry. So the index will be computed below
using various techniques developed for 3d $\mathcal{N}=2$ theories.
From the $\mathcal{N}=2$ viewpoint, $R+r$ is the $SO(2)\sim U(1)$ R-charge, while
$R-r$ is a flavor charge.
The index on $\mathbb{R}^2\times S^1$ can also be regarded as
the index on $D_2\times S^1$, where $D_2$ is a disk. One should impose
suitable boundary conditions at the edge of $D_2$, which should be chosen to
allow the nonzero Higgs VEV for the partition function on
$\mathbb{R}^2\times S^1$. The alternative formulation of this partition function
on $D_2\times S^1$ will have a technical advantage, when one studies the grand
partition function summing over all vortex particles. The integral form of the
$\mathcal{N}=2$ gauge theory index on $D_2\times S^1$ was derived in \cite{Yoshida:2014ssa}. We summarize the results of \cite{Yoshida:2014ssa},
focussing on our model. See \cite{Yoshida:2014ssa} for more details on
SUSY QFTs on $D_2\times S^1$.

We first explain the boundary conditions on $D_2$.
To realize the boundary conditions which admit nonzero VEV for
$q$ and $\phi$, we impose Neumann boundary conditions for them: see
eqn.(2.18) of \cite{Yoshida:2014ssa} for the full boundary conditions for
the corresponding chiral multiplets. As for the $\mathcal{N}=4$ vector multiplet,
we decompose it into $\mathcal{N}=2$ vector multiplet (containing $A_\mu$, $\Phi^3$)
and an adjoint chiral multiplet (containing $\Phi_1+i\Phi_2$). We impose the
boundary condition given by eqn.(2.10) of \cite{Yoshida:2014ssa} for the
$\mathcal{N}=2$ vector multiplet.
We further need to specify the boundary conditions for: the anti-fundamental
chiral multiplet containing $\tilde{q}$, the adjoint chiral multiplet containing
$\tilde\phi$, and another chiral multiplet containing $\Phi_1+i\Phi_2$ which
originates from the $\mathcal{N}=4$ vector multiplet.
Once the boundary conditions are given for $q,\phi$ and the $\mathcal{N}=2$ vector
as above, the boundary conditions for the remaining fields
can be naturally fixed as in section 6.4 of \cite{Yoshida:2014ssa}. Namely,
we give Dirichlet boundary conditions for the chiral multiplets
$\tilde{q},\tilde\phi$, and Neumann boundary condition for the chiral multiplet
$\Phi_1+i\Phi_2$. This choice naturally guarantees the cancelation
of boundary gauge anomaly. We shall assume these boundary conditions below.
The partition function with these boundary conditions will also naturally appear
as a holomorphic block of the factorized index on $S^2\times\mathbb{R}$.\footnote{We
also tried to define the $D_2\times S^1$
function of the ABJM theory \cite{Aharony:2008ug}. However, we were not sure about
the natural and simple anomaly-free boundary conditions. However, see section 4.3 for related discussions.}

The contour integral form of our index on $D_2\times S^1$ is given by
\cite{Yoshida:2014ssa}
\begin{equation}\label{index-integral}
  Z=
  \frac{1}{N!}\oint\prod_{a=1}^N\left[\frac{ds_a}{2\pi is_a}
  s_a^{-2\pi r\zeta}\right]\prod_{a=1}^N
  \frac{(s_at^{-\frac{1}{2}}q^{\frac{3}{2}};q^2)_\infty}
  {(s_at^{\frac{1}{2}}q^{\frac{1}{2}};q^2)_\infty}
  \cdot\frac{\prod_{a\neq b}(s_as_b^{-1};q^2)_\infty}
  {\prod_{a,b=1}^N(s_as_b^{-1}t^{-1}q;q^2)_\infty}\cdot
  \prod_{a,b=1}^N\frac{(s_as_b^{-1}zt^{-\frac{1}{2}}q^{\frac{3}{2}};q^2)_\infty}
  {(s_as_b^{-1}zt^{\frac{1}{2}}q^{\frac{1}{2}};q^2)_\infty}
\end{equation}
where
\begin{equation}
  (a;q)_\infty\equiv\prod_{n=0}^\infty(1-aq^n)
\end{equation}
is the q-Pochhammer symbol. The second/third/fourth product in the integrand
come from the fundamental
hypermultiplet, $\mathcal{N}=4$ vector multiplet, adjoint hypermultiplet,
respectively. All q-Pochhammer symbols in the denominator come from scalars assuming Neumann boundary conditions, while those in the numerator come from fermions whose superpartner bosons assume Dirichlet boundary conditions.
(The argument $t^{-1}q$ in the factor $(s_as_b^{-1}t^{-1}q;q^2)_\infty$ corrects a typo
in \cite{Yoshida:2014ssa}.)
$s_a$ are $N$ holonomy variables of the vector multiplet on $S^1$. Their integration
contours are given by unit circles, $|s_a|=1$.
Here, we note a subtle phenomenon that the FI parameter on $D_2\times S^1$ is
quantized, $2\pi r\zeta\in\mathbb{Z}$, where $r$ is the radius
of the hemisphere $D_2$. This is because the standard FI term is accompanied by
a $r^{-1}$ curvature correction given by a 1d Chern-Simons term along the time
direction \cite{Yoshida:2014ssa}, which demands the quantization of $\zeta$.
Clearly, the factor $s_a^{-2\pi r\zeta}$ in (\ref{index-integral})
makes sense only with this quantization.\footnote{More precisely,
the chemical potential $t$ induces a mixed anomaly with the $U(1)\subset U(N)$ gauge
symmetry. To make the system free of gauge anomaly including this effect, one has
to quantize $\zeta$ after shifting it suitably by the chemical potentials. $\zeta$
appearing in (\ref{index-integral}) is the shifted FI parameter.}
The extra parameter $2\pi r\zeta>0$ still admits one to introduce another
fugacity-like parameter $Q\equiv q^{4\pi r\zeta}$, which will be the fugacity for
the vortex number.
The quantization of $\zeta$ is an artificial constraint
as we regulate our problem on $\mathbb{R}^2\times S^1$ to that on
$D_2\times S^1$. After all the computation is done for the integral, we can
continue $\zeta$ back to an arbitrary parameter.

If $Q$ is small enough, one can write the integral as a
residue sum by evaluating $s_a$ integrals one by one.
For $2\pi r\zeta>0$, since the factors from $s_a^{-2\pi r\zeta}$
damp to zero at $s_a=\infty$, there is no pole at $s_a=\infty$.
We take residues from poles
outside the unit circle. We assume $|tq|<1$, $|t^{-1}q|<1$,
$|zt^{\frac{1}{2}}q^{\frac{1}{2}}|<1$,
$|q|<1$, for convenience. The poles contributing to the residue sum
take the following form, up to $N!$ permutations
which cancel the overall $\frac{1}{N!}$ factor of (\ref{index-integral}):
\begin{eqnarray}\label{pole-location-1}
  s_1&=&t^{-\frac{1}{2}}q^{-\frac{1}{2}-2n_1}\ \ \ (n_1\geq 0)\ ,\\
  s_a&=&s_{a-1}z^{-1}t^{-\frac{1}{2}}q^{-\frac{1}{2}-2n_a}\ \ \
  (a=2,\cdots,N; n_a\geq 0)\ .\nonumber
\end{eqnarray}
The value of $s_1$ is determined by the poles from the fundamental hypermultiplet,
while other $s_a$'s are determined by the adjoint hypermultiplet. If poles are
chosen from other denominators than the above, one can show that
the numerator vanishes so that they are actually not poles.
Iterating the second line of (\ref{pole-location-1}) to decide $s_a$'s,
and defining $k_a\equiv \sum_{i=1}^a n_i$, one finds
\begin{equation}
  s_a=u^{-1}v^{-a+1}q^{-2k_a}
\end{equation}
for $a=1,\cdots,N$, where $u\equiv (tq)^{\frac{1}{2}}$, $v\equiv z(tq)^{\frac{1}{2}}$, and $k_1\leq k_2\leq\cdots\leq k_N$. $k_a$'s labeling the poles
will turn out to be the $U(1)^N$ vortex charges $k_1,\cdots,k_N$ that we
introduced in the context of classical solitons. This
correspondence can be understood by noting that $n_a$ in the pole
(\ref{pole-location-1}) originates
from a factor $\frac{1}{(a;q^2)_\infty}\sim\frac{1}{1-aq^{2n_a}}$, which comes
from the mode of a bosonic field with winding number $n_a$. Residue of this pole
corresponds to a partition function with vortex defect inserted \cite{Gaiotto:2012xa},
confirming the vortex interpretation.  The residue sum for
(\ref{index-integral}) is given by
\begin{eqnarray}\label{vortex-residue-temp}
  Z&=&\frac{1}{(q^2;q^2)_\infty^{\ N}}\sum_{0\leq k_1\leq\cdots\leq k_N}^\infty
  \prod_{a=1}^N\left(ua^{a-1}q^{2k_a}\right)^{2\pi r\zeta}
  \frac{(u^{-2}v^{-a+1}q^{2-2k_a};q^2)_\infty}
  {(v^{-a+1}q^{-2k_a};q^2)^\prime_\infty}\nonumber\\
  &&\times\prod_{a,b=1}^N\frac{(v^{-a+b}q^{-2k_a+2k_b};q^2)_\infty^\prime}
  {(u^{-2}v^{-a+b}q^{2-2k_a+2k_b};q^2)_\infty}
  \frac{(u^{-2}v^{1-a+b}q^{2-2k_a+2k_b};q^2)_\infty}
  {(v^{1-a+b}q^{-2k_a+2k_b};q^2)_\infty^\prime}\ ,
\end{eqnarray}
where $(a;q^2)_\infty^\prime$ means $(a;q^2)_\infty$ if $a\neq q^{-2n}$
with any non-negative integer $n$, and
\begin{equation}\label{poch-prime}
  (q^{-2n};q^2)_\infty^\prime=
  \lim_{a\rightarrow q^{-2n}}\frac{(a;q^2)_\infty}{(1-aq^{2n})}\ .
\end{equation}
Using
\begin{equation}\label{finite-infinite-poch}
  (a;q)_n=\frac{(a;q)_\infty}{(aq^n;q)_\infty}
\end{equation}
for $n\geq 0$ and
\begin{equation}
  (a;q)_{-n}\equiv\frac{1}{(aq^{-n};q)_n}=\frac{(a;q)_\infty}{(aq^{-n};q)_\infty}
\end{equation}
for $-n<0$, one finds that (\ref{finite-infinite-poch}) is true for any integer
$n$. Using this, the second line of (\ref{vortex-residue-temp}) can be
rearranged as
\begin{eqnarray}\label{finite-poch-simplify}
  \hspace*{-.5cm}&&\prod_{a,b=1}^N\frac{(v^{-a+b}q^{-2k_a+2k_b};q^2)_\infty^\prime}
  {(u^{-2}v^{-a+b}q^{2-2k_a+2k_b};q^2)_\infty}
  \frac{(u^{-2}v^{1-a+b}q^{2-2k_a+2k_b};q^2)_\infty}
  {(v^{1-a+b}q^{-2k_a+2k_b};q^2)_\infty^\prime}\\
  \hspace*{-.5cm}&&=\prod_{a,b=1}^N\frac{(v^{-a+b};q^2)_\infty^\prime}
  {(u^{-2}v^{-a+b}q^{2};q^2)_\infty}
  \frac{(u^{-2}v^{1-a+b}q^{2};q^2)_\infty}{(v^{1-a+b};q^2)_\infty^\prime}
  \cdot\frac{(u^{-2}v^{b-a}q^2;q^2)_{k_b-k_a}}{(v^{b-a};q^2)_{k_b-k_a}}
  \cdot\frac{(v^{1-a+b};q^2)_{k_b-k_a}}{(u^{-2}v^{1-a+b}q^2;q^2)_{k_b-k_a}}
  \nonumber\\
  \hspace*{-.5cm}
  &&=\prod_{a,b=1}^N\frac{(u^{-2}v^{b-a}q^2;q^2)_{k_b-k_a}}{(v^{b-a};q^2)_{k_b-k_a}}
  \cdot\frac{(v^{1-a+b};q^2)_{k_b-k_a}}{(u^{-2}v^{1-a+b}q^2;q^2)_{k_b-k_a}}
  \cdot\prod_{a=1}^N\frac{(v^{-a+1};q^2)^\prime_\infty}{(v^{-a+N+1};q^2)_\infty}
  \cdot\frac{(u^{-2}v^{-a+N+1}q^2;q^2)_\infty}{(u^{-2}v^{-a+1}q^2;q^2)_\infty}
  \ .\nonumber
\end{eqnarray}
The product over $a=1,\cdots,N$ on the first line of (\ref{vortex-residue-temp})
and that on the last line of (\ref{finite-poch-simplify}) combine and get
rearranged as
\begin{eqnarray}
  &&\prod_{a=1}^N\left(ua^{a-1}q^{2k_a}\right)^{2\pi r\zeta}
  \frac{(u^{-2}v^{-a+1}q^{2-2k_a};q^2)_\infty}
  {(v^{-a+1}q^{-2k_a};q^2)^\prime_\infty}
  \frac{(v^{-a+1};q^2)^\prime_\infty}{(v^{-a+N+1};q^2)_\infty}
  \cdot\frac{(u^{-2}v^{-a+N+1}q^2;q^2)_\infty}{(u^{-2}v^{-a+1}q^2;q^2)_\infty}
  \nonumber\\
  &&=\left[u^Nv^{\frac{N(N-1)}{2}}\right]^{2\pi r\zeta}
  Q^{k_1+\cdots+k_N}\prod_{a=1}^N
  \frac{(v^{-a+1};q^2)_{-k_a}}{(u^{-2}v^{-a+1}q^2;q^2)_{-k_a}}\cdot
  \frac{(u^{-2}v^aq^2;q^2)_\infty}{(v^a;q^2)_\infty}
  \ ,
\end{eqnarray}
where $Q\equiv q^{4\pi r\zeta}$. So one obtains
\begin{eqnarray}\label{vortex-residue-final}
  Z&=&\frac{(u^Nv^{\frac{N(N-1)}{2}})^{2\pi r\zeta}}{(q^2;q^2)_\infty^{\ N}}
  \prod_{a=1}^N\frac{(u^{-2}v^aq^2;q^2)_\infty}{(v^a;q^2)_\infty}
  \sum_{0\leq k_1\leq\cdots\leq k_N}Q^{k_1+\cdots+k_N}
  \prod_{a=1}^N\frac{(v^{-a+1};q^2)_{-k_a}}{(u^{-2}v^{-a+1};q^2)_{-k_a}}
  \nonumber\\
  &&\cdot\prod_{a,b=1}^N\frac{(v^{-a+b+1};q^2)_{-k_a+k_b}
  (u^{-2}v^{-a+b}q^2;q^2)_{-k_a+k_b}}
  {(v^{-a+b};q^2)_{-k_a+k_b}(u^{-2}v^{-a+b+1}q^2;q^2)_{-k_a+k_b}}\ .
\end{eqnarray}
In the last expression, one can relax
the condition $2\pi r\zeta\in\mathbb{Z}_+$, so we can now regard $Q$
as an independent continuous parameter. Here, let us decompose $Z$ into three
factors, $Z=Z_{\rm prefactor}Z_{\rm pert}Z_{\rm vortex}$, where each
factor is given as follows:
\begin{eqnarray}\label{vortex-final-2}
  Z_{\rm prefactor}&=&\frac{(u^Nv^{\frac{N(N-1)}{2}})^{2\pi r\zeta}}
  {(q^2;q^2)_\infty^{\ N}}\ \ ,\ \ \
  Z_{\rm pert}=\prod_{a=1}^N\frac{(u^{-2}v^aq^2;q^2)_\infty}{(v^a;q^2)_\infty}
  \nonumber\\
  Z_{\rm vortex}&=&\sum_{0\leq k_1\leq\cdots\leq k_N}
  Q^{k_1+\cdots+k_N}Z_{k_1,\cdots,k_N}\\
  Z_{k_1,\cdots,k_N}&\equiv&
  \prod_{a=1}^N\frac{(v^{-a+1};q^2)_{-k_a}}{(u^{-2}v^{-a+1};q^2)_{-k_a}}
  \cdot\prod_{a,b=1}^N\frac{(v^{-a+b+1};q^2)_{-k_a+k_b}
  (u^{-2}v^{-a+b}q^2;q^2)_{-k_a+k_b}}
  {(v^{-a+b};q^2)_{-k_a+k_b}(u^{-2}v^{-a+b+1}q^2;q^2)_{-k_a+k_b}}\ .\nonumber
\end{eqnarray}
Here, $Z_{0,\cdots,0}=1$ by definition.
In the rational function $Z_{k_1,\cdots,k_N}$ appearing in
(\ref{vortex-final-2}), one finds further cancelations
between denominator and numerator. In fact, since $k_1\leq\cdots \leq k_N$
define a Young diagram with $k$ boxes, $Z_{k_1,\cdots,k_N}$
admits a simple expression in terms of this Young diagram $Y=(k_N,k_{N-1},\cdots,k_1)$
after cancelation. To explain the final result after the cancelation, let us
introduce the following `distance functions' on the Young diagram:
\begin{equation} \label{yfunc}
  \begin{array}{ll}
    a(s): & \textrm{arm (horizontal) length = number of boxes to the right of }s \\
    l(s): &  \textrm{leg (vertical) length = number of the boxes below }s \\
    x(s): & \textrm{horizontal position = number of boxes to the left of }s \\
    y(s): & \textrm{vertical position = number of the boxes above }s
  \end{array}
\end{equation}
Here, $s$ labels the boxes of the Young diagram. For instance,
for the two boxes $s_1$, $s_2$ of $Y=(6,5,3,2)$ below, they are given by
\begin{equation}
\ydiagram[*(yellow)s_1]{1+1}*[*(yellow)s_2]{0,2+1}*[*(white)]{6,5,3,2}
  \ \ \ \longrightarrow\ \ \
  \begin{array}{l}
    a(s_1)=4, \, l(s_1)=3, \, x(s_1)=1, \, y(s_1)=0 \\
    a(s_2)=2, \, l(s_2)=1, \, x(s_1)=2, \, y(s_1)=1
\end{array}\ .
\end{equation}
Using these notations,
$Z_{\rm vortex}$ is given by
\begin{equation}\label{vortex-Young}
  Z_{\rm vortex}= \sum_Y Q^{|Y|}
  \prod_{s \in Y} \frac{(1-u^{-2} q^{-2a(s)} v^{-l(s)})(1-u^{-2} v q^2 q^{2a(s)} v^{l(s)})(1-v^N q^{2x(s)} v^{-y(s)})}{(1-q^{-2} q^{-2a(s)} v^{-l(s)})(1-v q^{2a(s)} v^{l(s)})(1-u^{-2} q^2 v^N q^{2x(s)} v^{-y(s)})}\ .
\end{equation}
We checked this expression up to $Q^{11}$ order, till $N\leq 10$.
One can also prove (\ref{vortex-Young}) analytically, which we explain in
appendix B.

We also explain other factors, $Z_{\rm prefactor}$ and
$Z_{\rm pert}$. The factor $(u^Nv^{\frac{N(N-1)}{2}})^{2\pi r\zeta}$
in $Z_{\rm prefactor}$ is the `zero-point energy' factor, weighting the
`ground state' if one expands $Z$ in fugacities.
The factor $(q^2,q^2)_\infty^{\ -N}$ of $Z_{\rm prefactor}$
comes from $N$ chiral multiplets containing the $N$ complex scalars, which form
the Higgs branch moduli. These scalars are the massless fluctuations from
the reference point (\ref{higgs-reference}).
This part will not play any important role in the rest
of our works. For instance, $Z_{\rm prefactor}$ will not appear in the factorization
formula on $S^2\times S^1$ later. (More precisely, one can regard it as the two
$Z_{\rm prefactor}$'s canceling in the factorization formula.)
So $Z_{\rm prefactor}$ will be mostly neglected.
$Z_{\rm pert}$ comes from `perturbative' massive particles' contribution in
the Higgs branch, which will be important later. Normally, the Higgs branch
partition function on $\mathbb{R}^2\times S^1$ refers to
$Z_{\mathbb{R}^2\times S^1}=Z_{\rm pert}Z_{\rm vortex}$.

Now we have two alternative expressions for the index,
the integral form (\ref{index-integral}) and the residue sum
(\ref{vortex-residue-final}), (\ref{vortex-Young}).
The latter expression is a series which is useful for sufficiently
small $|Q|$, but (\ref{index-integral}) can be used
more generally.

Before closing this subsection,
we study the case with $N=1$, for single M2-brane. In this case,
the index given by the residue sum becomes simplified.
This is because the CFT on one M2-brane is expected to be a free QFT,
consisting of four free $\mathcal{N}=2$ chiral multiplets.
In fact, studying (\ref{vortex-Young}) to
certain high orders in $Q$, we find that (\ref{vortex-residue-final})
can be written as
\begin{equation}\label{abelian-exact}
  Z_{N=1}=\frac{(tq)^{\pi r\zeta}}{(q^2;q^2)_\infty}
  \cdot\frac{(zt^{-\frac{1}{2}}q^{\frac{3}{2}};q^2)_\infty}
  {(zt^{\frac{1}{2}}q^{\frac{1}{2}};q^2)_\infty}
  \cdot\frac{(q^2Q;q^2)_\infty}{(t^{-1}qQ;q^2)_\infty}
  =\frac{(tq)^{\pi r\zeta}}{(q^2;q^2)_\infty}
  \cdot\frac{(zt^{-\frac{1}{2}}q^{\frac{3}{2}};q^2)_\infty}
  {(zt^{\frac{1}{2}}q^{\frac{1}{2}};q^2)_\infty}
  \cdot\frac{(q^{\frac{3}{2}}t^{\frac{1}{2}}\hat{Q};q^2)_\infty}
  {(t^{-\frac{1}{2}}q^{\frac{1}{2}}\hat{Q};q^2)_\infty}
\end{equation}
at $N=1$. This can also be shown analytically
by using the infinite $q$-binomial theorem. Here we defined
$\hat{Q}\equiv q^{\frac{1}{2}}t^{-\frac{1}{2}}Q$.
The first factor of (\ref{abelian-exact}) is simply $Z_{\rm prefactor}$,
which we ignore. The second factor
$Z_{\rm pert}=\frac{(zt^{-\frac{1}{2}}q^{\frac{3}{2}};q^2)_\infty}
{(zt^{\frac{1}{2}}q^{\frac{1}{2}};q^2)_\infty}$ comes from the adjoint hypermultiplet
of the $\mathcal{N}=4$ theory, which is free at $N=1$. The factors in the
denominator/numerator come from the chiral multiplets with Neumann/Dirichlet boundary
conditions, respectively. The last factor $Z_{\rm vortex}=
\frac{(\hat{Q}t^{\frac{1}{2}}q^{\frac{3}{2}};q^2)_\infty}
{(\hat{Q}t^{-\frac{1}{2}}q^{\frac{1}{2}};q^2)_\infty}$ makes the contribution from
another free hypermultiplet, where two chiral multiplets in it are given
Neumann/Dirichlet boundary conditions, respectively. In fact it is well known
that the `vortex field' makes a free hypermultiplet in this case.
To see this, first note that with the adjoint hypermultiplet decoupled at $N=1$,
this theory is simply
an $\mathcal{N}=4$ SQED with $N_f=1$ flavor. In \cite{Gaiotto:2008ak},
$\mathcal{N}=4$ $U(N)$ SQCD with $N_f=2N-1$ flavors was studied. It was argued
that a monopole operator becomes free and decouples in IR. The
remaining system in IR was argued to be the $U(N-1)$ SQCD with
same number $N_f=2N-1$ of $U(N-1)$ fundamental flavors. Since the last theory
is void at $N=1$, SQED at $N_f=1$ in IR is dual to the free hypermultiplet.
Indeed, the vortex partition function of this SQED was shown to be precisely
that of a free hypermultiplet \cite{Kim:2012uz}. Defining $t_I$ ($I=1,2,3,4$) as
\begin{equation}
  (t_1,t_2,t_3,t_4)\equiv(t^{\frac{1}{2}}z,t^{\frac{1}{2}}z^{-1},
  t^{-\frac{1}{2}}\hat{Q},t^{-\frac{1}{2}}\hat{Q}^{-1})\ ,
\end{equation}
satisfying $t_1t_2t_3t_4=1$, the Abelian index can be written as
\begin{equation}
  \left.\frac{}{}Z_{\rm pert}Z_{\rm vortex}\right|_{N=1}=
  \frac{(t_2^{-1}q^{\frac{3}{2}};q^2)_\infty(t_4^{-1}q^{\frac{3}{2}};q^2)_\infty}
  {(t_1q^{\frac{1}{2}};q^2)_\infty(t_3q^{\frac{1}{2}};q^2)_\infty}\ .
\end{equation}

In section 4, we shall be interested in the large $N$ free energy of the index,
in the limit $\beta\rightarrow 0^+$ where $q\equiv e^{-\beta}$. Here, we make such a
study at $N=1$ as a warming up. We shall first study the limit $\beta\rightarrow 0$
from the exact expression (\ref{abelian-exact}), and then discuss how to
recover the same result from the saddle point analysis of the contour integral
expression (\ref{index-integral}).

To perform the $\beta\rightarrow 0$ approximation, one should
understand the $\beta\rightarrow 0$ limit of $(a;e^{-2\beta})_\infty$.
We are interested in taking $\beta\rightarrow 0$ while keeping it complex,
with ${\rm Re}(\beta)>0$. Also, other fugacities $t_I$ are kept as
pure phases: $|t_I|=1$, while satisfying $t_1t_2t_3t_4=1$. It is important
that these phases can be substantially away from $1$. This defines our
`Cardy limit' of the index. The importance of these phases was noticed
in \cite{Choi:2018hmj,Choi:2018vbz}, which will be seen again in our
later sections. In this set-up, one obtains
\begin{equation}\label{pochhamer-Li2}
  (a;q^2)_\infty=\prod_{n=0}^\infty(1-aq^{2n})
  =\exp\left[-\sum_{n=1}^\infty\frac{1}{n}\frac{a^n}{1-q^{2n}}\right]
  \stackrel{\beta\rightarrow 0}{\longrightarrow}\exp
  \left[-\frac{1}{2\beta}\sum_{n=1}^\infty\frac{a^n}{n^2}\right]
  =\exp\left[-\frac{{\rm Li}_2(a)}{2\beta}\right]
\end{equation}
when $a$ is a phase, $|a|=1$.
Therefore, in our Cardy limit,
the index (\ref{abelian-exact}) is given by
\begin{equation}
  \log Z_{N=1}\sim
  \frac{1}{2\beta}\left[{\rm Li}_2(\hat{Q}t^{-\frac{1}{2}})
  -{\rm Li}_2(\hat{Q}t^{\frac{1}{2}})+{\rm Li}_2(zt^{\frac{1}{2}})
  -{\rm Li}_2(zt^{-\frac{1}{2}})+{\rm Li}_2(1)\right]
  +\pi r\zeta\log t\ .
\end{equation}
Here, we define $\xi$ by $2\pi r\zeta\equiv\frac{\xi}{2\beta}$
($Q\equiv e^{-\xi}$), and keep
$\xi$ fixed as one takes $\beta\rightarrow 0$.
Then, defining $\mathcal{F}$ by
\begin{equation}
  \log Z\sim-\frac{\mathcal{F}}{2\beta}
\end{equation}
in the $\beta\rightarrow 0$ limit, one obtains
\begin{equation}\label{abelian-exact-asymp}
  \mathcal{F}_{N=1}={\rm Li}_2(zt^{-\frac{1}{2}})
  -{\rm Li}_2(zt^{\frac{1}{2}})
  +{\rm Li}_2(\hat{Q}t^{\frac{1}{2}})-{\rm Li}_2(\hat{Q}t^{-\frac{1}{2}})
  -{\rm Li}_2(1) -\frac{\xi}{2}\log t\ .
\end{equation}

Now we make the saddle point analysis of the integral expression
(\ref{index-integral}), at $N=1$ and in the limit $\beta\rightarrow 0$.
(\ref{index-integral}) in this setting becomes
\begin{equation}\label{abelian-D2-integral}
  Z_{N=1}\sim
  \int \frac{ds}{2\pi is}\exp\left[-\frac{\xi}{2\beta}\log s
  +\frac{1}{2\beta}\left({\rm Li}_2(zt^{\frac{1}{2}})
  -{\rm Li}_2(zt^{-\frac{1}{2}})+{\rm Li}_2(t^{-1})
  +{\rm Li}_2(t^{\frac{1}{2}}s)-
  {\rm Li}_2(t^{-\frac{1}{2}}s)\right)\right]
\end{equation}
where the contour is over the unit circle $|s|=1$.
In the Cardy limit, we can ignore the quantization condition of $\zeta$
and keep general complex $\xi$. Taking
$\xi$ to be purely imaginary, and $t,z$ to be phases, we try to find
the saddle point for $s$ at $|\beta|\ll 1$.
One needs to extremize
\begin{equation}
  \xi \log s+{\rm Li}_2(st^{-\frac{1}{2}})
  -{\rm Li}_2(st^{\frac{1}{2}})\ .
\end{equation}
The saddle point should satisfy
\begin{equation}
 0=\xi+{\rm Li}_1(st^{-\frac{1}{2}})-{\rm Li}_1(st^{\frac{1}{2}})
 =\xi-\log\frac{1-st^{-\frac{1}{2}}}{1-st^{\frac{1}{2}}}\ .
\end{equation}
The solution is given by
\begin{equation}\label{abelian-saddle}
  s_0=\frac{e^\xi-1}{e^\xi t^{\frac{1}{2}}-t^{-\frac{1}{2}}}
  =\frac{\sinh\frac{\xi}{2}}{\sinh\frac{\xi+T}{2}}
\end{equation}
with $t\equiv e^T$. $s_0$ is real for purely imaginary $\xi,T$.
Plugging in this value to the integrand of (\ref{abelian-D2-integral}),
$s_0=t^{-\frac{1}{2}}\frac{1-t^{\frac{1}{2}}\hat{Q}}{1-t^{-\frac{1}{2}}\hat{Q}}$,
one obtains precisely the same $\mathcal{F}$ as (\ref{abelian-exact-asymp}).
The last statement can be shown analytically by using the identity
\begin{equation}
\textrm{Li}_2 (xy) -\textrm{Li}_2 (x) -\textrm{Li}_2 (y) +\textrm{Li}_2 (1) = \textrm{Li}_2 \left(\frac{1-x}{1-xy}\right) -\textrm{Li}_2 \left(y\frac{1-x}{1-xy}\right) + \log (x) \log \left(\frac{1-x}{1-xy}\right)\ .
\end{equation}

\subsection{Factorization on $S^2\times S^1$}

So far, we examined the vortex partition function $Z_\text{\rm vortex}$
that is captured as a part of the $\mathbb R^2 \times S^1$ index, or equivalently the
$D_2 \times S^1$ index with a certain boundary condition at the edge. In the literature,
it was discussed that the vortex partition function can be a building block of many other supersymmetric partition functions on compact 3d manifolds such as $S^2 \times S^1$ and $S^3_b$ \cite{Pasquetti:2011fj}. We shall develop a similar factorization formula along
the line of \cite{Hwang:2018uyj}. More precisely, once we consider an $S^1$ fibration on $S^2$ where the angular momentum fugacity is turned on, the fields are effectively localized at the poles of $S^2$ and probe local $\mathbb R^2$ geometry. Thus, the supersymmetric partition functions on those manifolds are written in terms of the vortex partition function as the following universal form:
\begin{align}
\label{eq:univ}
Z =
\sum_\text{Higgs vacua} Z_\text{pert} Z_\text{vortex} \overline Z_\text{vortex}.
\end{align}
Only differences are the perturbative contribution $Z_\text{pert}$ and how to glue two pieces of the vortex partition functions, i.e., how to define $\overline Z_\text{vortex}$, which has the same functional form as $Z_\text{vortex}$ up to redefinitions of variables depending on the background geometry.
For our example, the index on $S^2\times S^1$
will take the form of
\begin{equation}
  Z_{S^2\times S^1}(\hat Q,t,z,q)=\sum_{\mathcal Y \in{\rm Higgs}}Z_{\rm pert}^{\mathcal Y}(t,z,q)
  Z_{\rm vortex}^{\mathcal Y}(\hat Q,t,z,q)Z_{\rm vortex}^{\mathcal Y}(\hat Q^{-1},t^{-1},z^{-1},q^{-1})\ ,
\end{equation}
where the points $\mathcal Y$ in the Higgs branch will be specified below.

\begin{figure}[tbp]
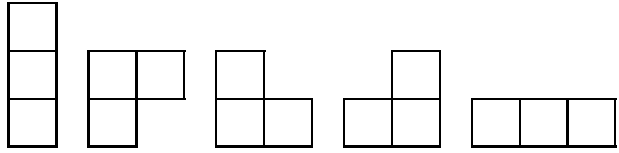

\centering
\ydiagram{1,1,1} \quad \ydiagram{0,2,1} \quad \ydiagram{0,1,2} \quad \ydiagram{0,1+1,2} \quad \ydiagram{0,0,3}
\caption{\label{fig:box} The Higgs vacua of the (massive) $\mathcal N = 2$ $U(3)$ theory with one fundamental and two adjoint chirals are represented by 2-dimensional box diagrams due to the D-term conditions. If there is a superpotential, they are further restricted.}
\end{figure}

Our theory of interest includes one fundamental and one adjoint hypermultiplets. Since
a 3d $\mathcal N = 4$ vector multiplet contains an $\mathcal N = 2$ chiral multiplet as
well, we have in total three $\mathcal N = 2$ chirals in the adjoint representation. In
the previous section, we showed that the chiral from the $\mathcal N = 4$ vector does not
yield any contributing pole. Thus, the factorization of our partition function mimics that of a theory with two adjoints. The factorization of a 3d $\mathcal N = 2$ theory with two adjoints is recently discussed in \cite{Hwang:2018uyj}. It was shown that the D-term equations of the $\mathcal N = 2$ theory restrict its Higgs vacua such that they are represented by 2-dimensional box diagrams; e.g., see figure \ref{fig:box}.
Furthermore, if the theory has a superpotential, there will be extra conditions from the F-term equations. In our case, we have the following F-term condition:
\begin{align}
q \tilde q+[\phi,\tilde \phi] = 0,
\end{align}
which is a part of the $\mathcal N = 4$ D-term conditions. As we have shown in the previous section, the vacuum solutions have vanishing $\tilde q$ and accordingly vanishing $[\phi,\tilde \phi]$. The condition $[\phi,\tilde \phi] = 0$ demands that only the first, third and fifth diagrams in figure \ref{fig:box} are allowed; in general, only the Young diagram types are allowed.

To establish the factorization formula with
the structures outlined in the previous paragraph,
we start from the known expression for the index on $S^2 \times S^1$
\cite{Bhattacharya:2008zy,Bhattacharya:2008bja}, which is
\cite{Kim:2009wb,Imamura:2011su,Gang:2011xp}:
\begin{align}
\label{eq:SCI}
& Z_{S^2\times S^1}(\hat{Q},z,t,q) = \\
& \sum_{\{m\} = -\infty}^\infty \frac{1}{{\rm Weyl(\{m\})}} \oint \left(\prod_{a = 1}^N
\frac{d s_a}{2 \pi i s_a} \hat{Q}^{m_a} t^{-|m_a|/2} q^{|m_a|/2}\right) \times \nonumber \\
&\left(\prod_{1 \leq a \neq b \leq N} \left(1-s_a s_b^{-1} q^{|m_a-m_b|}\right)\right) \left(\prod_{a = 1}^N \frac{(s_a^{-1} t^{-\frac{1}{2}} q^{\frac{3}{2}+|m_a|};q^2)}{(s_a t^\frac{1}{2} q^{\frac{1}{2}+|m_a|};q^2)} \frac{(s_a t^{-\frac{1}{2}} q^{\frac{3}{2}+|m_a|};q^2)}{(s_a^{-1} t^\frac{1}{2} q^{\frac{1}{2}+|m_a|};q^2)}\right) \times \nonumber \\
&\left(\prod_{a,b = 1}^N \frac{(s_a^{-1} s_b t q^{1+|-m_a+m_b|};q^2) (s_a^{-1} s_b z^{-1} t^{-\frac{1}{2}} q^{\frac{3}{2}+|-m_a+m_b|};q^2) (s_a^{-1} s_b z t^{-\frac{1}{2}} q^{\frac{3}{2}+|-m_a+m_b|};q^2)}{(s_a s_b^{-1} t^{-1} q^{1+|m_a-m_b|};q^2) (s_a s_b^{-1} z t^\frac{1}{2} q^{\frac{1}{2}+|m_a-m_b|};q^2) (s_a s_b^{-1} z^{-1} t^\frac{1}{2} q^{\frac{1}{2}+|m_a-m_b|};q^2)}\right).\nonumber
\end{align}
Here the integration contour for each $s_a$ is taken to be the unit circle. ${\rm Weyl(\{m\})}$ is the order of the Weyl group remaining unbroken for given magnetic flux $\{m\} \in \mathbb Z^N/S_N$. In the following computation, however, it will be more convenient to distinguish the permutations in $\{m\}$ and to take the symmetry factor $N!$ instead of ${\rm Weyl(\{m\})}$. In other words, we replace the flux summation by
\begin{align}
\sum_{\{m\} = -\infty}^\infty \frac{1}{{\rm Weyl(\{m\})}} \qquad \rightarrow \qquad \frac{1}{N!} \sum_{\{m\} \in \mathbb Z^N}.
\end{align}
From here, we also use a shorthand expression $(a;q) \equiv (a;q)_\infty$
in the rest of this paper.

We are aiming to evaluate this integral using the residue theorem. Assuming $|q| < 1$ and $|t| = |z| = 1$, we take the poles outside the unit circle, which are given by the intersections of the following hyperplanes:
\begin{align}
\begin{aligned}
\label{eq:hyperplanes}
s_a &= t^{-\frac{1}{2}} q^{-\frac{1}{2}} q^{-|m_a|-2 k_a}, \\
s_a &= s_b z^{-1} t^{-\frac{1}{2}} q^{-\frac{1}{2}} q^{-|m_a-m_b|-2 k_a}, \\
s_a &= s_b z t^{-\frac{1}{2}} q^{-\frac{1}{2}} q^{-|m_a-m_b|-2 k_a}, \\
s_a &= s_b t q^{-1} q^{-|m_a-m_b|-2 k_a}
\end{aligned}
\end{align}
where $k_a \geq 0$. However, poles sitting at the hyperplanes of the
fourth type have vanishing residues. In the set-up of the previous paragraph, this implies that there are no poles from the adjoint chiral in the $\mathcal{N}=4$ vector multiplet.
The relevant poles are only determined by hyperplanes of the other types. Thus, as we noted already, the residue evaluation of our theory resembles that of the two adjoint theory. While a pole is typically determined by $N$ hyperplanes, for a general two adjoint theory, it may happen that a set of hyperplanes degenerate such that more than $N$ hyperplanes meet at the point. In such cases, one encounters a double or higher order pole when the $N$-dimensional integral is evaluated iteratively. Nevertheless, a particular choice of the superpotential sometimes yield extra zeros by imposing conditions on the fugacities so that the higher order poles become simple. Indeed, our $\mathcal N = 4$ SYM example turns out to be such a case.

At first let us forget about the issue of higher order poles and just focus on how we organize $N$ linearly independent hyperplanes. Once we pick up $N$ hyperplanes intersecting at a pole, they can be represented by a binary tree graph of $N$ nodes where each node is accompanied by a label of three parameters $(a,z_a,k_a)$. While the meanings of the tree and the labels $(a,z_a,k_a)$ are rather clear from \eqref{eq:hyperplanes}, let us explain them briefly. The first parameter $a$, which is an integer in
the range $1\leq a\leq N$ without repetition, can be used to label the nodes. Namely, we will refer to the node with $(a,z_a,k_a)$ as the $a$th node. Then one can represent a tree graph using a map $p:\{1,\dots,N\} \rightarrow \{0,\dots,N\}$. $p$ is defined such that $p(a) = b$ if the $b$th node is the parent node of the $a$th node. If the $a$th node is the root node, which doesn't have a parent node, $p(a) = 0$. The other two parameters are chosen such that
\begin{align}
z_a = \left\{\begin{array}{ll}
1, \quad & p(a) = 0, \\
z,z^{-1}, \quad & p(a) \neq 0,
\end{array}\right. \qquad
k_a \geq 0.
\end{align}
Note that $z_a$ distinguishes whether the $a$th node is the left child or the right child of the parent node, which are two available choices in a binary tree. Once a tree $p$ and $(a,k_a,z_a)$ for each node are given, they specify the hyperplanes as follows:
\begin{align}
s_a = \left\{\begin{array}{ll}
t^{-\frac{1}{2}} q^{-\frac{1}{2}} q^{-|m_a|-2 k_a} \, , \qquad & p(a) = 0, \\
s_{p(a)} z_{a}^{-1} t^{-\frac{1}{2}} q^{-\frac{1}{2}} q^{-|m_a-m_{p(a)}|-2 k_{a}} \, , \qquad & p(a) \neq 0.
\end{array}\right.
\end{align}
Consequently, each $s_a$ at the pole is given by
\begin{align}
\label{eq:pole}
s_a &= \left(\prod_{n = 0}^{l_a-1} z_{p^n(a)}^{-1}\right) (t q)^{-\frac{l_a}{2}} q^{-\sum_{n = 0}^{l_a-1} |m_{p^n (a)}-m_{p^{n+1} (a)}|-2 \sum_{n = 0}^{l_a-1} k_{p^n(a)}}
\end{align}
where $l_a$ is the integer satisfying $p^{l_a}(a) = 0$.
For example, the root note has $l_a = 1$. We also define $m_0 = 0$.

Now let us evaluate the residue at the pole \eqref{eq:pole}. First we consider $N < 4$, in which case, the pole \eqref{eq:pole} is always simple. We have to sum the residues for all possible $p$ and $(a,z_a,k_a)$. Combined with the flux summation, they give rise to the expression for the index factorized into the perturbative part and the vortex parts sketched earlier. In particular, the perturbative part can be extracted out by evaluating the residue for $m_a = k_a = 0$, which is given by
\begin{align}
\label{eq:pert}
Z^p_\text{pert}(z,t,q) &= \left(\prod_{1 \leq a \neq b \leq N} \left(1-v_a^{-1} v_b\right)\right) \left(\prod_{a = 1}^N \frac{(v_a q^2;q^2)}{(v_a^{-1};q^2)'} \frac{(v_a^{-1} u^{-2} q^2;q^2)}{(v_a u^2;q^2)}\right) \nonumber \\
&\quad \times \left(\prod_{a,b = 1}^N \frac{(v_a v_b^{-1} u^2;q^2) (v_a v_b^{-1} v^{-1} q^2;q^2) (v_a v_b^{-1} u^{-2} v q^2;q^2)}{(v_a^{-1} v_b u^{-2} q^2;q^2) (v_a^{-1} v_b v;q^2)' (v_a^{-1} v_b u^2 v^{-1};q^2)'}\right)
\end{align}
where
\begin{gather}
    \begin{gathered}
    v_a = \left(\prod_{n = 0}^{l_a-1} z_{p^n(a)}\right) (t q)^{\frac{l_a-1}{2}}, \\
    u = t^\frac{1}{2} q^\frac{1}{2}, \qquad v = z t^\frac{1}{2} q^\frac{1}{2}.
    \end{gathered}
\end{gather}
Note that $v_a$ reduces to $v^{l_a-1}$ if $z_b = z$ for all $b$. $(a;q^2)'$ is defined
around (\ref{poch-prime}).
Namely, it is defined as an ordinary q-Pochhammer symbol up to the vanishing factors discarded. Note that there are $N$ such vanishing factors, which arise due to the pole we have taken.

The expression \eqref{eq:pert} is specified by a binary tree $p$. Note that a binary tree of $N < 4$ nodes can be represented by a 2-dimensional box diagram; e.g., see figure \ref{fig:box} for $N = 3$. The left child node is placed on top of the parent node and the right child node is placed at the right side of the parent node. Among the box diagrams in figure \ref{fig:box}, the second and the fourth diagrams have vanishing residues due to the factor $\prod_{a,b = 1}^N (v_a v_b^{-1} u^2;q^2)$ in \eqref{eq:pert}. This factor gives an extra zero whenever we have diagonally adjacent boxes along the top-right direction. Thus, for $N = 3$, only the first, third and fifth diagrams contribute.

Such box diagrams can label the residues for higher $N$ as well. One may worry that the correspondence between the binary trees and the 2-dimensional box diagrams is not one-to-one for $N \geq 4$. Indeed, there are two such cases. First, there exist tree graphs that do not have box diagram counterparts. That happens only if two nodes of the binary tree are overlapped when they are represented in the 2-dimensional box diagram. However, such a tree with overlapping nodes has the vanishing residue due to the first factor $\prod_{a \neq b} \left(1-v_a^{-1} v_b\right)$ of \eqref{eq:pert}. Thus, one can always find the corresponding box diagram unless the tree graph has the vanishing residue.

\begin{figure}[tbp]
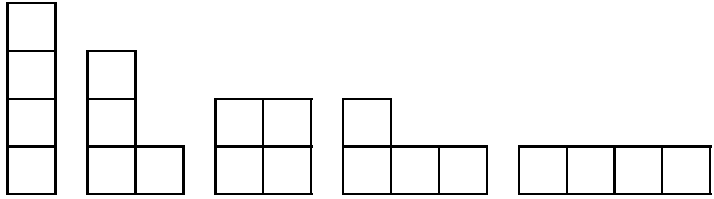

\centering
\ydiagram{1,1,1,1} \quad \ydiagram{0,1,1,2} \quad \ydiagram{0,0,2,2} \quad \ydiagram{0,0,1,3} \quad \ydiagram{0,0,0,4}
\caption{\label{fig:ydiagrams} For our $\mathcal N = 4$ SYM example, the contributing poles are labeled by Young diagrams. For $N = 4$ there are five diagrams, among which the third diagram corresponds to a degenerate singularity where five hyperplanes intersect rather than four.}
\end{figure}

Second, there can be multiple tree graphs that are mapped to the same box diagram. This is related to the possibility of higher order poles, which will demand us to modify
the formula (\ref{eq:pert}).
In that case, more than $N$ vanishing factors appear in the denominator of \eqref{eq:pert} if we forget about the $'$ symbol for a moment. Such a case, for instance, happens for the third diagram in figure \ref{fig:ydiagrams}.
One can associate two different tree graphs to this box diagram because the top-right box can be either the right child of the top-left node or the left child node of the bottom-right node. This is exactly due to the fact that the five hyperplanes intersect at this pole rather than four. Although the singularity is unique, there are two ways of picking up four linearly independent hyperplanes defining this singularity.
Therefore, (\ref{eq:pert}) is wrong if $p$ does not uniquely label the poles.
Instead, we should seek for a formula in which the box diagrams rather than $p$
label the residues.

Now recall that if there are diagonally adjacent boxes along the top-right direction, they yield an extra zero. For the third diagram in figure \ref{fig:ydiagrams}, this extra zero cancels out the extra pole from the degenerate hyperplanes such that the singularity becomes a simple pole. Thus, a simple modification of \eqref{eq:pert} will
give the right residue formula if we discard the extra vanishing factors in the numerator and the denominator simultaneously. In our $\mathcal N = 4$ SYM example,
for arbitrary box diagrams, an extra
vanishing factor in the denominator is always accompanied by an extra zero in the
numerator. Furthermore, a pole corresponding to a non-Young diagram has the vanishing residue as we have demonstrated for $N = 3$.
Thus, the contributing poles are all simple and labeled by Young diagrams.

Collecting all, we write down a modification of \eqref{eq:pert} in terms of the Young diagrams. For a Young diagram $\mathcal Y$, the perturbative part is written as follows:
\begin{align}
Z^{\mathcal Y}_\text{pert}(z,t,q) &= \left(\prod_{\mathsf a\neq \mathsf b \in \mathcal Y} \left(1-v_\mathsf a^{-1} v_{\mathsf b}\right)\right) \left(\prod_{\mathsf a \in \mathcal Y} \frac{(v_\mathsf a q^2;q^2)}{(v_\mathsf a^{-1};q^2)'} \frac{(v_\mathsf a^{-1} u^{-2} q^2;q^2)}{(v_\mathsf a u^2;q^2)}\right) \nonumber \\
&\quad \times \left(\prod_{\mathsf a,\mathsf b \in \mathcal Y} \frac{(v_\mathsf a
v_{\mathsf b}^{-1} u^2;q^2)' (v_\mathsf a v_{\mathsf b}^{-1} v^{-1} q^2;q^2) (v_\mathsf a v_{\mathsf b}^{-1} u^{-2} v q^2;q^2)}{(v_\mathsf a^{-1} v_{\mathsf b} u^{-2} q^2;q^2) (v_\mathsf a^{-1} v_{\mathsf b} v;q^2)' (v_\mathsf a^{-1}
v_{\mathsf b} u^2 v^{-1};q^2)'}\right).
\end{align}
$v_\mathsf a$ is now given by
\begin{align}
v_\mathsf a = z^{\mathsf i (\mathsf a)-\mathsf j (\mathsf a)} (t q)^{\frac{1}{2} (\mathsf i (\mathsf a)+\mathsf j (\mathsf a)-2)}
\end{align}
where $(\mathsf i (\mathsf a),\mathsf j (\mathsf a))$ is the position of box $\mathsf a$ in the Young diagram $\mathcal Y$. Again $'$ denotes that the vanishing factors are discarded. Note that the label $a=1,\cdots,N$ of each node that we began with is now irrelevant. It will turn out that this is also true for the vortex parts, so we have $N!$ identical contributions, which are canceled by the symmetry factor $1/N!$.

Now we move on to the vortex parts. After evaluating the integral by taking the non-vanishing residues, we are left with the summation over Young diagrams as well as the two summations over $m_a$ and $k_a$. The latter sums over $m_a, k_a$ can be reorganized into the sums over the vorticity and the anti-vorticity, which are completely factorized for given Young diagram $\mathcal Y$.
The detailed computation of the vortex parts is similar to what is done in \cite{Hwang:2018uyj}. It turns out that the result is simply given by making the following replacements in $Z_\text{vortex}$ in (\ref{vortex-final-2}),
which we obtained from the $D_2 \times S^1$ index in the previous section:
\begin{equation}
\prod_{a = 1}^N \rightarrow \prod_{\mathsf a \in \mathcal Y}\ \ ,\ \ \
v^{a-1} \rightarrow v_\mathsf a = z^{\mathsf i (\mathsf a)-\mathsf j (\mathsf a)} (t q)^{\frac{1}{2} (\mathsf i (\mathsf a)+\mathsf j (\mathsf a)-2)}\ \ ,\ \ \
k_a \rightarrow k_\mathsf a\ \ ,\ \ \
Q \rightarrow \hat{Q} t^\frac{1}{2} q^{-\frac{1}{2}}\ .
\end{equation}
$k_\mathsf a$ is a non-negative integer assigned to each $\mathsf a \mathsf \in \mathcal Y$ such that those integers are non-decreasing in each row and column of $\mathcal Y$. This resembles the standard Young tableau, in which the associated integers are strictly increasing rather than non-decreasing. Taking into account those modifications, we have the following expression of $Z_\text{vortex}^{\mathcal Y}$ for the Young diagram $\mathcal Y$:
\begin{eqnarray}
\label{eq:vort}
Z_\text{vortex}^\mathcal Y(\hat Q,z,t,q) &=& \sum_{k_{\mathsf a}}
(\hat Q t^\frac{1}{2} q^{-\frac{1}{2}})^{\sum_{\mathsf a \in \mathcal Y} k_{\mathsf a}} \left(\prod_{\mathsf a \in \mathcal Y} \frac{(v_{\mathsf a}^{-1};q^2)_{-k_{\mathsf a}}}{(u^{-2} v_{\mathsf a}^{-1} q^2;q^2)_{-k_{\mathsf a}}}\right) \nonumber \\
&& \times \left(\prod_{\mathsf a \neq \mathsf b \in \mathcal Y} \frac{(v v_{\mathsf a}^{-1} v_{\mathsf b};q^2)_{-k_{\mathsf a}+k_{\mathsf b}} (u^{-2} v_{\mathsf a}^{-1} v_{\mathsf b} q^2;q^2)_{-k_{\mathsf a}+k_{\mathsf b}} }{(v_{\mathsf a}^{-1} v_{\mathsf b};q^2)_{-k_{\mathsf a}+k_{\mathsf b}} (u^{-2} v v_{\mathsf a}^{-1} v_{\mathsf b} q^2;q^2)_{-k_{\mathsf a}+k_{\mathsf b}}}\right)\ .
\end{eqnarray}
If we take $\mathcal Y = (1^N)$, \eqref{eq:vort} reduces to $Z_\text{vort}$ in the previous section.

In the end, combining the perturbative part and the vortex parts,
\eqref{eq:SCI} is written in the following factorized form
\begin{equation}\label{eq:S2S1}
  Z_{S^2\times S^1}(\hat{Q},z,t,q) =
  \sum_{|\mathcal Y| = N} Z^\mathcal Y_\text{pert}(z,t,q) \, Z^\mathcal Y_\text{vortex} (\hat Q,z,t,q) \, Z^\mathcal Y_\text{vortex} (\hat Q^{-1},z^{-1},t^{-1},q^{-1})\ .
\end{equation}
The expression \eqref{eq:S2S1} is also checked numerically up to
$N=3$ as a series expansion in $q$ up to $q^3$, and also at $N = 4$ up to $q^2$.

\section{Cardy limit of the index on $S^2\times S^1$: set-up}

In this section, we set up a direct framework of making the Cardy limit
approximation of the index on $S^2\times\mathbb{R}$. The result will be
connected to the vortex partition function that we studied in the
previous section. Although we focus on the $\mathcal{N}=4$ Yang-Mills theory
for M2-branes introduced in the previous section, the framework applies
to other 3d QFTs. We shall provide similar analysis for the ABJM theory in
section 4.3.

The index of our $\mathcal{N}=4$ gauge theories on $S^2\times S^1$ is
given by ($e^{-\hat\xi}\equiv\hat{Q}$) \cite{Gang:2011xp}
\begin{eqnarray}\label{N=4-index}
  Z&=&\!\!\sum_{\{m\}=-\infty}^\infty\frac{1}{{\rm Weyl}(\{m\})}
  \oint\frac{d\alpha_a}{2\pi}
  e^{-\hat\xi\sum_{a=1}^N{m_a}}
  \prod_{a=1}^N (qt^{-1})^{\frac{|m_a|}{2}}
  \frac{(e^{-i\alpha_a}q^{|m_a|}t^{-\frac{1}{2}}q^{\frac{3}{2}};q^2)
  (e^{i\alpha_a}q^{|m_a|}t^{-\frac{1}{2}}q^{\frac{3}{2}};q^2)}
  {(e^{i\alpha_a}q^{|m_a|}t^{\frac{1}{2}}q^{\frac{1}{2}};q^2)
  (e^{-i\alpha_a}q^{|m_a|}t^{\frac{1}{2}}q^{\frac{1}{2}};q^2)}\nonumber\\
  &&\times\prod_{a\neq b}q^{-\frac{|m_{ab}|}{2}}(1-e^{i\alpha_{ab}}q^{|m_{ab}|})
  \prod_{a,b=1}^N t^{\frac{|m_{ab}|}{2}}
  \frac{(e^{i\alpha_{ab}}tq^{1+|m_{ab}|};q^2)}
  {(e^{i\alpha_{ab}}t^{-1}q^{1+|m_{ab}|};q^2)}\\
  &&\times\prod_{a,b=1}^N (qt^{-1})^{\frac{|m_{ab}|}{2}}
  \frac{(e^{i\alpha_{ab}}q^{|m_{ab}|}z^{-1}t^{-\frac{1}{2}}q^{\frac{3}{2}};q^2)
  (e^{i\alpha_{ab}}q^{|m_{ab}|}zt^{-\frac{1}{2}}q^{\frac{3}{2}};q^2)}
  {(e^{i\alpha_{ab}}q^{|m_{ab}|}zt^{\frac{1}{2}}q^{\frac{1}{2}};q^2)
  (e^{i\alpha_{ab}}q^{|m_{ab}|}z^{-1}t^{\frac{1}{2}}q^{\frac{1}{2}};q^2)}\nonumber
\end{eqnarray}
Here, the factor $\prod_{a\neq b}(1-e^{i\alpha_{ab}}q^{|m_{ab}|})$ coming from
the Haar measure and the $\mathcal{N}=2$ vector multiplet may be written as
\begin{equation}\label{N=2-vector-decompose}
  \prod_{a\neq b}(1-e^{i\alpha_{ab}}q^{|m_{ab}|})=
  \prod_{a\neq b}\frac{(e^{i\alpha_{ab}}q^{|m_{ab}|};q^2)}
  {(e^{i\alpha_{ab}}q^{2+|m_{ab}|};q^2)}\ ,
\end{equation}
which was relevant in section 2 when we discussed the factorization of
this index into vortex partition functions.

We would first like to rewrite the index in the following way.
Each chiral multiplet contributes the following factor
to the contour integrand:
\begin{equation}
  \left(e^{-i\rho(\alpha)}q^{1-R}y^{-1}\right)^{\frac{|\rho(m)|}{2}}
  \frac{(e^{-i\rho(\alpha)}q^{2-R+|\rho(m)|}y^{-1};q^2)}
  {(e^{i\rho(\alpha)}q^{R+|\rho(m)|}y;q^2)}\ .
\end{equation}
For the chiral multiplets in our $\mathcal{N}=4$ theory, $R=\frac{1}{2}$
and $y$ is given by a suitable combination of $t$ and $z$.
For the adjoint chiral multiplet in the $\mathcal{N}=4$ vector multiplet,
this formula applies with $R=1$ and $y=t^{-1}$.
Even for the $\mathcal{N}=2$ vector multiplet, inverse of
this expression applies at $R=0$ and $y=1$ if one uses the decomposition
(\ref{N=2-vector-decompose}). One can show that \cite{Dimofte:2011py}
\begin{equation}\label{abs-remove-1}
  \left(e^{-i\rho(\alpha)}q^{1-R}y^{-1}\right)^{\frac{|\rho(m)|}{2}}
  \frac{(e^{-i\rho(\alpha)}q^{2-R+|\rho(m)|}y^{-1};q^2)}
  {(e^{i\rho(\alpha)}q^{R+|\rho(m)|}y;q^2)}=
    \left(e^{-i\rho(\alpha)}q^{1-R}y^{-1}\right)^{-\frac{\rho(m)}{2}}
  \frac{(e^{-i\rho(\alpha)}q^{2-R-\rho(m)}y^{-1};q^2)}
  {(e^{i\rho(\alpha)}q^{R-\rho(m)}y;q^2)}\ .
\end{equation}
This identity states that one can replace all $|\rho(m)|$'s by $-\rho(m)$.
(Of course one can have a similar identity replacing
$|\rho(m)|\rightarrow+\rho(m)$.) One also finds
\begin{equation}\label{abs-remove-2}
  \left(e^{i\rho(\alpha)}q^{1-R}\tilde{y}^{-1}\right)^{\frac{|\rho(m)|}{2}}
  \frac{(e^{i\rho(\alpha)}q^{2-R+|\rho(m)|}\tilde{y}^{-1};q^2)}
  {(e^{-i\rho(\alpha)}q^{R+|\rho(m)|}\tilde{y};q^2)}=
    \left(e^{i\rho(\alpha)}q^{1-R}\tilde{y}^{-1}\right)^{-\frac{\rho(m)}{2}}
  \frac{(e^{i\rho(\alpha)}q^{2-R-\rho(m)}\tilde{y}^{-1};q^2)}
  {(e^{-i\rho(\alpha)}q^{R-\rho(m)}\tilde{y};q^2)}\ .
\end{equation}
In our $\mathcal{N}=4$ theory, one obtains a product of the two left hand sides
of (\ref{abs-remove-1}) and (\ref{abs-remove-2})
for each hypermultiplet. The above identities state that this factor can be replaced by
\begin{equation}
  \left(q^{1-R}y^{-\frac{1}{2}}\tilde{y}^{-\frac{1}{2}}\right)^{-\rho(m)}
  \frac{(e^{\rho(\bar{u})}q^{2-R}y^{-1};q^2)(e^{\rho(u)}q^{2-R}\tilde{y}^{-1};q^2)}
  {(e^{\rho(u)}q^Ry;q^2)(e^{\rho(\bar{u})}q^R\tilde{y};q^2)}
\end{equation}
where $q=e^{-\beta}$ and $u\equiv \beta m + i\alpha$. We shall apply this
formula for all weights $\rho$ in a representation ${\bf R}$, so that there is a
product $\prod_{\rho\in{\bf R}}$ which comes with holomorphic
$\rho(u)$, while $\prod_{-\rho\in\overline{\bf R}}$ comes with anti-holomorphic
$\rho(\bar{u})$. In other words, one obtains the following factorization of
the integrand into `holomophic' and `anti-holomorphic' parts:
\begin{equation}
  \prod_{a=1}^N(t/q)^{\frac{u_a}{4\beta}}
  \frac{(e^{u_a}t^{-\frac{1}{2}}q^{\frac{3}{2}};q^2)}
  {(e^{u_a}t^{\frac{1}{2}}q^{\frac{1}{2}};q^2)}
  \prod_{a,b=1}^N\frac{(e^{u_{ab}}zt^{-\frac{1}{2}}q^{\frac{3}{2}};q^2)}
  {(e^{u_{ab}}zt^{\frac{1}{2}}q^{\frac{1}{2}};q^2)}
  \ \cdot\ \prod_{a=1}^N(t/q)^{\frac{\bar{u}_a}{4\beta}}
  \frac{(e^{\bar{u}_a}t^{-\frac{1}{2}}q^{\frac{3}{2}};q^2)}
  {(e^{\bar{u}_a}t^{\frac{1}{2}}q^{\frac{1}{2}};q^2)}
  \prod_{a,b=1}^N\frac{(e^{\bar{u}_{ab}}z^{-1}t^{-\frac{1}{2}}q^{\frac{3}{2}};q^2)}
  {(e^{\bar{u}_{ab}}z^{-1}t^{\frac{1}{2}}q^{\frac{1}{2}};q^2)}\ .
\end{equation}
Here, we inserted $R=\frac{1}{2}$ for all hypermultiplet fields,
$y=\tilde{y}=t^{\frac{1}{2}}$ for fundamental hyper, and
$y=t^{\frac{1}{2}}z$, $\tilde{y}=t^{\frac{1}{2}}z^{-1}$ for adjoint
hyper. The `holomorphic' part depending on $u_a$
is part of the integrand appearing in the $D_2\times S^1$ index (\ref{index-integral})
after setting $e^{u_a}=s_a$,
except the factor $(t/q)^{\frac{u_a}{4\beta}}$ that will be accounted for shortly.
The `anti-holomorphic part' will also have a similar
interpretation on $D_2\times S^1$.
As for the integrand coming from the $\mathcal{N}=4$ vector multiplet,
\begin{eqnarray}\label{vector-abs}
  &&\prod_{a\neq b}(e^{-i\alpha_{ab}}q)^{-\frac{|m_{ab}|}{2}}
  (1-e^{i\alpha_{ab}}q^{|m_{ab}|})
  \prod_{a,b=1}^N(e^{-i\alpha_{ab}}t)^{\frac{|m_{ab}|}{2}}
  \frac{(e^{-i\alpha_{ab}}tq^{1+|m_{ab}|};q^2)}
  {(e^{i\alpha_{ab}}t^{-1}q^{1+|m_{ab}|};q^2)}\\
  &&=\prod_{a\neq b}(e^{-i\alpha_{ab}}q)^{-\frac{|m_{ab}|}{2}}
  \frac{(e^{i\alpha_{ab}}q^{|m_{ab}|};q^2)}{(e^{-i\alpha_{ab}}q^{2+|m_{ab}|};q^2)}
  \prod_{a,b=1}^N(e^{-i\alpha_{ab}}t)^{\frac{|m_{ab}|}{2}}
  \frac{(e^{-i\alpha_{ab}}tq^{1+|m_{ab}|};q^2)}
  {(e^{i\alpha_{ab}}t^{-1}q^{1+|m_{ab}|};q^2)}\ ,\nonumber
\end{eqnarray}
the first factor comes from the
$\mathcal{N}=2$ vector multiplet, and
the second factor from the $\mathcal{N}=2$ adjoint chiral multiplet within the
$\mathcal{N}=4$ vector multiplet.
Applying (\ref{abs-remove-1}), one obtains
\begin{equation}
  \prod_{a\neq b}(e^{-i\alpha_{ab}}q)^{\frac{m_{ab}}{2}}
  \frac{(e^{u_{ab}};q^2)}{(e^{\bar{u}_{ab}}q^{2};q^2)}
  \prod_{a,b=1}^N(e^{-i\alpha_{ab}}t)^{-\frac{m_{ab}}{2}}
  \frac{(e^{\bar{u}_{ab}}tq;q^2)}
  {(e^{u_{ab}}t^{-1}q;q^2)}=
  \frac{\prod_{a\neq b}(e^{u_{ab}};q^2)}{\prod_{a,b}(e^{u_{ab}}t^{-1}q;q^2)}
  \cdot\frac{\prod_{a,b}(e^{\bar{u}_{ab}}tq;q^2)}
  {\prod_{a\neq b}(e^{\bar{u}_{ab}}q^{2};q^2)}\ .
\end{equation}
The holomorphic part is again part of the integrand appearing in the
$D_2\times S^1$ index (\ref{index-integral}).

Finally, the fugacity factor $e^{-\hat\xi\sum_a m_a}$ for the topological $U(1)_T$
can be written as
\begin{equation}
  e^{-\hat\xi\sum_a m_a}=e^{-\frac{\hat\xi}{2\beta}\sum_a u_a}
  e^{-\frac{\hat\xi}{2\beta}\sum_a\bar{u}_a}\ ,
\end{equation}
which again factorizes to holomorphic and anti-holomorphic part. Combined with
the factor $(t/q)^{\sum_a\frac{u_a}{4\beta}}$ from hypermultiplets,
one obtains
\begin{equation}
  e^{-\frac{\xi}{2\beta}\sum_au_a}e^{-\frac{\xi}{2\beta}\sum_a\bar{u}_a}
\end{equation}
where
\begin{equation}
  e^{-\xi}\equiv e^{-\hat\xi}(t/q)^{\frac{1}{2}}
\end{equation}
is the FI parameter that appeared in the $D_2\times S^1$ index.
(Recall that $2\pi r\zeta=\frac{\xi}{2\beta}$.)
So one obtains a `formal factorization' of the
integrand of the index on $S^2\times S^1$. This is not a true factorization
yet, because $u_a,\bar{u}_a$ have to be partly integrated (imaginary part)
while partly summed over discretely (real part).

Before proceeding, with the formula for the index with all absolute values of
$\rho(m)$ removed as above, we identify the periods of the chemical
potentials and present a natural basis. This will be useful later for understanding
the precise structures of the saddle point free energy. From the integrand including the flux-dependent
zero point energy factor, one identifies the following periodicities:
\begin{eqnarray}
  \hat\xi&\sim&\hat\xi+2\pi i\\
  (T,\beta)&\sim&(T\pm 2\pi i,\beta+2\pi i)\nonumber\\
  f&\sim&f+2\pi i\nonumber\\
  (\hat\xi,f,\beta;\alpha_a)&\sim&
  (\hat\xi\pm \pi i,f\pm \pi i,\beta+2\pi i;\alpha_a+\pi )\\
  (T,\hat\xi,f;\alpha_a)&\sim&(T+2\pi i,\hat\xi\pm\pi i,f\pm\pi i;\alpha_a+\pi )\nonumber
\end{eqnarray}
where $t=e^T$, $z=e^f$.
The $\pm$ signs appearing on the right hand sides are independent.
Note that the shift $\alpha_a\rightarrow\alpha_a+\pi $ of
the integral variables is sometimes required to see that the integrand is invariant.
Now let us define the variables,
\begin{equation}\label{SO8-chemical}
  \Delta_1\equiv-\hat\xi+\frac{T}{2}+\frac{\beta}{2}\ ,\ \
  \Delta_2\equiv\hat\xi+\frac{T}{2}+\frac{\beta}{2}\ ,\ \
  \Delta_3\equiv f-\frac{T}{2}+\frac{\beta}{2}\ ,\ \
  \Delta_4\equiv-f-\frac{T}{2}+\frac{\beta}{2}\ .
\end{equation}
Note that these four variables can be regarded as four independent chemical
potentials of the index. They are related to $\beta$ as
$\Delta_1+\Delta_2+\Delta_3+\Delta_4-2\beta=0$, so that the sum over them
is approximately zero in the Cardy limit $\beta\rightarrow 0$.
In terms of these variables, the $12$ periodicities identified above can be rephrased as
\begin{equation}\label{SO8-period}
  (\Delta_I,\Delta_J)\sim (\Delta_I+2\pi i,\Delta_J\pm 2\pi i)
\end{equation}
shifts for $6$ possible pairs $\Delta_I,\Delta_J$ among $\Delta_{1,2,3,4}$.
These are the basic periodicities expected for the $SO(8)$ chemical potentials,
coupling to vector, spinors or their product representations.
In terms of $\Delta_I$'s, the index can be written as
\begin{equation}\label{index-Delta}
  Z_{S^2\times S^1}(\Delta_I)={\rm Tr}\left[(-1)^F
  e^{-\sum_{I=1}^4\Delta_I(Q_I+J)}\right]
\end{equation}
where $Q_I$'s are the $U(1)^4\subset SO(8)$ Cartans, and $J$ is the angular momentum
on $S^2$. $\Delta_I$ should satisfy
\begin{equation}\label{Delta-properties}
  {\rm Re}(\Delta_I)>0\ \ ,\ \ \ \sum_{I=1}^4\Delta_I=2\beta\ ,
\end{equation}
as they are conjugate the charges $Q_I+J$ which are non-negative in the BPS
sector and furthermore can grow to $+\infty$.

Let us now take the $\beta\rightarrow 0$ Cardy limit
of the index, keeping small complex $\beta$ with ${\rm Re}(\beta)>0$.
The idea \cite{Pasquetti:2019uop}
is to now regard $u_a=\beta m_a+i\alpha_a$ as a continuum
complex variable, and replace the sum over $m_a$ by integration.
The $N$ dimensional integral over $\alpha_a$ and sum over $m_a$ are replaced by
a $2N$ dimensional integral over $u_a$, $\bar{u}_a$. One obtains
\begin{eqnarray}\label{semi-factorization}
  Z&\sim&\int \prod_{a=1}^Ndu_a e^{-\frac{\xi}{2\beta}\sum_{a=1}^Nu_a}
  \prod_{a=1}^N\frac{(e^{u_a}t^{-\frac{1}{2}}q^{\frac{3}{2}};q^2)}
  {(e^{u_a}t^{\frac{1}{2}}q^{\frac{1}{2}};q^2)}
  \prod_{a,b=1}^N\frac{(e^{u_{ab}}zt^{-\frac{1}{2}}q^{\frac{3}{2}};q^2)}
  {(e^{u_{ab}}zt^{\frac{1}{2}}q^{\frac{1}{2}};q^2)}
  \frac{\prod_{a\neq b}(e^{u_{ab}};q^2)}{\prod_{a,b}(e^{u_{ab}}t^{-1}q;q^2)}\\
  &&\times\int \prod_{a=1}^Nd\bar{u}_a e^{-\frac{\xi}{2\beta}\sum_{a=1}^N\bar{u}_a}
  \prod_{a=1}^N\frac{(e^{\bar{u}_a}t^{-\frac{1}{2}}q^{\frac{3}{2}};q^2)}
  {(e^{\bar{u}_a}t^{\frac{1}{2}}q^{\frac{1}{2}};q^2)}
  \prod_{a,b=1}^N\frac{(e^{\bar{u}_{ab}}z^{-1}t^{-\frac{1}{2}}q^{\frac{3}{2}};q^2)}
  {(e^{\bar{u}_{ab}}z^{-1}t^{\frac{1}{2}}q^{\frac{1}{2}};q^2)}
  \frac{\prod_{a,b}(e^{\bar{u}_{ab}}tq;q^2)}
  {\prod_{a\neq b}(e^{\bar{u}_{ab}}q^{2};q^2)}\ .\nonumber
\end{eqnarray}
Here, we have formally separated the integrands into $u$ dependent parts
and $\bar{u}$ dependent parts. Note that, with complex $\beta$ (which will play
crucial roles later in this paper), $u_a$ and $\bar{u}_a$ are not complex conjugate
to each other. As we took $\beta\rightarrow 0$ limit to
make the continuum approximation for the summation of $m_a$, the q-Pochhammer symbols
appearing in the integrand should also be approximated to dilogarithm functions
as follows:
\begin{equation}
  (xq^a;q^2)\stackrel{\beta\rightarrow 0}{\longrightarrow}
  \exp\left[-\frac{{\rm Li}_2(x)}{2\beta}\right]\ .
\end{equation}
We shall seek for the saddle points of $u_a,\bar{u}_a$ which will approximate
the integral in the $\beta\rightarrow 0$ limit. While seeking for the saddle points,
one can separately consider the saddle points for $u_a$, $\bar{u}_a$ independently,
since the integrand factorizes.  During this course,
whenever any of the $x$ variables appearing in ${\rm Li}_2$ functions are larger
than $1$, i.e. $|x|>1$, analytic continuations are made for
those ${\rm Li}_2(x)$ functions.
Whenever $|x|$ is greater than $1$, one would have to worry about
the branch cut issues of ${\rm Li}_2(x)$ after making the analytic continuations.
This issue will be treated later when we discuss concrete problems.
(However, `branch cuts' here should always be understood as singularities of the Cardy
free energy rather than signaling multi-valued functions.)

Here, note that the first line of (\ref{semi-factorization}) is the
$\beta\rightarrow 0$ limit of the vortex partition function on $D_2\times S^1$,
considered in section 2, corresponding to the vertical Young diagram $(1^N)$.
The holomorphic integrand is given by the exponential of
\begin{eqnarray}\label{cardy-integrand-1}
  &&\frac{1}{2\beta}\left[-\xi\sum_a u_a+\sum_a\left(
  {\rm Li}_2(e^{u_a}t^{\frac{1}{2}})-{\rm Li}_2(e^{u_a}t^{-\frac{1}{2}})\right)
  \right.\\
  &&\hspace{1cm}\left.+\sum_{a,b}\left({\rm Li}_2(e^{u_{ab}}zt^{\frac{1}{2}})
  -{\rm Li}_2(e^{u_{ab}}zt^{-\frac{1}{2}})+{\rm Li}_2(e^{u_{ab}}t^{-1})\right)
  -\sum_{a\neq b}{\rm Li}_2(e^{u_{ab}})
  \right]\ .
  \nonumber
\end{eqnarray}
On the other hand, the $\beta\rightarrow 0$ limit of the integrand on the
second line of (\ref{semi-factorization}) can be obtained from
(\ref{cardy-integrand-1}) by flipping
$(\beta,\xi)\rightarrow(-\beta,-\xi)$ and
$(t,z)\rightarrow(t^{-1},z^{-1})$. This is the same as the Cardy limit
of the anti-vortex partition function of section 2. Therefore, at least in
the Cardy limit,
the two factors in (\ref{semi-factorization}) can be interpreted as the
vortex-anti-vortex factorization which refers to a particular point in the
Higgs branch (corresponding to the vertical Young diagram).
In particular, we have shown
that the particular vortex partition function chosen in
section 2.1 will provide the Cardy saddle point of the index on $S^2\times S^1$,
which is not clear at all in the factorization formula of section 2.2.

Note that, after the factorization, the periodicities (\ref{SO8-period}) of
the four chemical potentials $\Delta_I$ are not manifest in each integrand.
Therefore, when we study the Cardy (and large $N$) limits in the next section,
we shall first make a suitable period shifts of $\Delta_I$'s to bring them
into a canonical chamber, and then factorize using the setup of this section.

\section{Cardy limit: results}

In this section, we study the Cardy limit of our index on $S^2\times S^1$.
We shall discuss in sections 4.1 and 4.3 the large $N$ and Cardy limits
for our $\mathcal{N}=4$ Yang-Mills theory and the ABJM theory, respectively.
In section 4.2, we study the finite $N$ Cardy limit.
The Cardy limit is defined as
$\beta\rightarrow 0$ with other chemical potentials (e.g. $\Delta_I$'s)
imaginary and finite \cite{Choi:2018hmj}.

\subsection{Large $N$ Cardy free energy and black holes}

In this subsection, we study the large $N$ free energy of the index on
$S^2\times S^1$ in the Cardy limit. In section 3, we have seen its connection
to the partition function $Z_{D_2\times S^1}$ on
$D_2\times S^1\sim\mathbb{R}^2\times S^1$ at a particular
point on the Higgs branch.

The holomorphic factorization of section 3 obscures
the periodicities of chemical potentials if one pays attention to the holomorphic
factor only. So before performing the factorization of section 3, we should
first specify the ranges of the imaginary parts of $\xi,T,f$. (Recall that
$t=e^T$, $z=e^f$, $2\pi r\zeta=\frac{\xi}{2\beta}$.)
Note that these three variables are in the natural convention of the vortex
partition function of section 2. Especially, $\xi$ is related to the fugacity
of the topological charge (on $S^2\times S^1$) by
$\hat{Q}\equiv e^{-\hat\xi}=q^{1/2}t^{-1/2}Q=e^{-\xi-T/2-\beta/2}$.
Without losing generality, we take
\begin{eqnarray}\label{range-convenient}
  &&2\pi p_1<{\rm Im}(\xi)<2\pi(p_1+1)\ ,\ \
  2\pi p_2<{\rm Im}(T)<2\pi(p_2+1)\\
  &&2\pi p_3<{\rm Im}\left(f-\frac{T}{2}\right)<2\pi(p_3+1)\ ,\ \
  2\pi p_4<{\rm Im}\left(-f-\frac{T}{2}\right)<2\pi(p_4+1)\nonumber
\end{eqnarray}
for certain integers $p_1,\cdots,p_4$. (It will be convenient later to set
ranges as above.)
Although we gave ranges to the imaginary parts of chemical potentials, they
are (approximately) pure imaginary in the Cardy limit.
This is because, as we take $\beta\rightarrow 0$ with ${\rm Re}(\beta)>0$,
(\ref{Delta-properties}) demands ${\rm Re}(\Delta_I)\rightarrow 0^+$
for all $I$'s. Adding the last three
inequalities of (\ref{range-convenient}), one obtains
\begin{equation}\label{case-p}
  2\pi (p_1+p_2+p_3)<0<2\pi(p_2+p_3+p_4+3)\ \rightarrow\
  p_2+p_3+p_4=-1,-2\ .
\end{equation}
We also recall the periodicities of these variables that we explained in section 3.
Since we are now taking the Cardy limit $\beta\rightarrow 0$, we collect the
periodic shifts which leave small $\beta$ invariant:
\begin{equation}
  (\xi,T,f)\sim(\xi+2\pi i,T,f)\sim(\xi,T,f+2\pi i)
  \sim(\xi,T+2\pi i,f\pm \pi i)\ .
\end{equation}
In terms of the variables $\xi,T,f-\frac{T}{2},-f-\frac{T}{2}$, the four shifts
above are rewritten as
\begin{eqnarray}\label{shift-beta-invariant}
  \hspace*{-1cm}&&\left(\xi,T,f-\frac{T}{2},-f-\frac{T}{2}\right)\sim
  \left(\xi+2\pi i,T,f-\frac{T}{2},-f-\frac{T}{2}\right)\sim
  \left(\xi,T+2\pi i,f-\frac{T}{2}-2\pi i,-f-\frac{T}{2}\right)\nonumber\\
  \hspace*{-1cm}&&\sim\left(\xi,T+2\pi i,f-\frac{T}{2},-f-\frac{T}{2}-2\pi i\right)
  \sim\left(\xi,T,f-\frac{T}{2}+2\pi i,-f-\frac{T}{2}-2\pi i\right)\ .
\end{eqnarray}
Using these four period shifts, one can set $p_1,p_2,p_3,p_4$ to be one of the two
cases:
\begin{eqnarray}\label{two-cases-p}
  \textrm{region I}&:&p_1=-1\ ,\ \ p_2=-1\ ,\ \ p_3=0\ ,\ \ p_4=0\nonumber\\
  \textrm{region II}&:&p_1=0\ ,\ \ p_2=0\ ,\ \ p_3=-1\ ,\ \ p_3=-1\ .
\end{eqnarray}
Notice that, each of the last three shifts in (\ref{shift-beta-invariant}) picks
a pair in $p_2,p_3,p_4$, and shifts this pair by $(+1,-1)$. So the case I is for
$p_2+p_3+p_4=-1$ in (\ref{case-p}), while the case II is for
$p_2+p_3+p_4=-2$. $\xi$ has its own shift symmetry in (\ref{shift-beta-invariant}),
which we have suitably set as (\ref{two-cases-p}) for later convenience.
Collecting all, it suffices to consider the two cases of (\ref{two-cases-p}) only.

Here, recall that the index on $S^2\times S^1$ is related
to that on $D_2\times S^1$ as follows:
\begin{equation}\label{factor}
  \lim_{\beta\rightarrow 0}Z_{S^2\times S^1}(\beta,\hat\xi,f,T)\sim
  \lim_{\beta\rightarrow 0}Z_{D_2\times S^1}(\beta,\xi,f,T)
  Z_{D_2\times S^1}(-\beta,-\xi,-f,-T)\ .
\end{equation}
From \eqref{factor}, we (formally) find the following expression
\begin{equation} \label{-branch}
Z_{S^2\times S^1}(\beta, \hat\xi, f, T) \sim
Z_{S^2\times S^1}(-\beta, -\hat\xi, -f, -T)\ ,
\end{equation}
for the free energy in the Cardy limit.
Thus, from the Cardy index in the region I of (\ref{two-cases-p}),
one can easily generate that in the region II, since the two regions
\begin{equation}
\begin{aligned}
& \textrm{I} : -2\pi < \textrm{Im} (\xi)< 0, \;  -2\pi <\textrm{Im} (T)< 0, \; 0 < \textrm{Im} \left(f-\frac{T}{2}\right) < 2\pi, \;  0 < \textrm{Im} \left(-f-\frac{T}{2}\right) < 2\pi\\
& \textrm{II} : 0 < \textrm{Im} (\xi)<2\pi, \; 0<\textrm{Im} (T)< 2\pi, \; -2\pi < \textrm{Im} \left(f-\frac{T}{2}\right) < 0, \;  -2\pi < \textrm{Im} \left(-f-\frac{T}{2}\right) < 0
\end{aligned}
\end{equation}
are related to each other by the sign flips of
$(\beta,\xi,f,T)$. So from now on, we focus on the calculations in region I.
Then, from \eqref{factor}, in order to obtain the Cardy
index on $S^2\times S^1$ in region I, one should compute two Cardy indices on
$D_2 \times S^1$.
However, we need not compute them independently. To see this, first note that
the Cardy limit of the latter index takes the form of
$\log Z_{D_2\times S^1}\sim -\frac{\mathcal{F}(\xi,f,T)}{2\beta}$.
Now consider the complex conjugation of this free energy. By definition of this index,
which traces over the Hilbert space with integer coefficients and real charges,
the complex conjugated free energy can be obtained by simply complex conjugating
the chemical potentials. So one obtains
\begin{equation} \label{conjugate}
\overline{\log Z_{D_2\times S^1}(\beta, \xi, f, T)}
\sim-\frac{\mathcal{F}(\bar\xi,\bar{f},\bar {T})}{2\bar\beta}
\sim \log Z_{D_2\times S^1}(\bar\beta, -\xi, -f, -T)\ .
\end{equation}
At the last step, we used the fact that $\xi,f,T$ are all imaginary in our Cardy limit.
Therefore, the nontrivial part $\mathcal{F}(-\xi,-f,-T)$ of
$Z_{D_2\times S^1}(-\beta,-\xi,-f,-T)$ in (\ref{factor})
can be obtained once we compute $\mathcal{F}(\xi,f,T)$ in region I.

We compute the large $N$ and Cardy limit of
$\log Z_{D_2\times S^1}(\beta,\xi,f,T)$ in region I.
The Cardy limit $\beta \to 0^+$ of $Z_{D_2 \times S^1}$
can be evaluated by the saddle point method as
\begin{equation} \label{asyint}
Z_{D_2\times S^1} \sim
\exp \left(-\frac{1}{2\beta} \mathcal{W}^* \right)\ ,
\end{equation}
with $\mathcal{W}$ given by
\begin{equation} \label{eff-action}
\begin{aligned}
\hspace*{-1cm}
\mathcal{W} = &\,N \Big(\textrm{Li}_2(z t^{-1/2}q^{1/2}) - \textrm{Li}_2(z t^{1/2}q^{-1/2}) - \textrm{Li}_2(t^{-1}) \Big)
+ \displaystyle \sum_{a=1}^N \Big( \xi \log s_a + \textrm{Li}_2(s_a t^{-1/2}q^{1/2}) - \textrm{Li}_2(s_a t^{1/2}q^{-1/2}) \Big) \\
\hspace*{-1cm}
&+ \displaystyle \sum_{1 \leq a \neq b \leq N} \Big(\textrm{Li}_2(s_a {s_b}^{-1}q^{-1}) - \textrm{Li}_2(s_a {s_b}^{-1} t^{-1}) + \textrm{Li}_2(s_a {s_b}^{-1} z t^{-1/2}q^{1/2}) - \textrm{Li}_2(s_a {s_b}^{-1} z t^{1/2}q^{-1/2}) \Big)\ .
\end{aligned}
\end{equation}
Here, we used the asymptotic formula of the $q$-Pochhammer symbol \eqref{poch}.
$\mathcal{W}^*$ denotes the saddle point value of $\mathcal{W}$.
Saddle point equations are given by (no summation for $a$):
\begin{equation}\label{bethe}
\begin{aligned}
s_a \partial_{s_a} \mathcal{W} = &\;\xi + \textrm{Li}_1(s_a t^{-1/2}q^{1/2}) - \textrm{Li}_1(s_a t^{1/2}q^{-1/2}) + \displaystyle \sum_{b \neq a} \Big[\textrm{Li}_1(s_a {s_b}^{-1}q^{-1}) - \textrm{Li}_1(s_b {s_a}^{-1}q^{-1}) \\
& \qquad \, - \textrm{Li}_1(s_a {s_b}^{-1} t^{-1}) + \textrm{Li}_1(s_b {s_a}^{-1} t^{-1}) + \textrm{Li}_1(s_a {s_b}^{-1} z t^{-1/2}q^{1/2}) \\
& \qquad \, - \textrm{Li}_1(s_b {s_a}^{-1} z t^{-1/2}q^{1/2}) - \textrm{Li}_1(s_a {s_b}^{-1} z t^{1/2}q^{-1/2}) + \textrm{Li}_1(s_b {s_a}^{-1} z t^{1/2}q^{-1/2}) \Big]=0.
\end{aligned}
\end{equation}
Note that $s_a=0$ is a fake solution since the original equations $\partial_{s_a} \mathcal{W} = 0$ have $1/s_a$ factors. By redefining parameters and exponentiating both sides, one can see that the above saddle point equations \eqref{bethe} take the
form of the Bethe ansatz equations
\cite{Yoshida:2014ssa}.\footnote{In (\ref{eff-action}) and (\ref{bethe}), we have
no essential need to keep
$q\rightarrow 1^-$ in our Cardy limit. In fact we shall
insert $q=1$ in these formulae shortly, except that we temporarily
need $q^{-1}$ factors for the terms
${\rm Li}_2(s_as_b^{-1}q^{-1})$ and ${\rm Li}_1(s_as_b^{-1}q^{-1})$,
as natural regulators to keep the saddle point slightly
away from the branch cuts.}
Finally, combining \eqref{factor}, \eqref{conjugate}, \eqref{asyint}, one obtains
\begin{equation} \label{Cardy-index}
\begin{aligned}
\log Z_{S^2\times S^1}(\beta, \hat\xi, f, T) & \sim
-\frac{\mathcal{W}^*(\xi, f, T)-\overline{\mathcal{W}^*(\xi, f, T)}|_{(\bar\xi,\bar{f},\bar{T})=(-\xi,-f,-T)}}{2\beta} \\
& = -\frac{2i\,\textrm{Im} \left[\mathcal{W}^*(\xi, f, T)\right]|_{(\bar\xi,\bar{f},\bar{T})=(-\xi,-f,-T)}}{2\beta}\ ,
\end{aligned}
\end{equation}
where $\xi, f, T$ are taken to be pure imaginary while taking complex
conjugations.

We now analytically
find the relevant solution of \eqref{bethe}. We will basically follow the
procedures used in \cite{Herzog:2010hf}.
Based on the discussions made so far, we consider the region I of (\ref{two-cases-p}),
\begin{equation}
  -2\pi < \textrm{Im} (\xi)<0, \; -2\pi<\textrm{Im} (T)< 0, \; 0 < \textrm{Im} \left(f-\frac{T}{2}\right) < 2\pi, \;  0 < \textrm{Im} \left(-f-\frac{T}{2}\right) < 2\pi\ ,
\end{equation}
where $\xi, T, f$ are imaginary. Our ansatz for the eigenvalue distribution is given by
\begin{equation} \label{ansatz}
s_a=s_0 e^{N^{\alpha} x_{(a)}+i y(x_{(a)})} \quad (x_1 \leq x_{(a)} \leq x_2)  \ ,
\end{equation}
where $s_0>0$ is a positive real constant. Here $x_{(a)}$ and $y(x)$
are real, which we take to be at $\mathcal{O}(N^0)$.
We introduced a factor $N^{\alpha}$ with $0<\alpha<1$.
The constant $\alpha$ will be determined later. Also, we assumed that the eigenvalues are distributed in $[x_1, x_2]$ for some $x_1<x_2$. Then, we
introduce the continuum variable $x_{(a)}\rightarrow x$
assuming that we ordered the eigenvalues to make $x$
to be an increasing function of $a$. This particular ordering cancels out the Weyl factor $N!$. In addition, we introduce the density function of the eigenvalues as $\rho(x)=\frac{1}{N}\frac{da}{dx}$. Here, we further assume a connected distribution of eigenvalues where $\rho$ is always positive in $(x_1, x_2)$.

In this setting, we first take the continuum limit of $\mathcal{W}$. We
will only consider the leading contribution at small $\beta$, plugging in
$q=1$ in (\ref{eff-action}) and (\ref{bethe}). $\mathcal{W}$ can be divided into two parts, $\mathcal{W}=\mathcal{W}_{ext}+\mathcal{W}_{int}$. $\mathcal{W}_{ext}$ denotes the contribution from the external potential:
\begin{equation}
\begin{aligned}
\mathcal{W}_{ext}&=N \int_{x_1}^{x_2} dx \rho (x) \Big(  \xi \log s(x) + \textrm{Li}_2(s(x) t^{-1/2}) - \textrm{Li}_2(s(x) t^{1/2})  \Big)\ .
\end{aligned}
\end{equation}
$\mathcal{W}_{int}$ comes from the interactions of
eigenvalue pairs:
\begin{align}
\mathcal{W}_{int}=& N^2 \int_{x_1}^{x_2}  dx \rho(x) \int_x^{x_2} dx' \rho(x') \Big(\textrm{Li}_2(s(x) {s(x')}^{-1}) + \textrm{Li}_2(s(x') {s(x)}^{-1}) \nonumber\\
&- \textrm{Li}_2(s(x) {s(x')}^{-1} t^{-1}) - \textrm{Li}_2(s(x') {s(x)}^{-1} t^{-1}) + \textrm{Li}_2(s(x) {s(x')}^{-1} z t^{-1/2}) \nonumber\\
&+ \textrm{Li}_2(s(x') {s(x)}^{-1} z t^{-1/2}) - \textrm{Li}_2(s(x) {s(x')}^{-1} z t^{1/2}) - \textrm{Li}_2(s(x') {s(x)}^{-1} z t^{1/2}) \Big)\ .
\end{align}
The main strategy to extract the leading order contribution at in large $N$
is to use the following integral formula \cite{Benini:2015eyy}:
\begin{equation}\label{short-range-formula}
\begin{aligned}
\int_0^{x>0} dx \rho(x) \textrm{Li}_s (e^{-N^{\alpha}x +iy(x)}) \; &=
\int_0^x dx \rho(x) \sum_{n=1}^\infty \frac{e^{n\left(-N^\alpha x +iy(x)\right)}}{n^s} \\
&= \sum_{n=0}^\infty \frac{1}{n^s} \left[ \left. - \rho(x) e^{iny(x)} \frac{e^{-nN^\alpha x}}{nN^\alpha} \right|_0^x + \int_0^x dx (\rho(x) e^{iny(x)})' \frac{e^{-nN^\alpha x}}{nN^\alpha} \right] \\
&= N^{-\alpha} \rho(0) \textrm{Li}_{s+1} (e^{iy(0)}) +O(N^{-2\alpha})\ ,
\end{aligned}
\end{equation}
where we used the power series definition of the polylogarithm function on the first line. One can see that the integral on the second line is suppressed by a factor of $1/N^{\alpha}$ compared to the boundary term, by performing integration by parts repeatedly. Note that we assumed $d\rho/dx, \, \left|dy/dx\right| < N^\alpha$.

Applying
\begin{eqnarray}\label{invm}
  &&{\rm Li}_n(a)+(-1)^n{\rm Li}_n(a^{-1})=-\frac{(2\pi i)^n}{n!}
  B_n\left(\frac{\log a}{2\pi i}-p\right)\ \ \ (2\pi p<{\rm Im}(\log a)<2\pi(p+1),\
  a\notin(0,1))\nonumber\\
  &&B_1(x)=x-\frac{1}{2}\ ,\ \ B_2(x)=x^2-x+\frac{1}{6}\ ,\ \
  B_3(x)=x^3-\frac{3}{2}x^2+\frac{1}{2}x\ ,\ \ \cdots\ ,
\end{eqnarray}
$\mathcal{W}_{ext}$ is approximated at large $N$ as
\begin{equation}\label{W-ext}
\begin{aligned}
&\mathcal{W}_{ext} = N^{1+\alpha} \Bigg[\xi \int_{x_1}^{x_2} dx \rho(x) x  \\
& + \left(T-2\pi i \left(\left\lfloor \frac{2y(x)+\textrm{Im}(T)}{4\pi}\right\rfloor -\left\lfloor \frac{2y(x)-\textrm{Im}(T)}{4\pi}\right\rfloor \right) \right) \int_{\textrm{max}(0,x_1)}^{\textrm{max}(x_2,0)} dx \rho(x) x  \Bigg] +O(N^1) \\
& \qquad \equiv  N^{1+\alpha} \Bigg[\xi \int_{x_1}^{x_2} dx \rho(x) x + \left(T-2\pi i p_2'\right) \int_{\textrm{max}(0,x_1)}^{\textrm{max}(x_2,0)} dx \rho(x) x  \Bigg] +O(N^1)
\ .
\end{aligned}
\end{equation}
Here, $\lfloor a\rfloor$ means the unique integer $n$ satisfying $n\leq a<n+1$.
The last step is the definition of the integer $p_2^\prime$, whose values
will be specified in a moment.
One can see that the specific form of $y(x)$ does not affect to the leading order. Only the range of $y(x)$ contributes because it appears in the
$\lfloor \ldots \rfloor$ symbols. (Its specific form may affect the sub-leading order in $1/N$, which is not of our interest here.) As part of our extremization problem,
one should extremize $\mathcal{W}$ with respect to $y(x)$. However, it seems hard
to make a fully general extremization of the functional containing a discrete function $\lfloor \ldots \rfloor$. To further manipulate, we assume that $y(x)$ does not pass across the branch cuts which cause the discrete jumps. One can regard it as part
of our ansatz.
There are two branch cuts from two $\lfloor \ldots \rfloor$'s. Hence, we should demand $y(x)$ to be within a specific region bounded by the two branch cuts.
There are two possible regions, with the following values of
$p_2^\prime$:
\begin{eqnarray} \label{branch-con}
&(\textrm{i})& \; \frac{\textrm{Im} (T)}{2}  <y(x)< -\frac{\textrm{Im} (T)}{2} \quad (\textrm{mod } 2\pi) \ :\ \  p_2' = -1\ , \nonumber\\
&(\textrm{ii})&\; - \frac{\textrm{Im} (T)}{2} <y(x)<2\pi + \frac{\textrm{Im} (T)}{2}  \quad (\textrm{mod } 2\pi) \ :\ \  p_2' =0\ .
\end{eqnarray}
Later, we will determine which case yields non-trivial large $N$ solutions.

We then consider the large $N$ approximation of $\mathcal{W}_{int}$. Again to
simplify the manipulations after using (\ref{invm}), (\ref{short-range-formula}),
we assume that $y(x')-y(x)$ at $x'>x$ does not pass across the branch cuts.
In particular, during this manipulation, one apparently encounters terms at order $O(N^{2+\alpha})$ whose coefficient is nonzero unless
\begin{equation}
  \left\lfloor \frac{y(x')-y(x)+\textrm{Im} (\beta)}{2\pi} \right\rfloor=-1
  \ \ \ (x'>x)\ .
\end{equation}
(Here, we restored the subleading $\mathcal{O}(\beta)$ correction
by not strictly plugging in $q=1$ in (\ref{eff-action}), which is a
convenient and natural regularization.)
As we just started from a QFT with $N^2$ degrees of freedom, There will not
be physical saddle points whose free energies scale like $N^{2+\alpha}$. So
we impose the condition above on $y(x)$. There are other conditions
for $y(x)$ so that no branch cuts are crossed at all. Collecting them all,
one obtains the following conditions for $x^\prime > x$:
\begin{eqnarray}\label{y-sol-preliminary}
&&-2\pi<y(x')-y(x)+\textrm{Im} (\beta) <0\\
&&0<y(x^\prime)-y(x) - \textrm{Im} (T)<2\pi\nonumber\\
&&0<y(x^\prime)-y(x) + \textrm{Im} (f-T/2)<2\pi\nonumber\\
&&-2\pi<y(x^\prime)-y(x) + \textrm{Im} (f+T/2)<0\ .\nonumber
\end{eqnarray}
Here we quote a result that it we shall eventually pay attention to
small $\beta$ satisfying ${\rm Im}(\beta)<0$.
This is because, once we compute the free energy and go back to
the microcanonical ensemble by the Legendre transformation,
the dominant saddle point of our interest will always satisfy
${\rm Im}(\beta)<0$. This is basically the result of \cite{Choi:2018fdc},
which we shall briefly review later in this subsection. With this assumed in
foresight, the right inequality of the first line of (\ref{y-sol-preliminary})
says that $y(x)$ is a non-increasing function, i.e.
$y(x^\prime)-y(x)\leq 0$ at $x^\prime >x$. Here, the equality in
$\leq$ is allowed because of the regularization with ${\rm Im}(\beta)<0$.
With this non-increasing property assumed, all the right inequalities of
the second, third, fourth lines of (\ref{y-sol-preliminary}) are automatically
satisfied. Also, the left inequality on the first line of (\ref{y-sol-preliminary}) is
a consequence of the left inequality on the fourth line.
Finally, the left inequalities of the second, third, fourth lines
take the form of $y(x^\prime)-y(x)>A$
with negative real numbers $A$. With $y(x)$ being a
non-increasing function in the interval $(x_1,x_2)$,
such a condition is equivalent to $y(x_2)-y(x_1)>A$, since $y(x_2)-y(x_1)$ is
the minimum of $y(x^\prime)-y(x)$. So collecting all,
(\ref{y-sol-preliminary}) can be rephrased as
\begin{eqnarray}\label{y-sol}
&&y(x')-y(x)\leq 0 \ \ \ (\textrm{for }x'>x)\ ,\ \
y(x_2)-y(x_1) - \textrm{Im} (T)>0\ ,\\
&&y(x_2)-y(x_1) + \textrm{Im} (f-T/2) >0\ ,\ \
y(x_2)-y(x_1) + \textrm{Im} (f+T/2)>-2\pi\ .\nonumber
\end{eqnarray}
A particularly important possibility for $y(x)$ would be
\begin{equation} \label{y-sol2}
y(x) = \textrm{constant}\ .
\end{equation}
Indeed, in the next subsection, we will numerically see that the Cardy saddle
point solutions satisfy $y(x)=0$ at arbitrary finite $N$.
With the conditions \eqref{y-sol}, one obtains the following result for
$\mathcal{W}_{\rm int}$ after some calculations:
\begin{equation}\label{W-int}
\begin{aligned}
\mathcal{W}_{int}&=-\frac{N^2}{2}\left(T\!+\!2\pi i\right)\left(\!-f-\!\frac{T}{2}\right)
 \!-\!\frac{N^{2-\alpha}}{2}\left(T\!+\!2\pi i\right)
 \left(\!f-\frac{T}{2} \right)\left(\!-f-\frac{T}{2}\right)
 \int_{x_1}^{x_2}\!\! dx \rho(x)^2 + O(N^{2-2\alpha})\ .
\end{aligned}
\end{equation}
Here, we used $\int_{x_1}^{x_2} \rho(x)dx  =1$.
$\mathcal{W}_{\rm int}$ in (\ref{W-int}) shows
short-ranged interactions only between nearby eigenvalues.
One can see again that the specific form of $y(x)$ does not matter at the leading order. Since we are only interested in the leading free energy in $N$, we will not care about $y(x)$ below.

As a side remark before proceeding, we comment on
the first term of (\ref{W-int}) proportional to $N^2$,
which does not depend on the eigenvalue distribution $\rho(x)$. The terms in $\mathcal{W}_{\rm ext}$, $\mathcal{W}_{\rm int}$
which depend on $\rho(x)$ will be soon extremized below at $\alpha=\frac{1}{2}$,
with the expected
$N^{\frac{3}{2}}$ scaling for M2-branes. However, the first term of (\ref{W-int})
proportional to $N^2$ might apparently look contradictory to the expected
M2-brane behaviors. Here, we note that there is a very natural interpretation of
such a term in the context of of the partition function on $D_2\times S^1$.
Namely, if one considers a 3d QFT on $D_2\times \mathbb{R}$ or $D_2\times S^1$,
boundary chiral anomalies may be induced on $S^1\times \mathbb{R}$ or $T^2$.
We chose the boundary conditions so that the $U(N)$ gauge anomaly is canceled.
But there are boundary 't Hooft anomalies for the global symmetries which are
probed by the chemical potentials $T,f$. Since these boundary anomalies are
proportional to $N^2$, the spectrum on $D_2\times \mathbb{R}$ should contain
such light degrees of freedom at the boundary. So even if
the bulk physics would only see $N^{\frac{3}{2}}$ degrees of freedom,
$\log Z_{D_2\times S^1}$ will see certain terms at $N^2$ order. This is
our interpretation of the first term of (\ref{W-int}). If
one combines two vortex partition functions to make an index on $S^2\times S^1$
without any boundary using (\ref{Cardy-index}), the two terms proportional to
$N^2$ indeed cancel,
\begin{equation}
  -\frac{N^2}{2}(T+2\pi i)\left(-f-\frac{T}{2}\right)
  -(\textrm{complex conjugate})=0\ .
\end{equation}
This is consistent with our interpretation. Also, note that we have
no terms scaling like $N^2$ in (\ref{W-int}) which depend on the dynamical
gauge holonomy $x$ (and accordingly not $\rho(x)$). This is because our QFT in
section 2 has no boundary gauge anomaly. On the
other hand, as commented briefly in footnote 2, p.7, we found it quite tricky
(if not impossible) to provide simple boundary conditions for the
ABJM theory without gauge anomaly. This will make the large $N$ calculus
very difficult. In section 4.3, we will introduce a rather ugly factorization for
the ABJM index which breaks the $U(N)\times U(N)$ gauge symmetry,
to circumvent this problem.

Collecting (\ref{W-ext}) and (\ref{W-int}), one obtains
\begin{equation}
\begin{aligned}
\mathcal{W}\sim \, &N^{1+\alpha} \left[\xi \int_{x_1}^{x_2} dx \rho(x) x + \left(T-2\pi i p_2'\right) \int_{\textrm{max}(0,x_1)}^{\textrm{max}(x_2,0)} dx \rho(x) x  \right] \\
& - \frac{N^{2-\alpha}}{2}\left(T+2\pi i\right)\left(f-\frac{T}{2} \right)\left(-f-\frac{T}{2}\right) \int_{x_1}^{x_2} dx \rho(x)^2 + \mathcal{W}_0
\end{aligned}
\end{equation}
with $\mathcal{W}_0\equiv -\frac{N^2}{2}\left(T+2\pi i\right)\left(-f-\frac{T}{2}\right)$,
where $p_2'$ is either $-1$ or $0$, as shown in (\ref{branch-con}).
$\mathcal{W}_0$ can be ignored during our extremization problem. We extremize
$\mathcal{W}$ with $\rho(x)$ in the set
$\mathcal{C}=\Big\{\rho|\int_{x_1}^{x_2} \rho (x) dx =1 ; \; \rho(x) \geq 0 \; \textrm{pointwise} \Big\}$.
As $N \to \infty$, in order to get nontrivial solutions, $\mathcal{W}_{ext}$ and  $\mathcal{W}_{int}-\mathcal{W}_0$ should be at the same order in $N$.
So we will now set $\alpha = \frac{1}{2}$.
Introducing the Lagrange multiplier $\lambda$, we extremize the following functional,
where $\hat{\mathcal{W}}\equiv\mathcal{W}-\mathcal{W}_0$:
\begin{eqnarray}
\frac{\hat{\mathcal{W}}}{N^{\frac{3}{2}}}&= & \xi \int_{\textrm{min}(x_1,0)}^{\textrm{min}(0,x_2)} dx \rho(x) x    + \left(\xi+T-2\pi i p_2'\right) \int_{\textrm{max}(0,x_1)}^{\textrm{max}(x_2,0)} dx \rho(x) x  \\
&&-  \frac{1}{2}\left(T+2\pi i\right)\left(f-\frac{T}{2} \right)\left(-f-\frac{T}{2}\right) \int_{x_1}^{x_2} dx \rho(x)^2
 +\lambda \left(\int_{x_1}^{x_2} dx \rho(x) -1\right)\ .\nonumber
\end{eqnarray}
When $x_1 \leq x_2 \leq 0$ or $0 \leq x_1 \leq x_2$, one can see that
there are no solutions for $x_1, x_2$ extremizing $\hat{\mathcal{W}}$.
Thus, a non-trivial saddle point only exists when $x_1\leq 0 \leq x_2$.
In the last case, the extremal $\rho(x)$ is given by
\begin{equation}
\rho(x) = \left\{
\begin{array}{ll}
\frac{4\lambda + 4\xi x}{(T+2\pi i)\left(2f-T \right)\left(-2f-T\right)}\ ,
& x_1 \leq x \leq 0 \\
\frac{4\lambda+ 4(\xi+T-2\pi i p_2') x}{(T+2\pi i)\left(2f-T \right)\left(-2f-T\right)}\ ,  &  0 \leq x \leq x_2
\end{array}
\right.\ .
\end{equation}
From the normalization condition $\int_{x_1}^{x_2} \rho (x) dx = 1$, the Lagrange multiplier $\lambda$ is given by
\begin{equation}
\lambda = \frac{(T+2\pi i)\left(2f-T \right)\left(-2f-T\right) -2(\xi+T-2\pi i p_2')x_2^2 +2\xi x_1^2}{4(x_2-x_1)}\ .
\end{equation}
With pure imaginary $\xi, f, T$, $\rho(x)$ is automatically a real function.
Inserting the above $\rho (x)$ and $\lambda$ back to $\hat{\mathcal{W}}$, one obtains
\begin{equation}
\begin{aligned}
&\hat{\mathcal{W}}= N^{\frac{3}{2}} \frac{-12 \gamma^2+12\gamma(T' x_2^2-\xi x_1^2) + \xi^2 x_1^4 + T'^2x_2^4 - 4 x_1 x_2 (\xi^2 x_1^2 + T'^2 x_2^2)+6\xi T' x_1^2 x_2^2}{24 \gamma (x_2-x_1)}\ ,
\end{aligned}
\end{equation}
where $T' \equiv \xi + T -2\pi i p_2', \; \gamma \equiv (T+2\pi i)\left(f-\frac{T}{2}\right)\left(-f-\frac{T}{2}\right)$. Then, differentiating the above $\hat{\mathcal{W}}$ with $x_1, x_2$, the extremal $x_1,x_2$ satisfies
\begin{eqnarray}\label{extreme-x12}
  x_1 x_2&=& \frac{(T+2\pi i)\left(2f-T \right)\left(-2f-T\right)}{2(T-2\pi i p_2')} <0\\
  x_1^2&=&\frac{(T+2\pi i)\left(2f-T \right)\left(-2f-T\right)(\xi+T-2\pi i p_2')}{2\xi(T-2\pi i p_2')} >0,\nonumber\\
  x_2^2&=&\frac{(T+2\pi i)\left(2f-T \right)\left(-2f-T\right)\xi}{2(\xi+T-2\pi i p_2')(T-2\pi i p_2')} >0\ .\nonumber
\end{eqnarray}
The first condition is compatible with the product of last two, and we have
been careful so far not to make any square roots.
Here, negativity of $x_1x_2$ demands $p_2^\prime=-1$,
so that one should choose the case (i) of (\ref{branch-con}).
Also, the positivity of $x_1^2$, $x_2^2$ demands
$-2\pi<{\rm Im}(\xi+T)<0$. (Its range was originally
$-4\pi<{\rm Im}(\xi+T)<0$.) Otherwise, we do not find any large $N$ Cardy saddle
point for $s_a$'s in the region I of (\ref{two-cases-p}).
In this set up, non-negativity of $\rho(x)$ is guaranteed
in the whole region $(x_1,x_2)$. In particular, one finds
$\rho(x_1) = \rho(x_2) = 0$ at this saddle point.

Inserting the above saddle point solution,
the extremal value of $\mathcal{W}$ is given by
\begin{equation}\label{W-saddle}
\mathcal{W}^*\sim - N^{\frac{3}{2}} \frac{1}{3} \frac{(T+2\pi i)\left(2f-T \right)\left(-2f-T\right)}{x_2-x_1} + \mathcal{W}_0\ .
\end{equation}
We took no square-roots so far to avoid branch ambiguities. We now explain
this final step. One should simply remember that, while taking the square roots
of the expressions for $x_1^2$, $x_2^2$ in (\ref{extreme-x12}),
one takes the negative
root for $x_1$ and positive root for $x_2$. The final result can be phrased
in a simple manner by recalling the allowed ranges of chemical potentials,
\begin{equation} \label{D2-range}
\begin{aligned}
0<\textrm{Im}(-\xi), \, \textrm{Im} (\xi+T+2\pi i), \, \textrm{Im} \left(f-\frac{T}{2}\right), \, \textrm{Im} \left(-f-\frac{T}{2}\right)<2\pi\ .
\end{aligned}
\end{equation}
Especially, all expressions appearing in ${\rm Im}$ above are $i$ times positive numbers.
After plugging in the values of $x_1,x_2$ in (\ref{W-saddle}),
one obtains
\begin{equation} \label{D2-W}
\mathcal{W}^* \sim i \; \frac{2\sqrt{2} N^{\frac{3}{2}}}{3} \sqrt{(-\xi)(\xi+T+2\pi i)\left(f-\frac{T}{2}\right) \left(-f-\frac{T}{2}\right)}-\frac{N^2}{2}\left(T+2\pi i\right)\left(-f-\frac{T}{2}\right)\ .
\end{equation}
Here, the expression appearing in the square-root is the product of the
four numbers appearing in (\ref{D2-range}), where each of them is $i$ times
a positive real number in the Cardy limit. So the product of them is real and
positive. Our convention for the formulae involving square-roots,
starting from (\ref{D2-W}), is to take square roots of positive numbers only,
and to take the positive root. This applies to all our formulae below for
the free energies in the Cardy limit. Sometimes our formulae are used in
the non-Cardy regime, e.g. in \cite{Choi:2018fdc} to discuss dual AdS$_4$
black holes. In this case, one takes the unique root which reduces to the
positive root in the Cardy limit.\footnote{Equivalently but more concretely,
the rule for taking the square root $\sqrt{z}$ of a complex number $z$ in the
free energy of \cite{Choi:2018fdc} is to take $z$ in the principal branch
$-\pi<{\rm Arg}(z)<\pi$.}
Consequentially, the free energy
$\log Z_{D_2\times S^1}\sim-\frac{\mathcal{W}^\ast}{2\beta}$ is given by
\begin{eqnarray} \label{D2-index}
\log Z_{D_2\times S^1}&\!\sim\!& - i \, \frac{\sqrt{2} N^{\frac{3}{2}}}{3\beta} \sqrt{(-\xi)(\xi+T+2\pi i)\left(f-\frac{T}{2}\right) \left(-f-\frac{T}{2}\right)} + \frac{N^2}{4\beta}\left(T+2\pi i\right)\left(-f-\frac{T}{2}\right) \nonumber\\
&\!\equiv\!& - i \, \frac{\sqrt{2} N^{\frac{3}{2}}}{3\beta}
\sqrt{\left(\!-\hat\xi\!+\!\frac{T}{2}\right)\left(\hat\xi+\frac{T}{2}\!
 + 2\pi i\!\right)\left(f-\frac{T}{2}\right) \left(\!-f-\frac{T}{2}\right)}
+ \log Z_0\ ,
\end{eqnarray}
where $\hat\xi=\xi + \frac{T}{2} + \frac{\beta}{2},\; Z_0 \equiv e^{-\frac{\mathcal{W}_0}{2\beta}}$.

Based on the studies made on $Z_{D_2\times S^1}$, we now compute the
large $N$ and Cardy limit of the index on $S^2\times S^1$, using (\ref{Cardy-index}).
Recall that in this formula, we consider the imaginary part of $\mathcal{W}^*$ in \eqref{D2-W} at pure imaginary $\xi, f, T$. Using \eqref{D2-range}, the first term in \eqref{D2-W} is pure imaginary, while the second term is purely real. So
multiplying two $Z_{D^2 \times S^1}$'s in (\ref{Cardy-index}), $O(N^{\frac{3}{2}})$ term is doubled, while $O(N^2)$ term is canceled.
In fact at this stage, we can present both results in the two regions I and II
as defined in (\ref{two-cases-p}). The large $N$ Cardy free energies of
$Z_{S^2\times S^1}$ in the two cases are given by
\begin{equation} \label{free+}
\begin{aligned}
\log Z_{S^2\times S^1} (\beta, \hat\xi, f, T) & \sim \mp 2i \, \frac{\sqrt{2} N^{\frac{3}{2}}}{3\beta} \sqrt{\left(-\hat\xi+\frac{T}{2}\right)
\left(\hat\xi+\frac{T}{2} \pm 2\pi i\right)\left(f-\frac{T}{2}\right) \left(-f-\frac{T}{2}\right)}\ ,
\end{aligned}
\end{equation}
where the upper/lower signs are for the region I/II, respectively. The existence
of two regions will play a rather important physical role below.
We summarize again that in the two regions, the chemical potentials satisfy
\begin{eqnarray}
 \textrm{region I}&:&
 0<\textrm{Im}\left(-\hat\xi+\frac{T}{2}\right), \, \textrm{Im} \left(\hat\xi+\frac{T}{2}+2\pi i\right), \, \textrm{Im} \left(f-\frac{T}{2}\right), \, \textrm{Im} \left(-f-\frac{T}{2}\right)<2\pi\ , \nonumber\\
 \textrm{region II}&:&
 -2\pi<\textrm{Im}\left(-\xi_{ren}+\frac{T}{2}\right), \, \textrm{Im} \left(\hat\xi+\frac{T}{2}-2\pi i\right), \, \textrm{Im} \left(f-\frac{T}{2}\right), \, \textrm{Im} \left(-f-\frac{T}{2}\right)<0\ .\nonumber
\end{eqnarray}
To see the symmetry most transparently, we use the proper $SO(8)$ basis given by
(\ref{SO8-chemical}),
\begin{equation}\label{Delta-shift}
  \Delta_1 \equiv - \hat\xi+ \frac{T}{2} + \frac{\beta}{2}\ ,\ \
  \Delta_2 \equiv \hat\xi+ \frac{T}{2} + \frac{\beta}{2}\pm 2\pi i\ ,\ \
  \Delta_3 \equiv f- \frac{T}{2} + \frac{\beta}{2}\ , \ \
  \Delta_4 \equiv -f- \frac{T}{2} + \frac{\beta}{2}\ ,
\end{equation}
in the case I and II, respectively. This is an expression valid at finite $\beta$.
Compared to (\ref{SO8-chemical}), we have only made a $\pm 2\pi i$ shift for
$\Delta_2$ in the case I/II respectively.\footnote{Note that the index
${\rm Tr}\left[(-1)^Fe^{-\Delta_2 Q_2}\cdots\right]$ can be rewritten as
${\rm Tr}\left[e^{-(\Delta_2\pm 2\pi i)Q_2}\cdots\right]$ by absorbing $(-1)^F$
by $\pm 2\pi i $ shift of $\Delta_2$. So the shifted variables are chemical potentials
in the latter convention for the index.}
These chemical potentials satisfy $0<\pm {\rm Im}(\Delta_I)<2\pi$ in the Cardy limit,
and further satisfy
\begin{equation} \label{ind-con}
\sum_{I=1}^4 \Delta_I -2\beta = \pm 2\pi i\ ,
\end{equation}
where upper/lower signs are again for the case I/II, respectively.
In the two cases, the free energy is given by
\begin{equation} \label{large-N-free}
\log Z_{S^2\times S^1} \sim
    \mp i \, \frac{4\sqrt{2} N^{\frac{3}{2}}}{3} \frac{\sqrt{\Delta_1\Delta_2\Delta_3\Delta_4}}{2\beta}\ .
\end{equation}
This finishes the derivation of our large $N$ Cardy free energy on $S^2\times S^1$.
The free energy in other chambers of ${\rm Im}\Delta_I$ can be obtained by the period
shifts of $\Delta_I$'s.

\eqref{large-N-free} describes the deconfined phase of our gauge theory as it scales like $N^{3/2}$ at large $N$. Together with \eqref{ind-con}, \eqref{large-N-free} precisely matches the entropy function of electrically charged rotating supersymmetric black holes in AdS$_4 \times S^7$ \cite{Choi:2018fdc}. Namely, \cite{Choi:2018fdc} performed the Legendre transformation, which is extremizing
\begin{equation}\label{entropy-function}
  S(Q_I,J;\Delta_I,\beta)=\log Z_{S^2\times S^1}+\sum_{I=1}^4 Q_I\Delta_I+2\beta J
\end{equation}
with $\Delta_I, \beta$, subject to \eqref{ind-con}.
Then it was shown that the resulting microcanonical entropy agrees
with the Bekenstein-Hawking entropy of the BPS black holes in $AdS_4 \times S^7$ \cite{Cvetic:2005zi}, upon inserting a charge relation satisfied by known analytic
black hole solutions. Therefore, we have statistically accounted for the microstates
of the supersymmetric AdS$_4$ black holes by deriving this free energy.

One important fact which is perhaps not emphasized
in \cite{Choi:2018fdc} is the following. As one extremizes (\ref{entropy-function}),
the dominant saddle point has complex $\Delta_I,\beta$ as well as complex value
$S_\ast$ for the extremized `entropy.' Its interpretation is given as follows
(See also \cite{Choi:2019miv}). The exponential of the saddle point `entropy'
$S_\ast$ given by
\begin{equation}
  e^{S_\ast(Q_I,J)}=e^{i{\rm Im}S_\ast(Q_I,J)}e^{{\rm Re}S_\ast(Q_I,J)}
\end{equation}
should somehow represent the large charge and large $N$ degeneracies of BPS states.
Here we present an interpretation of the charge-dependent phase factor
$e^{i{\rm Im}S_\ast}$, as mimicking rapid oscillations between $\pm 1$ as the
macroscopic
angular momentum charges $Q_I,J$ are shifted by their minimal quantized units.
If the macroscopic bosonic and fermionic states are not completely cancelled at a
given charge order, the resulting integer after the
partial cancelation can be either positive or negative, depending on the precise
values of charges. Semi-classical Legendre transformation is not capable of
deciding these signs, which should depend on the precise quantized values of macroscopic
charges. Our interpretation is that, the macroscopic
Legendre transformation can at least imitate the rapid $\pm 1$ oscillation
by having an imaginary part of the saddle point entropy $S_\ast$ \cite{Choi:2019miv}.
However, to make this story more precise, one should recall that the unitarity of the
QFT demands the existence of complex conjugate pairs of saddle points if they are
not real. Indeed, the two cases I/II of (\ref{large-N-free}) guarantee that such
a pair exists for the physical saddle point.
Then, adding the contributions from the pair, one obtains
\begin{equation}
  \sim e^{{\rm Re}S_\ast}\cos\left({\rm Im}S_\ast+\cdots\right)\ ,
\end{equation}
where now one obtains a real entropy ${\rm Re}S_\ast$ and the $\cos$ factor is
interpreted as imitating the rapid oscillation between $\pm 1$.

Let us illustrate that the physical value of complex
$\beta$ that is relevant for the Legendre transformation satisfies
${\rm Im}(\beta)<0$, which was assumed during the computations.
The general studies are made in \cite{Choi:2018fdc}, so
we illustrate this fact in the case when all $U(1)^4\subset SO(8)$ charges are
equal: $Q_1=Q_2=Q_3=Q_4\equiv Q$. We therefore set
$\Delta_I\approx\frac{\pi i}{2}$ for all $I=1,\cdots,4$ in case I.
Then (\ref{large-N-free}) becomes
\begin{equation}
  \log Z_{S^2\times S^1}\sim-i\frac{\sqrt{2}\pi^2N^{\frac{3}{2}}}{6}\beta^{-1}
  \equiv-i\frac{c}{\beta}\ .
\end{equation}
$c$ is a positive number. For any positive number $c$, the Legendre transformation
will yield ${\rm Im}\beta<0$. This can be seen by considering the
extremization of (\ref{entropy-function}), which is now
\begin{equation}
  S(Q,J;\beta)\approx-i\frac{c}{\beta}+4Q\Delta+2\beta J\approx
  -i\frac{c}{\beta}+2J\beta+2\pi iQ\ .
\end{equation}
After extremization, one obtains
\begin{equation}
  S_\ast=2\sqrt{-2icJ}+2\pi iQ\ \ ,\ \ \
  \beta_\ast=\sqrt{\frac{-ic}{2J}}\ .
\end{equation}
The square roots are taken so that ${\rm Re}S_\ast>0$ and ${\rm Re}\beta_\ast>0$.
In particular, one obtains
\begin{equation}
  \beta_\ast=\sqrt{\frac{c}{2J}}\ e^{-\frac{\pi i}{4}}
\end{equation}
which indeed satisfies ${\rm Im}\beta_\ast<0$. So ${\rm Im}\beta<0$ is the
region of the chemical potential which is relevant for our
microstate counting, justifying this assumption made earlier in this section.
The set-up ${\rm Im}\beta <0$ will also be assumed in the rest of this section
even at finite $N$, which will be justified whenever the effective value of $c$
in the free energy is positive.

Before concluding this subsection, let us comment on the physics of
(de)confinement and the expectation value of the Wilson-Polyakov loops.
These discussions will shed more lights on the dynamics of
this system.

The reduction of the apparent $N^2$ degrees of freedom down to $N^{\frac{3}{2}}$
was triggered by the condensation of magnetic monopole operators
at the saddle point. Let us discuss the relation in more detail.
The condensation is measured by the eigenvalues
$u_a=\beta m_a+i\alpha_a$ deviating from the unit circle, $|s_a|=|e^{u_a}|\neq 1$.
The large $N$ condensation is macroscopic, $\max|\beta m_a|\sim N^{\frac{1}{2}}$.
More precisely, one finds
\begin{equation}
  M_{ab}\equiv |\beta m_{ab}+i\alpha_{ab}|
  \approx|{\rm Re}(\beta m_{ab})|=\sqrt{N}|x(a)-x(b)|\ \ ,\ \ \
  a(x)\equiv N\int_{x_1}^x\rho(x^\prime)dx^\prime\ ,
\end{equation}
with $x$ and $\rho(x)$ being $\mathcal{O}(N^0)$. The approximation $\approx$
is possible because
$u_a$ are close to the real axis in our saddle point ansatz.
$x(a)$ and $x(b)$ are given by the inverse function
of $a(x)$. Therefore, $M_{ab}$ becomes much
larger than $1$ when $|x(a)-x(b)|\gg N^{-\frac{1}{2}}$. From the fact that
$x$ and $\rho(x)$ do not scale with large $N$, one concludes that
$M_{ab}\gg 1$ if $|a-b|\gg\sqrt{N}$.

$M_{ab}$ is the effective mass of the off-diagonal mode at $a$'th row and
$b$'th column of the adjoint fields in our QFT, provided by the magnetic monopole
operator. This mass becomes much larger than $1$
if the mode is `deeply off-diagonal' $|a-b|\gg \sqrt{N}$. Therefore, the light modes
which can contribute to the free energy in this monopole background should
satisfy $|a-b|\lesssim\sqrt{N}$. These `near-diagonal' modes are a small fraction
of the $N^2$ matrix elements. Since the width of the near-diagonal region is
$\sqrt{N}$, the number of the near-diagonal modes
scales like $N\cdot N^{\frac{1}{2}}=N^{\frac{3}{2}}$, accounting for
the desired scaling. Technically, the two-body interaction potential
$\mathcal{W}_{\rm int}$ for the adjoint fields is approximated to a short-ranged
interaction (\ref{W-int}) after making the large $N$ approximation. This is because
only the near-diagonal modes remain light in the monopole background.
Therefore, we realize that
the $N^{\frac{3}{2}}$ scaling of the free energy is due to a partial confinement
triggered by the magnetic monopole condensation. This partial confinement
happens even in the high temperature limit of the CFT.

It is also interesting to consider the saddle point value of the
Polyakov loop in the fundamental representation, given by
\begin{equation}
  \mathcal{P}\equiv\frac{1}{N}\sum_{a=1}^N e^{u_a}\sim e^{\sqrt{N}x_2}
\end{equation}
with $x_2>0$ at $\mathcal{O}(N^0)$. $-\log \mathcal{P}$ measures the free energy
of an external quark running along the temporal circle, in the grand canonical ensemble
\cite{Aharony:2003sx}. The fact that $-\log\mathcal{P}\sim -\sqrt{N} x_2$ is
negative implies that the presence of such a quark loop is
thermodynamically preferred by the system.
Here, note that our $\mathcal{N}=4$ Yang-Mills theory
has dynamical fundamental fields. So at the saddle point with a large
expectation value for the Polyakov loop, the loop amplitude for
the dynamical fundamental fields will be amplified.
In fact, this amplification did happen in our calculus.
Namely, while approximating $\mathcal{W}_{\rm ext}$ to (\ref{W-ext}),
we encountered some ${\rm Li}_2(s_a\cdots)$ with $|s_a|\gg 1$.
These terms are the reason why $\mathcal{W}_{\rm ext}$ is amplified as
$N\rightarrow N^{1+\alpha}$ in (\ref{W-ext}).

To summarize, the loop amplification factor $N^{\alpha}$ for the
fundamental fields in $\mathcal{W}_{\rm ext}$ is balanced with the partial
confinement factor $N^{-\alpha}$ for the adjoint fields in
$\mathcal{W}_{\rm int}$, to yield the $N^{\frac{3}{2}}$ scaling at
$\alpha=\frac{1}{2}$. Both phenomena are triggered by the monopole condensation.

\subsection{Finite $N$ Cardy free energy}

In this subsection, we study $\log Z_{S^2\times S^1}$
in the finite $N$ Cardy regime.
We have already discussed in section 2 the Cardy limit at $N=1$, on
single M2-brane. Here we focus on the non-Abelian cases with
$N\geq 2$. The main goal of this subsection is to explore a finite $N$ version of the
$N^{\frac{3}{2}}$ degrees of freedom. Namely, we have obtained
\begin{equation}
  \log Z_{(0)}\sim-i\frac{2\sqrt{2}N^{\frac{3}{2}}}{3\beta}
  \sqrt{\Delta_1\Delta_2\Delta_3\Delta_4}
\end{equation}
as our large $N$ free energy (in what we called region I). We are interested in
the ratio $\frac{\log Z}{\log Z_{(0)}}$ of our finite $N$ free energy $\log Z$
and the fiducial one $\log Z_{(0)}$, to see whether the partial confinement
due to monopole condensation is stronger or weaker at finite $N$.
At $N=2$, we shall present an analytic solution
for the Cardy semi-classical approximation. At higher $N$'s,
we shall rely on numerical methods to find the Cardy saddle points.
Apparently, this might look similar to the numerical studies made on the `saddle points'
of the $S^3$ partition functions or the topological index
at finite $N$ \cite{Herzog:2010hf,Benini:2015eyy}. However,
in the previous studies in the literautre, there are no small parameters
to admit semi-classical saddle point approximations at finite $N$.
On the other hand, we do have
small $|\beta|$, which makes our finite $N$ results physical.
We will always find $\frac{\log Z}{\log Z_{(0)}}>1$.

For simplicity, we first consider the case with $\Delta_1=\Delta_2=\Delta_3=\Delta_4= \frac{\pi i}{2}$ (after shifting $\Delta_2$ by $2\pi i$ as (\ref{Delta-shift})),
which corresponds to the case with equal $U(1)^4\subset SO(8)$
R-charges, $Q_1=Q_2=Q_3=Q_4$. In terms of the variables of the Yang-Mills
theory, this amounts to setting $\xi=-\frac{\pi i}{2}, f= 0, T= -\pi i$.
Then, the saddle point equations \eqref{bethe} become
\begin{eqnarray} \label{trans-eq}
\hspace*{-.3cm}0\!&\!=\!&\!-\frac{\pi i}{2} + \textrm{Li}_1 (i s_a q^{\frac{1}{2}})
- \textrm{Li}_1 (-i s_a q^{-\frac{1}{2}}) + \sum_{b (\neq a)} \left[\frac{}{}\!\textrm{Li}_1 (s_a s_b^{-1} q^{-1}) - \textrm{Li}_1 (s_b s_a^{-1} q^{-1})  -\textrm{Li}_1 (-s_a s_b^{-1}) \right. \nonumber\\
\hspace*{-.3cm}&&\left. +\textrm{Li}_1 (-s_b s_a^{-1})
+ \textrm{Li}_1 (i s_a s_b^{-1} q^{\frac{1}{2}})
- \textrm{Li}_1 (i s_b s_a^{-1} q^{\frac{1}{2}})
-\textrm{Li}_1 (-i s_a s_b^{-1} q^{-\frac{1}{2}})
+ \textrm{Li}_1 (-i s_b s_a^{-1} q^{-\frac{1}{2}})  \!\frac{}{}\right]\ .
\hspace{1.2cm}
\end{eqnarray}
Here, we again temporarily included $q(\approx 1^-)$ to regularize
some variables sitting on top of the branch cuts, similar to the previous subsection.
Exponentiating both sides, one obtains
\begin{equation} \label{rat-eq}
\frac{1+iq^{-1/2}s_a}{1-iq^{1/2}s_a} \prod_{\substack{b=1\\b\neq a}}^N \frac{1-q^{-1}s_b s_a^{-1}}{1-q^{-1}s_a s_b^{-1}} \frac{1+s_a s_b^{-1}}{1+s_b s_a^{-1}} \frac{1-iq^{1/2} s_b s_a^{-1} }{1-iq^{1/2}s_a s_b^{-1}} \frac{1+iq^{-1/2} s_a s_b^{-1} }{1+iq^{-1/2}s_b s_a^{-1}} = i\ ,
\end{equation}
which are rational equations of $s_a$'s. Some solutions of \eqref{rat-eq} do not satisfy the original equations \eqref{trans-eq}. We are interested in the
solution of (\ref{trans-eq}).\footnote{We think that extra solutions to (\ref{rat-eq})
may also be valid saddle points, which
apparently look illegal in the current setting because we have replaced discrete magnetic
flux sums into continuum integrals. More carefully doing the flux sum along the line of
\cite{Jain:2013py}, we expect to reveal the relevance of these extra solutions.
However, it happens that a natural finite $N$ version of the saddle points encountered in section 4.1 solves (\ref{trans-eq}).} So after solving \eqref{rat-eq}, one should check whether the solutions satisfy \eqref{trans-eq} or not. Then, one should take $\beta \to 0$ (or $q\rightarrow 1$) limit on the solutions
to remove the branch cut regulator.

Before proceeding, let us comment on a `trivial solution' of (\ref{trans-eq}),
(\ref{rat-eq}), which is
\begin{equation}\label{maximal-deconfine}
  s_1=s_2=\cdots=s_N\equiv s_0\ \ ,\ \ \
  \frac{1+iq^{-\frac{1}{2}}s_0}{1-iq^{\frac{1}{2}}s_0}=i\ .
\end{equation}
$s_0$ is the Cardy saddle point solution to the Abelian M2-brane index,
(\ref{abelian-saddle}), which
in (\ref{maximal-deconfine}) is given by $s_0\rightarrow 1$. At $N=1$, we have shown in
section 2 that this is the one and only saddle point which yields the correct
free energy for single M2-brane. At higher $N\geq 2$, there are good reasons to
trust that they are forbidden saddle points, which we sketch now.

We first recall that a similar phenomenon was observed
for the 3d vector-Chern-Simons models \cite{Aharony:2012ns,Jain:2013py}, in which
one found an incompressible nature of the eigenvalue distribution for $s_a$ in the
high temperature limit. To understand this, one should first note that
partition functions of 4d gauge theories on $S^3\times S^1$ are also given
in terms of the holonomy integrals, over $\alpha_a$ (or $s_a$). At high temperature,
the general expectation is that these eigenvalues asymptotically approach
the same value, $s_1=\cdots =s_N$, so that the underlying gauge symmetry
is asymptotically unbroken. This is the `maximally deconfining' saddle point,
at which quarks and gluons are maximally liberated to a deconfined plasma.
However, in 3d gauge theories, partition functions are given by both
integrals over $\alpha_a$ and sums over the GNO charges $m_a$. In particular,
\cite{Jain:2013py} discussed the thermal partition functions of 3d vector-Chern-Simons
theories on $S^2\times S^1$ at high temperature. They showed that the discrete
sums over $m_a$ yield the following factor in the integrand for $\alpha_a$:
\begin{equation}\label{delta}
  \prod_{a=1}^N\delta\left(k\alpha_a\right)\ ,
\end{equation}
where $k$ is the Chern-Simons level for the $U(N)$ gauge symmetry, and
$\delta(x)$ is the periodic delta function satisfying $\delta(x)=\delta(x+2\pi)$.
It has
$k$ sharply peaked solutions for $N$ variables, $\alpha_a=\frac{2\pi n_a}{k}$,
where $n_a=0,1,\cdots,k-1$. Therefore, if $N$ is larger than $k$, more than one
eigenvalues should assume exactly the same value. Then \cite{Jain:2013py} argues
that the Haar measure $\prod_{a<b}\left(2\sin\frac{\alpha_a-\alpha_b}{2}\right)^2$
provides exact $0$, forbidding such a saddle point. To summarize, the GNO charge
sums and the Haar measure of 3d gauge theories may impose extra exclusion principles
on $\alpha_a$, forbidding them to assume same values.

Since our naive saddle point (\ref{maximal-deconfine}) also has
coinciding eigenvalues, one can suspect that similar exclusions may happen.
Indeed, by following the procedure of \cite{Jain:2013py} in our index,
we find such exclusions at $N\geq 2$. To explain this,
one should go one order beyond our Cardy approximation, which only keeps the leading
$\beta^{-1}$ order in the exponent. One starts from (\ref{N=4-index}), with
absolute values of the fluxes removed. Here, rather than making a continuum
approximation of the flux sum, one keeps the discrete sums (which is a
resolution needed to see the exclusion principle of coincident eigenvalues).
Then in the Cardy limit, one approximates
\begin{equation}
  (xe^{-\beta y};e^{-2\beta})
  \approx\exp\left[-\frac{{\rm Li}_2(x)}{2\beta}+\frac{1-y}{2}\log(1-x)+\cdots\right]\ ,
\end{equation}
keeping the subleading $\mathcal{O}(\beta^0)$ term.
In (\ref{N=4-index}), $x$ will contain $e^{i\alpha_a}$. $x$ will also
contain the macroscopic condensation of $m_a$ at the saddle point.
$y$ contains the fluctuation $l_a$ of the monopole flux $m_a=m_a^\ast+l_a$
around the saddle point value $m_a^\ast$. Following \cite{Jain:2013py},
we would like to sum over the discrete $l_a$, rather than making a continuum
approximation. Summing over $l_a$'s, one obtains
\begin{equation}\label{delta-our}
  \prod_{a=1}^N2\pi\delta\left(-\frac{i}{2}\log\frac{
  (1-e^{\beta m_a^\ast+i\alpha_a}t^{-\frac{1}{2}})
  (1-e^{\beta m_a^\ast-i\alpha_a}t^{-\frac{1}{2}})}
  {(1-e^{\beta m_a^\ast+i\alpha_a}t^{\frac{1}{2}})
  (1-e^{\beta m_a^\ast-i\alpha_a}t^{\frac{1}{2}})}+i\hat\xi-\frac{iT}{2}\right)
\end{equation}
in the integrand of $\alpha_a$ integrals. Here
we used $2\pi\delta(x)=\sum_{l=-\infty}^\infty e^{ilx}$ for the periodic
delta function $\delta(x)=\delta(x+2\pi)$. The argument of (\ref{delta-our})
is real. The delta function is peaked when $\alpha_a$ solves
\begin{equation}\label{peak-delta}
  e^{-\hat\xi+\frac{T}{2}}\left[
  \frac{(1-e^{\beta m_a^\ast+i\alpha_a}t^{-\frac{1}{2}})
  (1-e^{\beta m_a^\ast-i\alpha_a}t^{-\frac{1}{2}})}
  {(1-e^{\beta m_a^\ast+i\alpha_a}t^{\frac{1}{2}})
  (1-e^{\beta m_a^\ast-i\alpha_a}t^{\frac{1}{2}})}
  \right]^{\frac{1}{2}}=1\ .
\end{equation}
We are interested in the fate of the saddle point (\ref{maximal-deconfine}),
or more generally (\ref{abelian-saddle}).
In particular, plugging in the saddle point value of $\beta m_a^\ast$,
there is a unique solution $\alpha_a=0$ (mod $2\pi $) for (\ref{peak-delta}).
Therefore, following the arguments of \cite{Jain:2013py}, only
one eigenvalue can sit at this unique peak: otherwise, the Haar measure
will provide $0$. This leads to the conclusion that the naive saddle point
(\ref{maximal-deconfine}) will be relevant only at $N=1$.\footnote{However,
as commented in \cite{Jain:2013py}, this argument relies on the
fact that the delta functions like (\ref{delta}), (\ref{delta-our}) do not spread as one
includes further subleading corrections in $\beta$. To the best of our knowledge,
this issue is not completely clarified so far. We hope to completely resolve this
issue within the indices in the near future.}

With these understood, let us first consider the case with $N=2$.
Among 4 solutions of \eqref{rat-eq}, there are two solutions satisfying \eqref{trans-eq}.
One is given by (\ref{maximal-deconfine}), which is dismissed as explained.
Another solution is given by
\begin{equation}\label{N=2-solution}
  s_1= \frac{1}{2} \left(1-3^{1/4}\sqrt{2}+\sqrt{3}\right) \approx 0.435421\ ,\ \
  s_2(= s_1^{-1}) = \frac{1}{2} \left(1+3^{1/4}\sqrt{2}+\sqrt{3}\right) \approx 2.29663\ ,
\end{equation}
in $\beta \to 0$ limit with $\textrm{Im} (\beta)<0$, up to permutation.
It is important to keep the regulator $\beta$, with the correct sign for
${\rm Im}(\beta)<0$ as explained in section 4.1, to get this solution.
This is because of the presence of
$\textrm{Li}_1 (s_as_b^{-1}) = -\log (1-s_as_b^{-1})$ and
${\rm Li}_1(s_bs_a^{-1})$ in the saddle point equations at $q=1$, since the
solution (\ref{N=2-solution}) sits precisely at a branch cut.
This solution satisfies \eqref{trans-eq} only when $\textrm{Im}(\beta)<0$,
which is our physical region for complex $\beta$.
Finally, the Cardy free energy of $\log Z_{S^2\times S^1}$ at this
saddle point is given from (\ref{Cardy-index}) by
\begin{equation}\label{N=2-cardy-free}
\begin{aligned}
\left.\frac{}{}\!\!\!\log Z_{S^2\times S^1}\right|_{N=2}
& \sim \frac{i}{2\beta} \left[-8G -2 \, \textrm{Im} \left\{2 \textrm{Li}_2 (ix) + 2 \textrm{Li}_2 \left(\frac{i}{x}\right) + 2 \textrm{Li}_2 (ix^2) + 2 \textrm{Li}_2 \left(\frac{i}{x^2}\right) + \textrm{Li}_2 \left(\frac{1}{x}\right)\right\}\right] \\
& \approx -\frac{17.4771i}{2\beta}\ ,
\end{aligned}
\end{equation}
where $x\equiv s_1=s_2^{-1}= \frac{1}{2} \left(1-3^{1/4}\sqrt{2}+\sqrt{3}\right) \approx 0.435421$, and
\begin{equation}
  G\equiv\sum_{n=0}^\infty\frac{(-1)^n}{(2n+1)^2}
  =\frac{{\rm Li}_2(i)-{\rm Li}_2(-i)}{2i} \approx 0.915966
\end{equation}
is Catalan's constant.

When $N \geq 3$, we cannot solve \eqref{rat-eq} analytically since they are
sextic equations even at $N=3$. Thus, we numerically solve the saddle point equations at $\beta=0$. At $\beta=0$, \eqref{rat-eq} is simplified as
\begin{equation}\label{bethe-beta-0}
\frac{1+i s_a}{1-i s_a} \prod_{\substack{b=1 \\ b \neq a}}^N \left(\frac{1+ is_a s_b^{-1}}{1-i s_a s_b^{-1}}\right)^2  = i\ ,
\end{equation}
which is so-called the Bethe ansatz equations.
We first find numerical solutions of (\ref{bethe-beta-0}) at $N=3$.
Having set $\beta=0$, there could possibly be some solutions at finite $\beta$
that we miss. We
assume that the physical solution remains to solve (\ref{bethe-beta-0}) and proceed.
(This is obviously true at large $N$, as we confirm numerically below.)
Note also that, since we solve the exponentiated equation (\ref{bethe-beta-0}),
nonzero $\beta$ as a branch-cut regulator is unnecessary.
In this set-up, we found $13$ numerical solutions of (\ref{bethe-beta-0}). Among
them, there is only one solution (except (\ref{maximal-deconfine})) satisfying \eqref{trans-eq}, given by
\begin{equation}
s_1 \approx 0.230396, \; s_2 \approx 1, \; s_3 \approx 4.34035\ ,
\end{equation}
up to permutations of $s_a$'s. As before, this solution exactly lies on the branch cut of the vector multiplet part when $\beta=0$. The correct sign of $\textrm{Im} (\beta)$, which makes the above solution satisfy \eqref{trans-eq}, is $\textrm{Im}(\beta)<0$. The Cardy free energy from (\ref{Cardy-index}) is given by
\begin{equation}
  \left.\!\!\frac{}{}\log Z_{S^2\times S^1}\right|_{N=3}
  \approx \frac{- 29.8009 i}{2\beta}\ ,
\end{equation}
 assuming $\textrm{Im} (\beta)<0$.

For $N \geq 4$, we will directly solve \eqref{trans-eq} numerically, rather than
(\ref{rat-eq}). Since we have been obtaining solutions with real positive
$s_as_b^{-1}$ till $N\leq 3$, we should carefully treat the branch cuts on the real
axis of the $s_as_b^{-1}$ planes in the $\beta\rightarrow 0$ limit. The functions in
(\ref{trans-eq}) to be careful about
are $\textrm{Li}_1 (q^{-1}s_a s_b^{-1}) - \textrm{Li}_1 (q^{-1}s_b s_a^{-1})$,
as we take $q\rightarrow 1^-$ with ${\rm Re}(\beta)>0$, ${\rm Im}(\beta)<0$.
In the numerics, we plugged in very small ${\rm Im}(\beta)<0$ to get the solutions,
resolving the branch cut ambiguity. (On the other hand, we find no solutions after
plugging in very small ${\rm Im}(\beta)>0$.)

\begin{figure}[!t]
\centering
\begin{subfigure} [b]{0.48\textwidth}
\includegraphics[width=\textwidth]{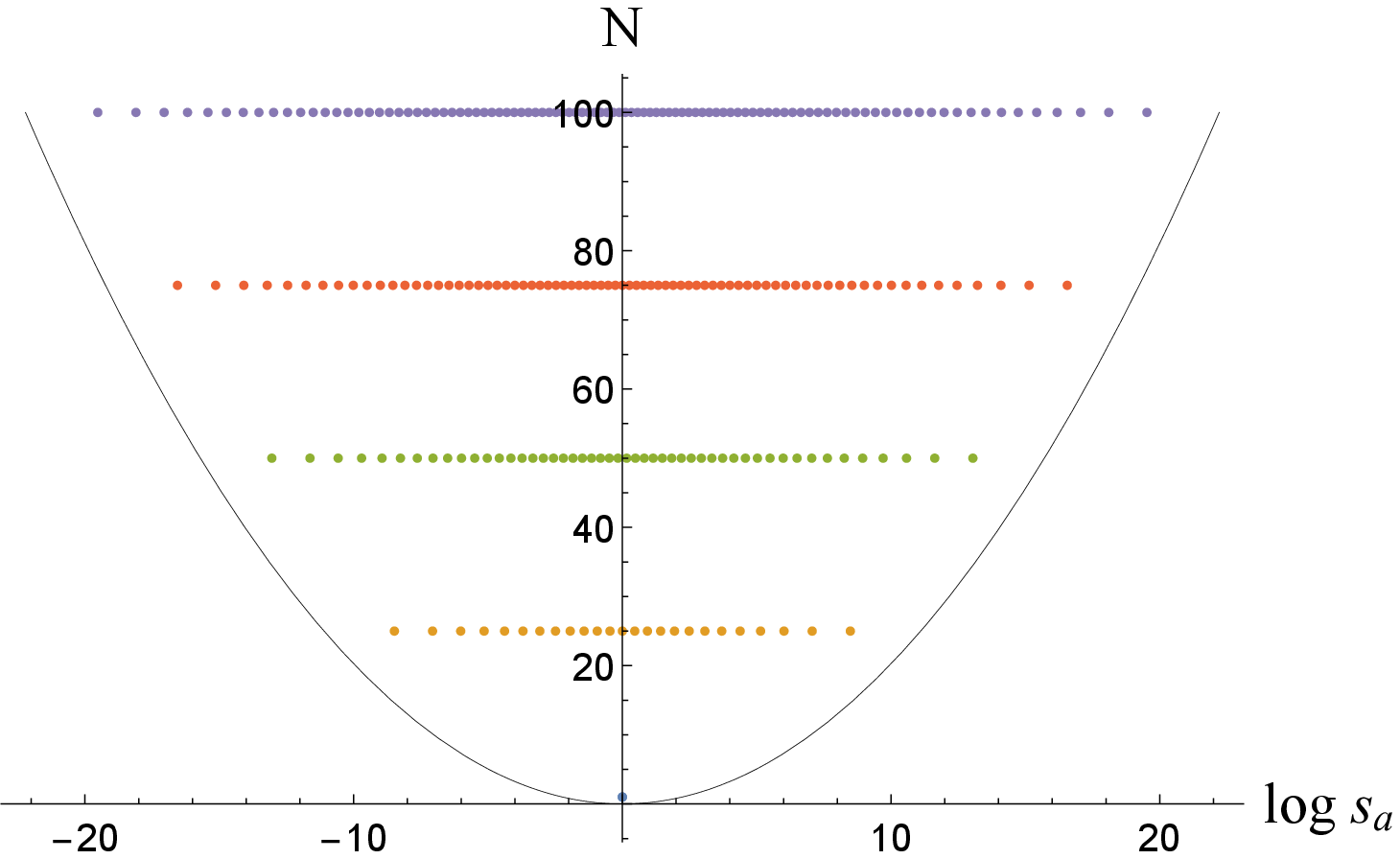}
\subcaption{The black solid line is the range of eigenvalue distribution extrapolated from large $N$.}
\end{subfigure}
\quad
\begin{subfigure} [b]{0.48\textwidth}
\includegraphics[width=\textwidth]{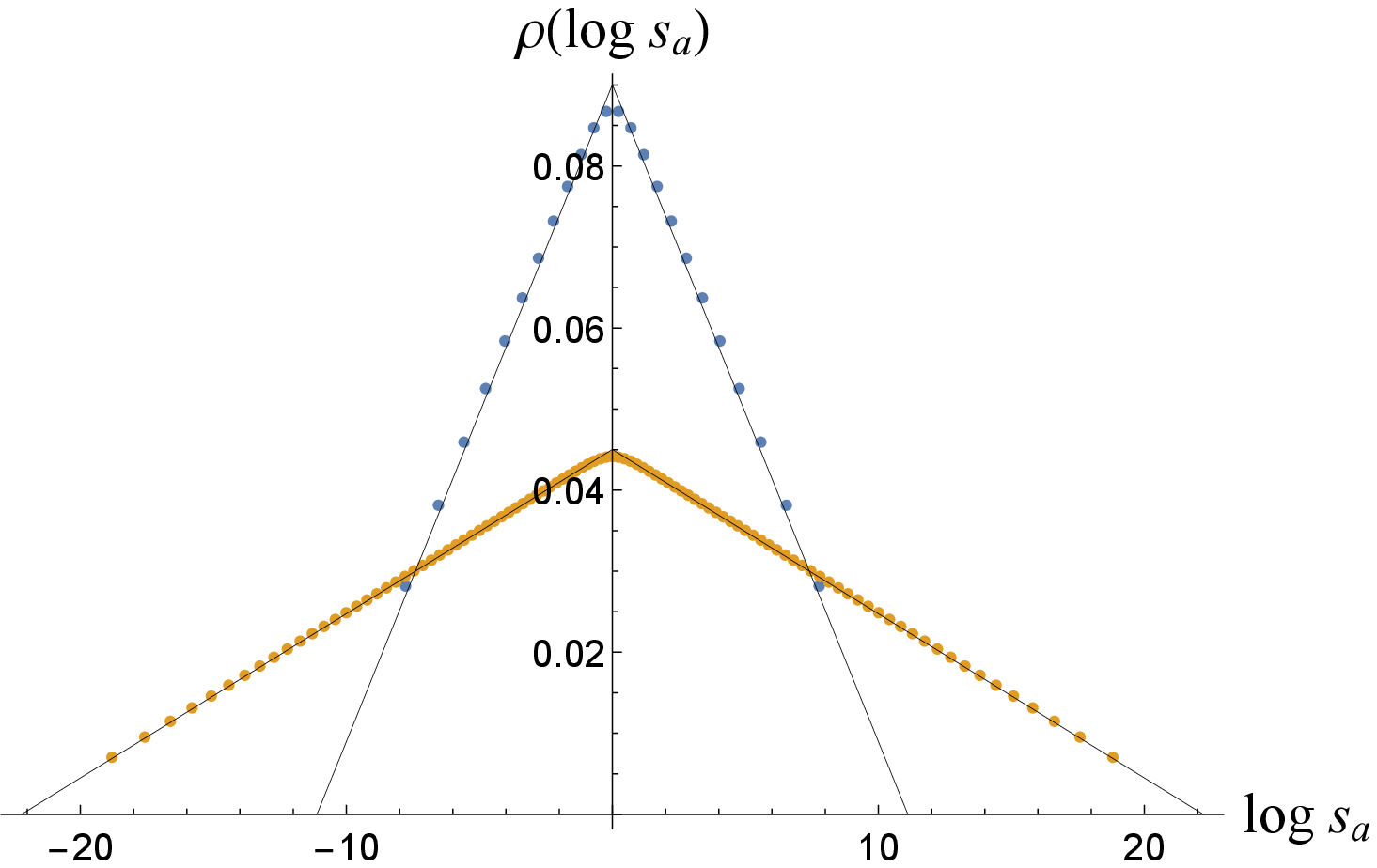}
\subcaption{Black solid lines are the density functions extrapolated from large $N$.}
\end{subfigure}
\caption{(a) Eigenvalue distributions at $N=25,50,75,100$,  (b) Densities of eigenvalues at $N=25$ (blue) and $N=100$ (yellow)}
\label{fig: eigenvalue-equal charges}
\end{figure}
Now we show the numerical results. We used Newton's method to find the roots of \eqref{trans-eq}.\footnote{The Newton method may in principle miss some solutions, as it
depends on the choice of initial values. However, even after trying many initial values,
we found no more solutions than those presented below.}
For $N \leq 100$, we found that all eigenvalues $s_a$ are positive real
in our solutions. These eigenvalues can be sorted in ascending order:
$y_a=0, \quad 0<s_1 < s_2 < \ldots < s_N$.
We also mention that our finite $N$ numerical solutions also satisfy all the assumptions
\eqref{branch-con}, \eqref{y-sol} made in section 4.1 for
large $N$ analysis, coming from the eigenvalue distributions not crossing branch cuts.
The $N$ eigenvalues spread out from $s_0$ ($\rightarrow 1$ in the Cardy limit)
with the width roughly proportional to $N^{1/2}$. The detailed eigenvalue distributions at various $N$ are given by Fig. \ref{fig: eigenvalue-equal charges}. The density
of the eigenvalues is defined as $\rho (x) = \frac{1}{N} \frac{da}{dx}$.

\begin{figure}[!t]
\centering
\begin{subfigure} [b]{0.48\textwidth}
\includegraphics[width=\textwidth]{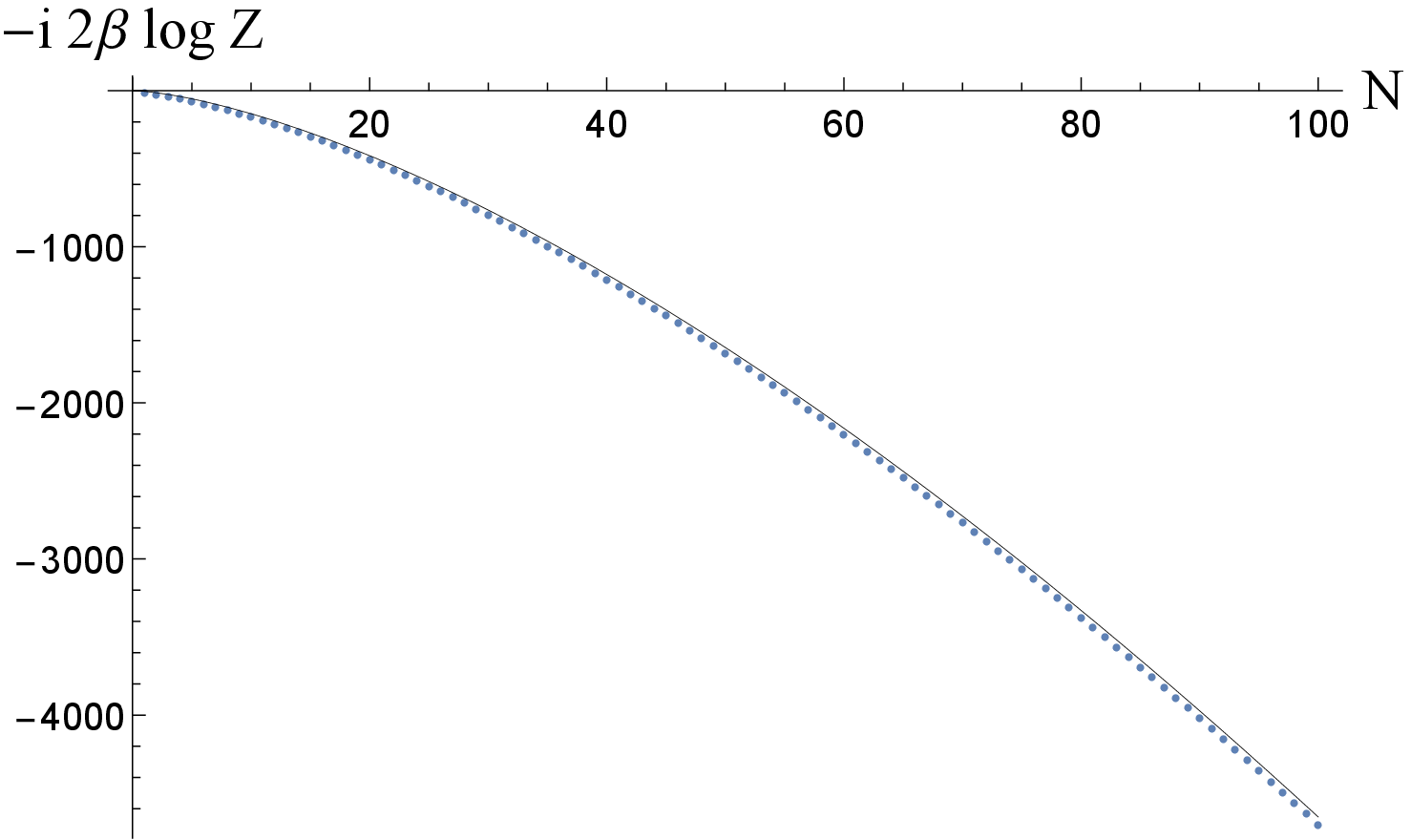}
\subcaption{}
\end{subfigure}
\quad
\begin{subfigure} [b]{0.48\textwidth}
\includegraphics[width=\textwidth]{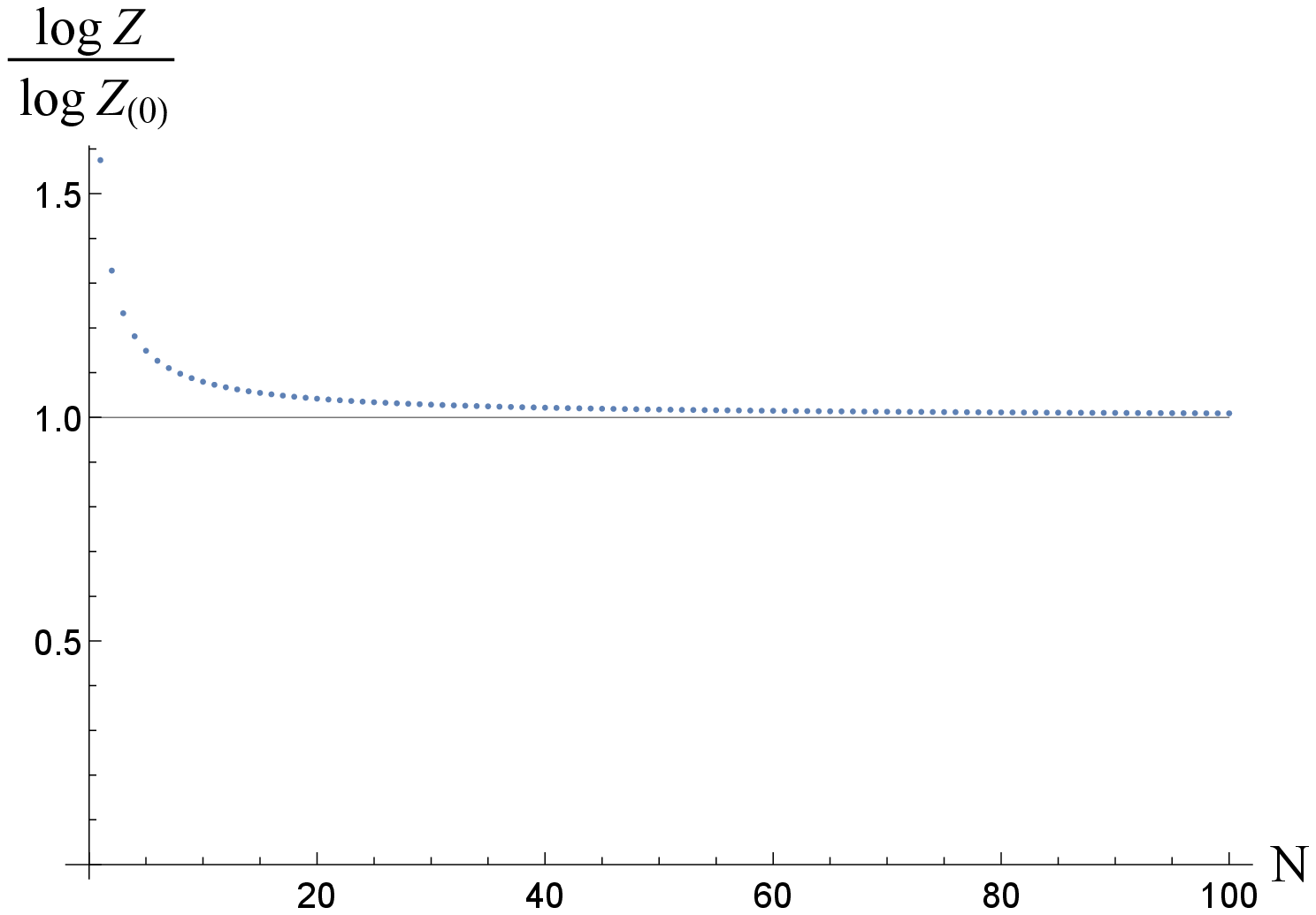}
\subcaption{}
\end{subfigure}
\caption{(a) Imaginary parts of $2\beta \log Z$ (dots) and
$2\beta \log Z_{(0)}$ (solid line).  (b) Ratio of the finite
$N$ free energy $\log Z$ and the fiducial free energy $\log Z_{(0)}$ (dots). Solid line
is drawn just as a reference line.}
\label{fig: free energy-equal charges}
\end{figure}
$\log Z_{S^2\times S^1}$ at various $N$ are given by
Fig. \ref{fig: free energy-equal charges}.
One finds that the large $N$ analytic approximation of section 4.1
is well-fitted with the numerical result at large enough $N$. The difference
between the numerical result and the fiducial one
in Fig. \ref{fig: free energy-equal charges}(a) increases as $N$ grows,
which seems to scale like $O(N^{\frac{1}{2}})$.
In addition, we find that the finite $N$ Cardy free energy ($F=-\textrm{Re}\,(\log Z)$) is always smaller than the fiducial one. Although we do not display the
relevant plot here, we found that the numerical result for $\textrm{Re}(\mathcal{W}^*)$ is also well-fitted to the analytically computed $\mathcal{W}_0$ at large enough $N$.

Our numerical solutions for $s_a$ are very simple, staying at
the positive real axis. One may wonder that such simple distributions
are due to the simplified setting $\Delta_1=\Delta_2=\Delta_3=\Delta_4$.
However, we found that the eigenvalues are positive real even at unequal
$\Delta_I$'s. As the qualitative behaviors are very similar, we shall not
plot the results for unequal $\Delta_I$'s here.

As long as we are aware of, our results are first quantitatively
explored finite $N$ versions of $N^{\frac{3}{2}}$ on
M2-branes. Especially, it will be interesting to see if there are any further
implications of the analytic coefficient of (\ref{N=2-cardy-free}),
which should be replacing $N^{\frac{3}{2}}$ at $N=2$.

\subsection{ABJM theory at large $N$}

In this subsection, we make a similar Cardy approximation with the ABJM model for
$N$ M2-branes. We reported some difficulties in section 2 to study the
vortex partition function for the ABJM theory on $D_2\times S^1$, due to the diverse
possibilities of anomaly-free boundary conditions. This will be closely related
the asymptotic factorization in the Cardy limit which we study here,
in the set-up of section 3.
Namely, we will have to factorize the integrand in a way that the `holomorphic'
and `anti-holomorphic' factors separately do not respect the
$U(N)\times U(N)$ Weyl symmetry.

The ABJM index on $S^2\times S^1$ is given by \cite{Kim:2009wb}
\begin{align}\label{index-ABJM}
Z_{S^2\times S^1}&=
 \frac{1}{(N!)^2}\!\!\sum_{m_a,\tilde{m}_a\in\mathbb{Z}^N}
 \oint \prod_{a=1}^N\left[
 \frac{ds_a}{2 \pi is_a} \frac{d\tilde s_a}{2 \pi i\tilde{s}_a} s_a^{k m_a}
 \tilde s_a^{-k \tilde m_a} \right]
 q^{\sum_{a,b}|m_a-\tilde m_b|-\frac{1}{2} \sum_{a, b} |m_a-m_b|-\frac{1}{2} \sum_{a,b} |\tilde m_a-\tilde m_b|} \nonumber \\
&\times \left[\prod_{a \neq b}^N \left(1-s_a s_b^{-1} q^{|m_a-m_b|}\right)\right] \left[\prod_{a \neq b}^N \left(1-\tilde s_a \tilde s_b^{-1} q^{|\tilde m_a-\tilde m_b|}\right)\right] \nonumber \\
 &\times \left[\prod_{A = 1}^2 \prod_{a = 1}^N \prod_{b = 1}^N \frac{(s_a^{-1} \tilde s_b t_A^{-1} q^{\frac{3}{2}+|m_a-\tilde m_b|};q^2)}{(s_a \tilde s_b^{-1} t_A q^{\frac{1}{2}+|m_a-\tilde m_b|};q^2)}\right] \left[\prod_{B = 3}^4 \prod_{a = 1}^N \prod_{b = 1}^N \frac{(s_a \tilde s_b^{-1} t_B^{-1} q^{\frac{3}{2}+|-m_a+\tilde m_b|};q^2)}{(s_a^{-1} \tilde s_b t_B q^{\frac{1}{2}+|-m_a+\tilde m_b|};q^2)}\right]
\end{align}
where $\prod_{I = 1}^4 t_I = 1$, and again $q=e^{-\beta}$.
Note that one of the charges conjugate to
$t_I$'s is the topological $U(1)$ charge $\sum_{a=1}^N(m_a+\tilde{m}_a)$,
so a priori it cannot by introduced as rotating elementary fields as shown in this
formula. However, by suitably rotating $s_a$ and $\tilde{s}_a$ by
$U(1)^2\subset U(N)^2$, one can absorb it in into a component of $t_A$'s,
as shown above.

Before proceeding, we need to comment on the periodicities of chemical potentials.
$t_I$ and $q=e^{-\beta}$ are related to our previous
chemical potentials $\Delta_I$ by
\begin{equation}
  t_I=e^{-\Delta_I+\frac{\beta}{2}}\ \ ,\ \
  2\beta=\sum_{I=1}^4\Delta_I\ (\textrm{mod }4\pi i)\ .
\end{equation}
One may first insert this expression to (\ref{index-ABJM}) to eliminate $t_I$'s.
Then, inserting $\beta=\frac{1}{2}\sum_{I=1}^4\Delta_I$, one can eliminate $\beta$
to express $Z_{S^2\times S^1}$ as a function of four independent $\Delta_I$'s.
After this insertion, one can again identify the expected periodicities
(\ref{SO8-period}), i.e. shifting any chosen pair of $\Delta_I$'s by
$(\Delta_I,\Delta_J)\rightarrow(\Delta_I+2\pi i,\Delta_J\pm 2\pi i)$.
These are the naturally expected periodicities from the kinematic considerations
of the M2-brane QFT. Namely, note from (\ref{index-Delta}) that
$\Delta_I$ is conjugate to the $SO(8)\times SO(3)$ angular momenta $Q_I+J$,
and also that observables in the spinor representation of $SO(8)$ are also
spinor in the spacetime $SO(3)$ in this QFT. These naturally demand
the periodicity (\ref{index-Delta}) for any thermal partition functions of this QFT,
since they are the intrinsic symmetries of the QFT.
However, the index (\ref{index-ABJM}) also has emergent periodicities.
Namely, let us define constrained variables $\lambda_I$'s by
$t_I\equiv e^{i\lambda_I}$. $\lambda_I$ satisfy $\sum_{I=1}^4\lambda_I=0$ (mod $2\pi $).
The emergent symmetry of (\ref{index-ABJM}) is given by the four independent shifts of
$\lambda_I$ by $2\pi $, holding $\beta$ fixed. Since (\ref{index-ABJM}) contains no
fractional powers of $t_I$, these shifts are obviously symmetries.
The symmetries and constraints are summarized as
\begin{equation}\label{lambda-properties}
  \lambda_I\sim\lambda_I+2\pi \ \ ,\ \ \
  \sum_{I=1}^4\lambda_I=0\ (\textrm{mod }2\pi)\ .
\end{equation}
The `mod $2\pi $' in the constraint is also an emergent one, related to the emergent
symmetries of $\lambda_I$. The symmetry has been used in \cite{Benini:2015eyy} in the
context of the topological index of the ABJM theory, to make a large $N$ analysis.
It will turn out below that similar procedures will be
applicable to our large $N$ Cardy free energy.

In the Cardy limit, $\Delta_I$'s are imaginary and $\lambda_I$'s are real.
Following \cite{Benini:2015eyy}, we first use the period shifts of $\lambda_I$'s to set
\begin{equation}
  0<\lambda_I<2\pi
\end{equation}
for all $I=1,\cdots,4$. Then from (\ref{lambda-properties}),
these variables should satisfy one of the following constraints:
\begin{equation}
  \sum_I\lambda_I=2\pi\ ,\ \ 4\pi\ ,\ \ 6\pi\ .
\end{equation}
If the right hand side is either $0$ or $8\pi$, the resulting free energy
is trivial. This is because
all $\lambda_I$'s are then $0$ up to $2\pi$ shifts, in which case
the Cardy behavior $\log Z_{S^2\times S^1}\propto\beta^{-1}$ is never visible
due to boson/fermion cancelations. It will turn out that, following the
studies of \cite{Benini:2015eyy}, only the two cases with the right hand side
being $2\pi$ and $6\pi$ have nontrivial large $N$ Cardy saddle points.
The case with $\sum_I\lambda_I=2\pi$ will turn out to be the case II of
(\ref{case-p}) in the section 4.1. The case with $\sum_I\lambda_I=6\pi$ is
equivalent to the case I in section 4.1, after shifting all $\lambda_I$'s
by $-2\pi$. The two cases yield mutually complex conjugate saddle points,
as in section 4.1. Below, we shall only consider the case II with
$\sum_I\lambda_I=2\pi$.

Assuming $0<\lambda_I<2\pi$, $\sum_I\lambda_I=2\pi$,
we again apply the identity (\ref{abs-remove-1}) to various factors in
(\ref{index-ABJM}), to remove the absolute values of $|\rho(m,\hat{m})|$.
As explained below (\ref{abs-remove-1}), there are two ways of removing
$|\rho(m)|$, either as $-\rho(m)$ or $+\rho(m)$.
For our purpose, the following choices turn out to be useful:
\begin{align}\label{abs-ABJM}
|m_a-\tilde m_b| &\rightarrow \left\{\begin{array}{ll}
-(m_a-\tilde m_b), & a \geq b, \\
+(m_a-\tilde m_b), & a < b,
\end{array}\right. \\
|-m_a+\tilde m_b| &\rightarrow \left\{\begin{array}{ll}
+(-m_a+\tilde m_b), & a \geq b, \\
-(-m_a+\tilde m_b), & a < b.
\end{array}\right.
\end{align}
We take the Cardy limit of this index assuming the above manipulations,
again making the continuum approximation of the monopole sums.
This leads to the following factorization into holomorphic and anti-holomorphic parts,
\begin{align}
Z_{S^2\times S^1} \sim Z_\text{hol} \, Z_\text{anti-hol}\ \ ,\ \ \
Z_\text{hol} = \int ds_a e^{-\frac{\mathcal W}{2 \beta}}, \qquad Z_\text{anti-hol} = \int d\bar s_a e^{\frac{\overline{\mathcal W}}{2 \beta}}
\end{align}
where
\begin{eqnarray}\label{eq:W}
\mathcal W &=& \frac{k}{2} \left(\sum_{a = 1}^N u_a^2-\sum_{b = 1}^N \tilde u_b^2\right)
+\sum_{a < b} \left[\mathrm{Li}_2(\tilde s_a s_b^{-1} t_{1,2}^{-1})-\mathrm{Li}_2(s_a \tilde s_b^{-1} t_{1,2})+\mathrm{Li}_2(s_a \tilde s_b^{-1} t_{3,4}^{-1})-\mathrm{Li}_2(\tilde s_a s_b^{-1} t_{3,4})\right] \nonumber \\
&&+\sum_{a = 1}^N \left[\mathrm{Li}_2(\tilde s_a s_a^{-1} t_{1,2}^{-1})-\mathrm{Li}_2(\tilde s_a s_a^{-1} t_{3,4})\right]
\end{eqnarray}
with the redefinition of the holonomy variable
\begin{align}
s_a q^{m_a} \rightarrow s_a.
\end{align}
Note that the real part of $u_a = \log s_a$ is identified with $-\beta m_a$.
We have made a Cardy factorization so that each
$\mathcal{W}$, $\overline{\mathcal{W}}$ does not respect $U(N)\times U(N)$
gauge symmetry. This is because we made inequivalent manipulations for
the upper-triangular and lower-triangular elements of the matrix-valued fields
in (\ref{abs-ABJM}). The reason for this ugly factorization will be clear shortly.

Taking large $N$ limit together with our Cardy limit, we introduce an ansatz
\begin{align}
u(x) = N^\alpha x+i y(x), \qquad \tilde u_a = N^\alpha x+i \tilde y(x)
\end{align}
with the density function $\rho(x) \geq 0$. Then the large $N$ approximation of \eqref{eq:W} is given by
\begin{align}
\label{eq:WlargeN}
\mathcal W &= -i k N^{1+\alpha} \int dx \rho(x) x (\tilde y-y)
+N^{2-\alpha} \int dx \rho(x)^2 \left[f(i (\tilde y-y+\lambda_{3,4}))-f(i (\tilde y-y-\lambda_{1,2}))\right] \nonumber \\
&-N \int dx \rho(x) \left[\mathrm{Li}_2(e^{i (\tilde y-y+\lambda_{3,4})})
-\mathrm{Li}_2(e^{i (\tilde y-y-\lambda_{1,2})})\right]
\end{align}
where
\begin{align}
f(x) = \frac{\{x\}^3}{6}-\frac{i \pi}{2} \{x\}^2-\frac{\pi^2}{3} \{x\},
\end{align}
\begin{align}
\{x\} = \left\{\begin{array}{ll}
x-2 \pi i \left\lfloor\frac{\mathrm{Im}(x)}{2 \pi}\right\rfloor, & \mathrm{Re}(x) \leq 0, \\
x-2 \pi i \left\lceil\frac{\mathrm{Im}(x)}{2 \pi}-1\right\rceil, & \mathrm{Re}(x) > 0.
\end{array}\right.
\end{align}
The last line of \eqref{eq:WlargeN} is subleading in $N$ but plays a role when we consider the equations of motion for the holonomy eigenvalues, i.e., the derivatives of $\mathcal W$ \cite{Benini:2015eyy}. Indeed, it gives rise to a repulsive force such that $\delta y(x) \equiv \tilde y(x)-y(x)$ cannot cross $\lambda_{1,2}, \, -\lambda_{3,4}$ and their periodic images. Nevertheless, it is not important for the final result once we obtain the extremization solution.

To obtain (\ref{eq:WlargeN}), it was important to make a gauge-non-invariant
factorization starting from (\ref{abs-ABJM}). Otherwise, there will be
a term in $\mathcal{W}$ proportional to $N^2$ which furthermore depends on
$\rho(x)$. Then it will be very difficult to set up the extremization problem
of $\rho(x)$ within the holomorphic part. This is because the leading
holomorphic part proportional to $N^2$ will cancel with the anti-holomorphic part,
so one should keep track of the subleading $N^{\frac{3}{2}}$ term
which does not cancel and becomes the leading term of $\log Z_{S^2\times S^1}$.
To summarize, the reason for our ugly factorization is to
avoid unwanted terms proportional to $N^2$ in our variational problem.

The saddle point is obtained by extremizing $\mathcal W$ with respect to $\rho(x)$ and $\delta y(x)$ under the constraint $\int dx \rho(x) = 1$. Our prime
interest is the case with $k=1$, which we consider first. Remarkably,
at $k = 1$, this extremization problem is exactly that addressed in
\cite{Benini:2015eyy} for the topologically twisted index. We choose
\begin{gather}
\begin{gathered}
0 < \delta y(x)+\lambda_{3,4} < 2 \pi, \quad -2 \pi < \delta y(x)-\lambda_{1,2} < 0, \\
\lambda_1 \leq \lambda_2, \qquad \lambda_3 \leq \lambda_4.
\end{gathered}
\end{gather}
In this range, the second term of \eqref{eq:WlargeN} proportional to
$N^{2-\alpha}$ becomes
\begin{align}
-i N^{2-\alpha} \int dx \rho(x)^2 \left[g_+\left(\delta y(x)+\lambda_{3,4}\right)-g_-\left(\delta y(x)-\lambda_{1,2}\right)\right]
\end{align}
where
\begin{align}
g_\pm (x) = \frac{x^3}{6}\mp\frac{\pi}{2} x^2+\frac{\pi^2}{3} x.
\end{align}
The nontrivial solution is obtained when the first line and the second lines are balanced. Thus, we take $\alpha = \frac{1}{2}$. Then we recognize that
\begin{align}
\mathcal W+i \mu N^{\frac{3}{2}} \left(\int dx \rho(x)-1\right)
\end{align}
is exactly $-\mathcal V$ in \cite{Benini:2015eyy} up to the definition of the
parameters. (See (2.60) in \cite{Benini:2015eyy}.)  By extremizing this functional
with  $\rho(x)$ and $\delta y(x)$, the solution is given by
\begin{align} \begin{array}{ll}
\begin{array}{l}
\rho(x) = \frac{\mu+x \lambda_3}{(\lambda_1+\lambda_3) (\lambda_2+\lambda_3) (\lambda_4-\lambda_3)}, \\
\delta y(x) = -\lambda_3,
\end{array} \qquad & x_\ll < x < x_<, \\
\\
\begin{array}{l}
\rho(x) = \frac{2 \pi \mu+x (\lambda_3 \lambda_4-\lambda_1 \lambda_2)}
{(\lambda_1+\lambda_3) (\lambda_2+\lambda_3) (\lambda_1+\lambda_4)
(\lambda_2+\lambda_4)}, \\
\delta y(x) = \frac{\mu (\lambda_1 \lambda_2-\lambda_3 \lambda_4)+x \sum_{A < B < C}
\lambda_A \lambda_B \lambda_C}{2 \pi \mu+x (\lambda_3 \lambda_4-\lambda_1 \lambda_2)},
\end{array} \qquad & x_< < x < x_>, \\
\\
\begin{array}{l}
\rho(x) = \frac{\mu-x \lambda_1}{(\lambda_1+\lambda_3)
(\lambda_1+\lambda_4) (\lambda_2-\lambda_1)}, \\
\delta y(x) = \lambda_1,
\end{array} \qquad & x_> < x < x_\gg,
\end{array}
\end{align}
where
\begin{align}
x_\ll = -\frac{\mu}{\lambda_3}, \qquad x_< = -\frac{\mu}{\lambda_4}, \qquad x_> = \frac{\mu}{\lambda_2}, \qquad x_\gg = \frac{\mu}{\lambda_1}, \qquad
\mu = \sqrt{2 \lambda_1 \lambda_2 \lambda_3 \lambda_4},
\end{align}
Substituting the solution back to $\mathcal W$, one obtains
\begin{align}
\mathcal W^\ast= -\frac{i N^\frac32 \mu^3}
{3 \lambda_1 \lambda_2 \lambda_3 \lambda_4} =
-\frac{2 \sqrt2 N^\frac32}{3} i \sqrt{\lambda_1 \lambda_2 \lambda_3 \lambda_4}
\end{align}
and moreover
\begin{equation}
\log Z_\text{hol} = -\frac{\mathcal W^\ast}{2 \beta} = \frac{\sqrt2 N^\frac32}{3 \beta} i \sqrt{\lambda_1 \lambda_2 \lambda_3 \lambda_4}\ ,\ \
\log Z_\text{anti-hol} = \frac{\overline{\mathcal W}^\ast}{2 \beta} = \frac{\sqrt2 N^\frac32}{3 \beta} i \sqrt{\lambda_1 \lambda_2 \lambda_3 \lambda_4}\ .
\end{equation}
Combining these two, one obtains the large $N$ Cardy  free energy given by
\begin{align}
\label{eq:ABJM}
\log Z_{S^2\times S^1} \sim \log Z_\text{hol}+\log Z_\text{anti-hol} \sim
\frac{2 \sqrt2 N^\frac32}{3 \beta} i \sqrt{\Delta_1 \Delta_2 \Delta_3 \Delta_4}
\end{align}
with $\sum_{I = 1}^4 \Delta_I -2\beta = -2 \pi i$.
This indeed agrees with the results in section 4.1 in the region II.
Similar studies can be made in the region I, providing the complex conjugate
results.

For generic $k$, one can easily obtain the result by replacing $\rho(x)$ by
$k \hat\rho(x)$ in the previous computation. Then one finds
\begin{equation}
  \mathcal W(x,y,\rho) = k^2 \mathcal W_{k = 1}(x,y,\hat\rho)\ .
\end{equation}
So most of the previous calculations at $k=1$ can be used here, except that
$\mu$ has to be tuned differently to match the constraint
$\int dx \hat\rho(x) = \frac{1}{k}$.
The modified value for $\mu$ turns out to be
$\mu = \sqrt{\frac{2 \lambda_1 \lambda_2 \lambda_3 \lambda_4}{k}}$.
Plugging all these into $\mathcal W$, one obtains
$\mathcal W^\ast = -\frac{i k^2 N^\frac32 \mu^3}
{3 \lambda_1 \lambda_2 \lambda_3 \lambda_4} =
-\frac{2 \sqrt2 k^\frac12 N^\frac32}{3} i \sqrt{\lambda_1 \lambda_2 \lambda_3 \lambda_4}$
and accordingly
\begin{align}
\log Z_{S^2\times S^1}\sim
\frac{2 \sqrt2 k^\frac12 N^\frac32}{3 \beta} i
\sqrt{\lambda_1 \lambda_2 \lambda_3 \lambda_4}
= \frac{2 \sqrt2 k^\frac12 N^\frac32}{3 \beta} i \sqrt{\Delta_1 \Delta_2 \Delta_3 \Delta_4}\ .
\end{align}

\section{Conclusion and remarks}

In this paper, we explored the Cardy limit of the index for the M2-brane SCFTs
on $S^2\times\mathbb{R}$. Our studies are made by analyzing the vortex partition
functions, and also suitably approximating the GNO charge sum for the magnetic monopole
operators. At large $N$, we have quantitatively shown that the deconfined free energy
scales like $N^{\frac{3}{2}}$. This free energy statistically accounts for the
Bekenstein-Hawking entropies of large BPS black holes in $AdS_4\times S^7$. We
discovered the important roles played by the condensation of magnetic monopole
operators, which provides a mechanism for partial confinement of $N^2$ degrees of
freedom. We have also
found finite $N$ versions of $N^{\frac{3}{2}}$ degrees of freedom by studying
the Cardy limit of the index.

We believe that these discoveries will shed very concrete lights on
the strongly interacting dynamics of 3d (S)CFTs, including the M2-brane
CFTs.

One important issue that has been treated rather briefly in this paper is
the exclusion behavior of eigenvalues in our index. This phenomenon has
been first explored in the 3d vector-Chern-Simons theories,
either using semi-classical arguments \cite{Aharony:2012ns}
or based on path integral approach \cite{Jain:2013py}. We employed the
strategy of \cite{Jain:2013py} and studied the index of our M2-brane QFT.
The key result is that the GNO charge sum forbids eigenvalues to assume same
values, not even asymptotically in the high temperature limit.
In the vector-Chern-Simons model, this phenomenon played important roles
to make certain dualities to hold. In our M2-brane QFT,
similar exclusion principle forbids the naive saddle point whose free energy
is proportional to $N^2$. Both in the study of \cite{Jain:2013py} and this paper,
there are further issues to clarify concerning the small spreading of
the delta functions of $\alpha_a$'s, as explained in the conclusion of \cite{Jain:2013py}.

As a technical remark, we mainly used the $\mathcal{N}=4$ Yang-Mills-matter theory
engineered on the D2-D6-brane system, rather than the ABJM theory. When we first
started our project, this was because we were aiming to use the vortex partition
function in the Higgs branch and the factorization of $Z_{S^2\times S^1}$.
In Chern-Simons-matter theories, studies of vortex partition functions are
more difficult. Apparently, this seems to be due to the difficulty in finding
natural anomaly-free boundary condition on $D_2\times S^1$.
More physically, with Chern-Simons terms, there may be so-called non-topological
vortices in the symmetric phase, apart from the topological vortices in the Higgs
phase. This is because once we have electrically charged configurations in the
symmetric phase, magnetic flux is induced due to the Gauss' law of
Chern-Simons-matter theory. It is natural that these non-topological vortices
may play roles in the factorization formulae of the ABJM theory, if there is one at all.
However, our alternative asymptotic factorization of section 3
(in the Cardy limit) can be applied to the ABJM theory, as we explained
in section 4.3.

\vskip 0.5cm

\hspace*{-0.8cm} {\bf\large Acknowledgements}
\vskip 0.2cm

\hspace*{-0.75cm} We thank Hee-Cheol Kim, Joonho Kim, Kimyeong Lee, Sungjay Lee,
Shiraz Minwalla, June Nahmgoong, Jaemo Park, Shuichi Yokoyama and
especially Dongmin Gang for helpful discussions.
This work is supported in part by the National
Research Foundation of Korea Grant 2018R1A2B6004914 (SC, SK),
NRF-2017-Global Ph.D. Fellowship Program (SC), the ERC-STG grant
637844-HBQFTNCER (CH) and the INFN (CH).

\appendix

\section{Asymptotic behavior of $q$-Pochhammer symbols} \label{sec : poch}
From \cite{Fredenhagen:2004cj}, we get the following asymptotic formulae of $q$-Pochhammer symbols for $|a| \leq 1$;
\begin{equation}\label{poch-formulae}
\begin{aligned}
&(aq^m;q^2)_{\infty} = \prod_{n=0}^{\infty}(1-aq^{m+2n}), \quad q=e^{-\beta} \; (|q|<1)\ , \\
&\lim_{\beta \to 0^+}(aq^m;q^2)_\infty = \,   (1-aq^m)^{1/2} \, \exp\Big[- \frac {1}{2\beta} \textrm{Li}_2 (aq^m)\Big]\big(1+o({\beta }^0)\big), \quad   |a| \leq 1\; \& \; a \neq 1 \; (a \in \mathbb{C}), \\
&\lim_{\beta \to 0^+}(q^m;q^2)_{\infty}= \frac{\sqrt{2\pi}}{\Gamma(m/2)} (2\beta)^{-(m-1)/2}  \exp \Big[-\frac{1}{2\beta} \textrm{Li}_2 (1) \Big]   \big(1+o({\beta }^0)\big).
\end{aligned}
\end{equation}
We will extend the above asymptotic formulae to the whole complex plane $\mathbb{C}$ with the help of the Jacobi theta function and the Dedekind eta function:
\begin{equation}\label{jacobi}
\begin{aligned}
&\theta_1(\tau, z)=-e^{\frac{i \pi}{2}} y^{\frac{1}{2}} \tilde{q}^{\frac{1}{8}}   \prod_{n=1}^{\infty} (1-\tilde{q}^n) (1-y\tilde{q}^{n})(1-y^{-1} \tilde{q}^{n-1})=-e^{\frac{i \pi}{2}} y^{\frac{1}{2}} \tilde{q}^{\frac{1}{8}} (\tilde{q};\tilde{q})_{\infty}(y\tilde{q};\tilde{q})_{\infty}(y^{-1};\tilde{q})_{\infty}, \\
&\eta(\tau)=\tilde{q}^{\frac{1}{24}} \prod_{n=1}^{\infty} (1-\tilde{q}^n)=\tilde{q}^{\frac{1}{24}} (\tilde{q};\tilde{q})_{\infty}, \quad \tilde{q}=e^{2\pi i \tau}\; (\tau \in \mathbb{C} \; \& \; \textrm{Im}(\tau)>0), \; y=e^{2\pi i z} \; (z \in \mathbb{C}).
\end{aligned}
\end{equation}
These functions have the following modular properties:
\begin{equation}
\begin{aligned}
&\theta_1(-\frac{1}{\tau},\frac{z}{\tau})=-i(-i \tau)^{1/2} \exp \Big(i \frac{\pi}{\tau} z^2 \Big) \theta_1(\tau, z), \\
&\eta(-\frac{1}{\tau})=(-i \tau)^{1/2} \eta(\tau).
\end{aligned}
\end{equation}
Now we relate the parameters appearing in (\ref{poch-formulae}) and (\ref{jacobi}) as
\begin{equation}
\begin{aligned}
&\tilde{q}=q^2=e^{-2\beta}, \; \beta>0 \; \Rightarrow \; \tau=\frac{i\beta}{\pi},  \; \textrm{Re}(\tau)=0, \; \textrm{Im}(\tau)>0, \\
&y=a \in \mathbb{C}, \; |a| \leq 1 \; \& \; a \neq 0, \, 1 \; \Rightarrow \; z=\frac{\log a}{2\pi i}, \; -\pi \leq \textrm{Arg}(a) \leq \pi.
\end{aligned}
\end{equation}
Then, one finds that
\begin{equation}
\begin{aligned}
(a^{-1};q^2)_{\infty}&=-e^{-\frac{i \pi}{2}} a^{-\frac{1}{2}} q^{-\frac{1}{6}} \frac{\theta_1(\tau,z)}{\eta(\tau) (aq^2;q^2)_{\infty}}  \\
&=- \frac{1-a}{a^{1/2}} q^{-\frac{1}{6}}  \exp \Big(\frac{(\log a)^2}{4\beta} \Big) \frac{\theta_1(-\frac{1}{\tau},\frac{z}{\tau})}{\eta(-\frac{1}{\tau}) (a;q^2)_{\infty}}.
\end{aligned}
\end{equation}

Taking $\beta \to 0^+$ limit, one obtains
\begin{equation}
\begin{aligned}
(a^{-1};q^2)_{\infty}=&- (a^{-1}-1)^{1/2} \frac{\theta_1(-\frac{1}{\tau},\frac{z}{\tau})}{\eta(-\frac{1}{\tau})} \, \exp\bigg[ \frac {1}{2\beta} \Big(\textrm{Li}_2 (a) +\frac{1}{2} (\log a)^2 \Big) +\frac{\beta}{6} \bigg]\big(1+o({\beta }^0)\big) \\
=& (a^{-1}-1)^{1/2} \exp\bigg[ \frac {1}{2\beta} \Big(\textrm{Li}_2 (a) +\frac{1}{2} (\log a)^2 -\frac{\pi^2}{3} \Big) \bigg]  \\
& 2 \frac{(e^{i\pi \frac{\log a}{2\beta}}-e^{-i\pi \frac{\log a}{2\beta}})}{2i} \prod_{n=1}^{\infty} (1-e^{-i\pi \frac{\log a}{\beta}-\frac{2n\pi^2}{\beta}})(1-e^{i\pi \frac{\log a}{\beta}-\frac{2n\pi^2}{\beta}})
\big(1+o({\beta }^0)\big) \\
=& (a^{-1}-1)^{1/2} \exp\bigg[ \frac {1}{2\beta} \Big(\textrm{Li}_2 (a) +\frac{1}{2} (\log a)^2 -\frac{\pi^2}{3} \Big) \bigg] \bigg(2 \frac{(e^{i\pi \frac{\log a}{2\beta}}-e^{-i\pi \frac{\log a}{2\beta}})}{2i} \bigg) \big(1+o({\beta }^0)\big).
\end{aligned}
\end{equation}
For $|a| \leq 1 \; \& \; a \notin [0,1] \; (a \in \mathbb{C})$, this is simplified as
\begin{equation}
\begin{aligned}
(a^{-1};q^2)_{\infty}=& (1-a^{-1})^{1/2} \exp\bigg[ \frac {1}{2\beta} \Big(\textrm{Li}_2 (a) +\frac{1}{2} (\log (-a))^2 + \frac{\pi^2}{6} \Big) \bigg]  \big(1+o({\beta }^0)\big)\\
=& (1-a^{-1})^{1/2} \exp\bigg[ - \frac {1}{2\beta} \textrm{Li}_2 (a^{-1}) \bigg]  \big(1+o({\beta }^0)\big), \quad |a| \leq 1 \; \& \; a \notin [0,1] \; (a \in \mathbb{C}).
\end{aligned}
\end{equation}
So we find the following asymptotic formulae of $q$-Pochhammer symbols:
\begin{equation}\label{poch}
\begin{aligned}
\lim_{\beta \to 0^+}(aq^m;q^2)_\infty &= \,   (1-aq^m)^{1/2} \, \exp\Big[- \frac {1}{2\beta} \textrm{Li}_2 (aq^m)\Big]\big(1+o({\beta }^0)\big)\\
&=\exp\left[-\frac{1}{2\beta}{\rm Li}_2(aq^{m-1})\right]\left(1+o(\beta^0)\right)
, \quad  a \in \mathbb{C} \; \& \;a \notin [1,\infty), \\
\lim_{\beta \to 0^+}(q^m;q^2)_{\infty} &= \, \frac{\sqrt{2\pi}}{\Gamma(m/2)} (2\beta)^{-(m-1)/2} \,  \exp \Big[-\frac{1}{2\beta} \textrm{Li}_2 (1) \Big]   \big(1+o({\beta }^0)\big) \quad (a=1).
\end{aligned}
\end{equation}

\section{Young diagram formula for $Z_{\rm vortex}$}

In section 2.1, we encountered the following formula
\begin{equation}\label{vortex-k}
  Z_{\rm vortex}=\!\!\!\!
  \sum_{0\leq k_1\leq\cdots\leq k_N}\!\!\!Q^{k_1+\cdots+k_N}
  \prod_{a=1}^N\frac{(v^{-a+1};q^2)_{-k_a}}{(u^{-2}v^{-a+1};q^2)_{-k_a}}
  \cdot\prod_{a,b=1\ (a\neq b)}^N\frac{(v^{-a+b+1};q^2)_{-k_a+k_b}
  (u^{-2}v^{-a+b}q^2;q^2)_{-k_a+k_b}}
  {(v^{-a+b};q^2)_{-k_a+k_b}(u^{-2}v^{-a+b+1}q^2;q^2)_{-k_a+k_b}}
\end{equation}
for the vortex partition function. In this appendix, we prove the Young diagram
formula (\ref{vortex-Young}) for $Z_{\rm vortex}$.

Note that $(k_N, \ldots, k_1)$ is a set of non-increasing non-negative integers. Thus, it can be represented by a Young diagram, $\lambda = (\lambda_1, \ldots, \lambda_N)$, where each $\lambda_i$ is defined by $\lambda_i = k_{N-i+1}$. In terms of $\lambda_i$,
(\ref{vortex-k}) is written as
\begin{align}
\label{eq:res2}
& \sum_{\lambda_1 \geq \lambda_2 \geq \ldots \geq \lambda_N \geq 0} Q^{|\lambda|} \left(\prod_{i = 1}^N \frac{(u^{-2} v^{-N+i};q^{-2})_{\lambda_i}}{(v^{-N+i}q^{-2};q^{-2})_{\lambda_i}}\right)
\left(\prod_{i < j}^N \frac{(v^{-(j-i)} q^{-2};q^{-2})_{\lambda_i-\lambda_j} (u^{-2} v^{-(j-i-1)};q^{-2})_{\lambda_i-\lambda_j}}{(u^{-2} v^{-(j-i)};q^{-2})_{\lambda_i-\lambda_j} (v^{-(j-i-1)} q^{-2};q^{-2})_{\lambda_i-\lambda_j}}\right) \nonumber \\
& \ \ \ \quad\quad\quad \times
\left(\prod_{i < j}^N \frac{(u^{-2} v^{j-i} q^2;q^2)_{\lambda_i-\lambda_j} (v^{j-i+1};q^2)_{\lambda_i-\lambda_j}}{(v^{j-i};q^2)_{\lambda_i-\lambda_j} (u^{-2} v^{j-i+1} q^2;q^2)_{\lambda_i-\lambda_j}}\right) \nonumber \\
&= \sum_{\lambda_1 \geq \lambda_2 \geq \ldots \geq \lambda_N \geq 0} Q^{|\lambda|}  \left(\prod_{i = 1}^N \frac{(u^{-2};q^{-2})}{(q^{-2};q^{-2})} \frac{(u^{-2} v q^2;q^2) (v^{N-i+1};q^2)}{(v;q^2) (u^{-2} v^{N-i+1} q^2;q^2)}\right)
\left(\prod_{i = 1}^N \frac{(v^{-N+i} q^{-2 (\lambda_i+1)};q^{-2})}{(u^{-2} v^{-N+i} q^{-2 \lambda_i};q^{-2})}\right) \nonumber \\
& \quad \times \left(\prod_{i < j}^N \frac{(u^{-2} v^{-(j-i)} q^{-2 (\lambda_i-\lambda_j)};q^{-2}) (v^{-(j-i-1)} q^{-2 (\lambda_i-\lambda_j+1)};q^{-2})}{(v^{-(j-i)} q^{-2 (\lambda_i-\lambda_j+1)};q^{-2}) (u^{-2} v^{-(j-i-1)} q^{-2 (\lambda_i-\lambda_j)};q^{-2})}\right) \nonumber \\
& \quad \times \left(\prod_{i < j}^N \frac{(v^{j-i} q^{2 (\lambda_i-\lambda_j)};q^2) (u^{-2} v^{j-i+1} q^{2 (\lambda_i-\lambda_j+1)};q^2)}{(u^{-2} v^{j-i} q^{2 (\lambda_i-\lambda_j+1)};q^2) (v^{j-i+1} q^{2 (\lambda_i-\lambda_j)};q^2)}\right)
\end{align}
where we have used
\begin{align}
(a;q)_n = \frac{(a;q)}{(a q^n;q)}.
\end{align}
Note that \eqref{eq:res2} includes $(a;q^{-2})$, whose analytic continuation should be understood for $|q| < 1$ using
\begin{align}
(a;q^{-2}) = \frac{1}{(a q^2;q^2)}.
\end{align}
Now one can use the following identity \cite{2001math.....12035R}
\begin{gather}
\left(\prod_{i = 1}^N \frac{(x;q)}{(q^{\lambda_i} t^{N-i} x;q)}\right) \left(\prod_{i < j} \frac{(q^{\lambda_i-\lambda_j} t^{j-i} x;q)}{(q^{\lambda_i-\lambda_j} t^{j-i-1} x;q)}\right) = \prod_{(i,j) \in \lambda} \left(1-q^{\lambda_i-j} t^{\lambda'_j-i} x\right), \\
\prod_{i = 1}^N \frac{(t^{1-i} x;q)}{(q^{\lambda_i} t^{1-i} x;q)} = \prod_{(i,j) \in \lambda} \left(1-q^{j-1} t^{1-i} x\right)
\end{gather}
to obtain
\begin{equation}
\label{eq:finiteN}
Z_\text{vortex}  = \sum_{\lambda} Q^{|\lambda|} \sum_{(i,j) \in \lambda}
\frac{\left(1-u^{-2} v^{-\lambda'_{j}+i} q^{-2 \lambda_{i}+2 j}\right) \left(1-u^{-2} v^{\lambda'_{j}-i+1} q^{2 \lambda_{i}-2 j+2}\right) \left(1-v^{N-i+1} q^{2 j-2}\right)}{\left(1-v^{-\lambda'_{j}+i} q^{-2 \lambda_{i}+2 j-2}\right) \left(1-v^{\lambda'_{j}-i+1} q^{2 \lambda_{i}-2 j}\right) \left(1-u^{-2} v^{N-i+1} q^{2 j}\right)}
\end{equation}
%\begin{align}
%\frac{\left(1-u^{-2} v^{-v_{j(s)}+i(s)} q^{-2 h_{i(s)}+2 j(s)}\right)}{\left(1-v^{-v_{j(s)}+i(s)} q^{-2 h_{i(s)}+2 j(s)-2}\right)} = \frac{\frac{(u^{-2};q^{-2})}{(q^{-2 \lambda_i} v^{-(N_c-i)} u^{-2};q^{-2})} \frac{(q^{-2 (\lambda_i-\lambda_j)} v^{-(j-i)} u^{-2};q^{-2})}{(q^{-2 (\lambda_i-\lambda_j)} v^{-(j-i-1)} u^{-2};q^{-2})}}{\frac{(q^{-2};q^{-2})}{(q^{-2 (\lambda_i+1)} v^{-(N_c-i)};q^{-2})} \frac{(q^{-2 (\lambda_i-\lambda_j+1)} v^{-(j-i)};q^{-2})}{(q^{-2 (\lambda_i-\lambda_j+1)} v^{-(j-i-1)};q^{-2})}} \\
%\frac{\left(1-u^{-2} v^{v_{j(s)}-i(s)+1} q^{2 h_{i(s)}-2 j(s)+2}\right)}{\left(1-v^{v_{j(s)}-i(s)+1} q^{2 h_{i(s)}-2 j(s)}\right)} = \frac{\frac{(q^2 v u^{-2};q^2)}{(q^{2 (\lambda_i+1)} v^{N_c-i+1} u^{-2};q^2)} \frac{(q^{2 (\lambda_i-\lambda_j+1)} v^{j-i+1} u^{-2};q^2)}{(q^{2 (\lambda_i-\lambda_j+1)} v^{j-i} u^{-2};q^2)}}{\frac{(v;q^{2})}{(q^{2 \lambda_i} v^{N_c-i+1};q^2)} \frac{(q^{2 (\lambda_i-\lambda_j)} v^{j-i+1};q^2)}{(q^{2 (\lambda_i-\lambda_j)} v^{j-i};q^2)}} \\
%\frac{\left(1-v^{N_c-i(s)+1} q^{2 j-2}\right)}{\left(1-u^{-2} v^{N_c-i(s)+1} q^{2 j}\right)} = \frac{\frac{\left(v^{N_c-i+1};q^2\right)}{\left(q^{2 \lambda_i} v^{N_c-i+1};q^2\right)}}{\frac{\left(q^2 v^{N_c-i+1} u^{-2};q^2\right)}{\left(q^{2 (\lambda_i+1)} v^{N_c-i+1} u^{-2};q^2\right)}}
%\end{align}
where $\lambda'$ is the conjugate of $\lambda$. Namely, $\lambda'_j$ is the vertical length of the $j$th column of $\lambda$ while $\lambda_i$ is the horizontal length of the $i$th row of $\lambda$. This proves (\ref{vortex-Young}).

Recall that the expression \eqref{eq:finiteN} is obtained under the conditions $|u^2| < 1, \, |u^{-2} q^2| < 1, \, |v| < 1, \, |Q| < 1$ and $|q| < 1$; i.e.,
\begin{align}
\label{eq:conf}
\pm T_*-\beta < 0, \quad f_*+\frac{T_*}{2}-\frac{\beta}{2} < 0, \quad \xi_* > 0, \quad \beta > 0
\end{align}
with $\xi = -\log Q$. $x_*$ denotes the real part of $x$. In particular, the first condition shows that $|T_*| < \beta$. We have numerically checked, by $q$-expansion, that \eqref{eq:finiteN} is not valid if any of the above inequalities is flipped. (We
have checked this at $N = 2$.)

As a simple application, let us study
study the large $N$ limit of the vortex partition function
(\ref{vortex-Young}). In this case, since $v<1$,
we can take $v^N$ to zero. Then the large $N$ limit of the vortex
partition function is given by
\begin{equation}
  Z_{\rm vortex}\stackrel{N\rightarrow\infty}{\longrightarrow}
  \sum_{Y}Q^{|Y|}\prod_{s\in Y}
  \frac{(1-u^{-2}(q^{2h(s)}v^{v(s)})^{-1})(1-u^{-2}vq^2\cdot q^{2h(s)}v^{v(s)})}
  {(1-q^{-2}(q^{2h(s)}v^{v(s)})^{-1})(1-vq^{2h(s)}v^{v(s)})}
\end{equation}
where $q=e^{-\beta}$, $u=(qt)^{\frac{1}{2}}=e^{\frac{T-\beta}{2}}$,
$v=z(qt)^{\frac{1}{2}}=e^{f+\frac{T-\beta}{2}}$. This series is well known.
Upon suitable parameter mappings, this is functionally
identical to the instanton partition function of 5d $\mathcal{N}=1^\ast$
super-Yang-Mills theory with $U(1)$ gauge group \cite{Nekrasov:2002qd}.
The $Q$ series of this partition function can be summed to
the following expression \cite{Iqbal:2008ra,Kim:2011mv},
\begin{equation}
  Z^{N\rightarrow\infty}_{\rm vortex}
  =PE\left[\frac{(1-u^{-2})(1-u^{-2}vq^2)}{(1-q^{-2})(1-v)}
  \frac{Q}{1-q^2Qu^{-2}}\right]\ ,
\end{equation}
where $PE\left[f(Q,q,u,v)\right]\equiv\exp\left[\sum_{n=1}^\infty\frac{1}{n}
f(Q^n,q^n,u^n,v^n)\right]$.
So $Z_{\rm vortex}$ has smooth large $N$ limit.
The perturbative part is given by
\begin{equation}
  Z_{\rm pert}=\prod_{a=1}^N\frac{(u^{-2}v^aq^2;q^2)_\infty}{(v^a;q^2)_\infty}
  =PE\left[\frac{(1-u^{-2}q^2)(v+v^2+\cdots+v^N)}{1-q^2}\right]\ .
\end{equation}
This also has a smooth large $N$ limit
\begin{equation}
  Z_{\rm pert}^{N\rightarrow\infty}=
  PE\left[-\frac{q^{-2}v(1-u^{-2}q^2)}{(1-q^{-2})(1-v)}\right]\ .
\end{equation}
Multiplying $Z_{\rm pert}$ and $Z_{\rm vortex}$,
one obtains
\begin{equation}
  Z^{N\rightarrow\infty}_{\rm pert}Z^{N\rightarrow\infty}_{\rm vortex}=
  PE\left[\frac{Q+vu^{-2}+Qvu^{-2}-Qu^{-2}-vq^{-2}-q^2Qvu^{-2}}
  {(1-q^{-2})(1-v)(1-q^2Qu^{-2})}\right]\ .
\end{equation}
As explained in section 2.1, we ignore the $Z_{\rm prefactor}$ factor.


\begin{thebibliography}{12345}

%\cite{Klebanov:1996un}
\bibitem{Klebanov:1996un}
  I.~R.~Klebanov and A.~A.~Tseytlin,
  %``Entropy of near extremal black p-branes,''
  Nucl.\ Phys.\ B {\bf 475}, 164 (1996)
  doi:10.1016/0550-3213(96)00295-7
  [hep-th/9604089].
  %%CITATION = doi:10.1016/0550-3213(96)00295-7;%%
  %364 citations counted in INSPIRE as of 22 Mar 2018



%\cite{Drukker:2010nc}
\bibitem{Drukker:2010nc}
  N.~Drukker, M.~Marino and P.~Putrov,
  %``From weak to strong coupling in ABJM theory,''
  Commun.\ Math.\ Phys.\  {\bf 306}, 511 (2011)
  doi:10.1007/s00220-011-1253-6
  [arXiv:1007.3837 [hep-th]].
  %%CITATION = doi:10.1007/s00220-011-1253-6;%%
  %324 citations counted in INSPIRE as of 22 Mar 2018

%\cite{Herzog:2010hf}
\bibitem{Herzog:2010hf}
  C.~P.~Herzog, I.~R.~Klebanov, S.~S.~Pufu and T.~Tesileanu,
  %``Multi-Matrix Models and Tri-Sasaki Einstein Spaces,''
  Phys.\ Rev.\ D {\bf 83}, 046001 (2011)
  doi:10.1103/PhysRevD.83.046001
  [arXiv:1011.5487 [hep-th]].
  %%CITATION = doi:10.1103/PhysRevD.83.046001;%%
  %165 citations counted in INSPIRE as of 22 Mar 2018




%\cite{Kim:2012ava}
\bibitem{Kim:2012ava}
  H.~C.~Kim and S.~Kim,
  %``M5-branes from gauge theories on the 5-sphere,''
  JHEP {\bf 1305}, 144 (2013)
  doi:10.1007/JHEP05(2013)144
  [arXiv:1206.6339 [hep-th]];
  %%CITATION = doi:10.1007/JHEP05(2013)144;%%
  %138 citations counted in INSPIRE as of 22 Mar 2018
%\cite{Kallen:2012zn}
%\bibitem{Kallen:2012zn}
  J.~Kallen, J.~A.~Minahan, A.~Nedelin and M.~Zabzine,
  %``$N^3$-behavior from 5D Yang-Mills theory,''
  JHEP {\bf 1210}, 184 (2012)
  doi:10.1007/JHEP10(2012)184
  [arXiv:1207.3763 [hep-th]];
  %%CITATION = doi:10.1007/JHEP10(2012)184;%%
  %56 citations counted in INSPIRE as of 22 Mar 2018
%\cite{Lockhart:2012vp}
%\bibitem{Lockhart:2012vp}
  G.~Lockhart and C.~Vafa,
  %``Superconformal Partition Functions and Non-perturbative Topological Strings,''
  arXiv:1210.5909 [hep-th];
  %%CITATION = ARXIV:1210.5909;%%
  %107 citations counted in INSPIRE as of 27 Jun 2018
%\cite{Kim:2012qf}
%\bibitem{Kim:2012qf}
  H.~C.~Kim, J.~Kim and S.~Kim,
  %``Instantons on the 5-sphere and M5-branes,''
  arXiv:1211.0144 [hep-th];
  %%CITATION = ARXIV:1211.0144;%%
  %99 citations counted in INSPIRE as of 27 Jun 2018
%\cite{Minahan:2013jwa}
%\bibitem{Minahan:2013jwa}
  J.~A.~Minahan, A.~Nedelin and M.~Zabzine,
  %``5D super Yang-Mills theory and the correspondence to AdS$_7$/CFT$_6$,''
  J.\ Phys.\ A {\bf 46}, 355401 (2013)
  doi:10.1088/1751-8113/46/35/355401
  [arXiv:1304.1016 [hep-th]];
  %%CITATION = doi:10.1088/1751-8113/46/35/355401;%%
  %43 citations counted in INSPIRE as of 04 Aug 2019
%\cite{Kim:2013nva}
%\bibitem{Kim:2013nva}
  H.~C.~Kim, S.~Kim, S.~S.~Kim and K.~Lee,
  %``The general M5-brane superconformal index,''
  arXiv:1307.7660 [hep-th].
  %%CITATION = ARXIV:1307.7660;%%
  %45 citations counted in INSPIRE as of 04 Aug 2019

%\cite{Harvey:1998bx}
\bibitem{Harvey:1998bx}
  J.~A.~Harvey, R.~Minasian and G.~W.~Moore,
  %``NonAbelian tensor multiplet anomalies,''
  JHEP {\bf 9809}, 004 (1998)
  doi:10.1088/1126-6708/1998/09/004
  [hep-th/9808060].
  %%CITATION = doi:10.1088/1126-6708/1998/09/004;%%
  %159 citations counted in INSPIRE as of 23 Apr 2019

%\cite{Maxfield:2012aw}
\bibitem{Maxfield:2012aw}
  T.~Maxfield and S.~Sethi,
  %``The Conformal Anomaly of M5-Branes,''
  JHEP {\bf 1206}, 075 (2012)
  doi:10.1007/JHEP06(2012)075
  [arXiv:1204.2002 [hep-th]].
  %%CITATION = doi:10.1007/JHEP06(2012)075;%%
  %38 citations counted in INSPIRE as of 23 Apr 2019

%\cite{Kim:2017zyo}
\bibitem{Kim:2017zyo}
  S.~Kim and J.~Nahmgoong,
  %``Asymptotic M5-brane entropy from S-duality,''
  JHEP {\bf 1712}, 120 (2017)
  doi:10.1007/JHEP12(2017)120
  [arXiv:1702.04058 [hep-th]].
  %%CITATION = doi:10.1007/JHEP12(2017)120;%%
  %3 citations counted in INSPIRE as of 23 Apr 2019


%\cite{Choi:2018hmj}
\bibitem{Choi:2018hmj}
  S.~Choi, J.~Kim, S.~Kim and J.~Nahmgoong,
  %``Large AdS black holes from QFT,''
  arXiv:1810.12067 [hep-th].
  %%CITATION = ARXIV:1810.12067;%%
  %14 citations counted in INSPIRE as of 23 Apr 2019

\bibitem{nahmgoong} J. Nahmgoong, arXiv:1907.12582 [hep-th].


%\cite{Benini:2015eyy}
\bibitem{Benini:2015eyy}
  F.~Benini, K.~Hristov and A.~Zaffaroni,
  %``Black hole microstates in AdS$_{4}$ from supersymmetric localization,''
  JHEP {\bf 1605}, 054 (2016)
  doi:10.1007/JHEP05(2016)054
  [arXiv:1511.04085 [hep-th]].
  %%CITATION = doi:10.1007/JHEP05(2016)054;%%
  %52 citations counted in INSPIRE as of 22 Mar 2018

%\cite{Benini:2016rke}
\bibitem{Benini:2016rke}
  F.~Benini, K.~Hristov and A.~Zaffaroni,
  %``Exact microstate counting for dyonic black holes in AdS4,''
  Phys.\ Lett.\ B {\bf 771}, 462 (2017)
  doi:10.1016/j.physletb.2017.05.076
  [arXiv:1608.07294 [hep-th]].
  %%CITATION = doi:10.1016/j.physletb.2017.05.076;%%
  %30 citations counted in INSPIRE as of 06 Jun 2018



%\cite{Cvetic:2005zi}
\bibitem{Cvetic:2005zi}
  M.~Cvetic, G.~W.~Gibbons, H.~Lu and C.~N.~Pope,
  %``Rotating black holes in gauged supergravities: Thermodynamics, supersymmetric limits, topological solitons and time machines,''
  hep-th/0504080.
  %%CITATION = HEP-TH/0504080;%%
  %91 citations counted in INSPIRE as of 22 Jun 2018

%\cite{Hristov:2019mqp}
\bibitem{Hristov:2019mqp}
  K.~Hristov, S.~Katmadas and C.~Toldo,
  %``Matter-coupled supersymmetric Kerr-Newman-AdS$_4$ black holes,''
  arXiv:1907.05192 [hep-th].
  %%CITATION = ARXIV:1907.05192;%%

%\cite{Kapustin:2010xq}
\bibitem{Kapustin:2010xq}
  A.~Kapustin, B.~Willett and I.~Yaakov,
  %``Nonperturbative Tests of Three-Dimensional Dualities,''
  JHEP {\bf 1010}, 013 (2010)
  doi:10.1007/JHEP10(2010)013
  [arXiv:1003.5694 [hep-th]].
  %%CITATION = doi:10.1007/JHEP10(2010)013;%%
  %162 citations counted in INSPIRE as of 06 Jun 2018

%\cite{Gang:2011xp}
\bibitem{Gang:2011xp}
  D.~Gang, E.~Koh, K.~Lee and J.~Park,
  %``ABCD of 3d ${\cal N}=8$ and 4 Superconformal Field Theories,''
  arXiv:1108.3647 [hep-th].
  %%CITATION = ARXIV:1108.3647;%%
  %19 citations counted in INSPIRE as of 06 Jun 2018


%\cite{Bhattacharya:2008zy}
\bibitem{Bhattacharya:2008zy}
  J.~Bhattacharya, S.~Bhattacharyya, S.~Minwalla and S.~Raju,
  %``Indices for Superconformal Field Theories in 3,5 and 6 Dimensions,''
  JHEP {\bf 0802}, 064 (2008)
  doi:10.1088/1126-6708/2008/02/064
  [arXiv:0801.1435 [hep-th]].
  %%CITATION = doi:10.1088/1126-6708/2008/02/064;%%
  %139 citations counted in INSPIRE as of 22 Jun 2018

%\cite{Bhattacharya:2008bja}
\bibitem{Bhattacharya:2008bja}
  J.~Bhattacharya and S.~Minwalla,
  %``Superconformal Indices for N = 6 Chern Simons Theories,''
  JHEP {\bf 0901}, 014 (2009)
  doi:10.1088/1126-6708/2009/01/014
  [arXiv:0806.3251 [hep-th]].
  %%CITATION = doi:10.1088/1126-6708/2009/01/014;%%
  %111 citations counted in INSPIRE as of 22 Jun 2018

%\cite{Kim:2009wb}
\bibitem{Kim:2009wb}
  S.~Kim,
  %``The Complete superconformal index for N=6 Chern-Simons theory,''
  Nucl.\ Phys.\ B {\bf 821}, 241 (2009)
  Erratum: [Nucl.\ Phys.\ B {\bf 864}, 884 (2012)]
  doi:10.1016/j.nuclphysb.2012.07.015, 10.1016/j.nuclphysb.2009.06.025
  [arXiv:0903.4172 [hep-th]].
  %%CITATION = doi:10.1016/j.nuclphysb.2012.07.015, 10.1016/j.nuclphysb.2009.06.025;%%
  %203 citations counted in INSPIRE as of 11 Jun 2018

%\cite{Choi:2018fdc}
\bibitem{Choi:2018fdc}
  S.~Choi, C.~Hwang, S.~Kim and J.~Nahmgoong,
  %``Entropy functions of BPS black holes in AdS$_4$ and AdS$_6$,''
  arXiv:1811.02158 [hep-th].
  %%CITATION = ARXIV:1811.02158;%%
  %3 citations counted in INSPIRE as of 23 Apr 2019


%\cite{Cabo-Bizet:2018ehj}
\bibitem{Cabo-Bizet:2018ehj}
  A.~Cabo-Bizet, D.~Cassani, D.~Martelli and S.~Murthy,
  %``Microscopic origin of the Bekenstein-Hawking entropy of supersymmetric AdS$_{\bf 5}$ black holes,''
  arXiv:1810.11442 [hep-th].
  %%CITATION = ARXIV:1810.11442;%%
  %16 citations counted in INSPIRE as of 23 Apr 2019

%\cite{Choi:2018vbz}
\bibitem{Choi:2018vbz}
  S.~Choi, J.~Kim, S.~Kim and J.~Nahmgoong,
  %``Comments on deconfinement in AdS/CFT,''
  arXiv:1811.08646 [hep-th].
  %%CITATION = ARXIV:1811.08646;%%
  %8 citations counted in INSPIRE as of 23 Apr 2019

%\cite{Benini:2018ywd}
\bibitem{Benini:2018ywd}
  F.~Benini and P.~Milan,
  %``Black holes in 4d $\mathcal{N}=4$ Super-Yang-Mills,''
  arXiv:1812.09613 [hep-th].
  %%CITATION = ARXIV:1812.09613;%%
  %10 citations counted in INSPIRE as of 23 Apr 2019

%\cite{Honda:2019cio}
\bibitem{Honda:2019cio}
  M.~Honda,
  %``Quantum Black Hole Entropy from 4d Supersymmetric Cardy formula,''
  arXiv:1901.08091 [hep-th].
  %%CITATION = ARXIV:1901.08091;%%
  %7 citations counted in INSPIRE as of 23 Apr 2019

%\cite{ArabiArdehali:2019tdm}
\bibitem{ArabiArdehali:2019tdm}
  A.~Arabi Ardehali,
  %``Cardy-like asymptotics of the 4d $\mathcal{N}=4$ index and AdS$_5$ blackholes,''
  arXiv:1902.06619 [hep-th].
  %%CITATION = ARXIV:1902.06619;%%
  %5 citations counted in INSPIRE as of 23 Apr 2019

%\cite{Kim:2019yrz}
\bibitem{Kim:2019yrz}
  J.~Kim, S.~Kim and J.~Song,
  %``A 4d N=1 Cardy Formula,''
  arXiv:1904.03455 [hep-th].
  %%CITATION = ARXIV:1904.03455;%%
  %1 citations counted in INSPIRE as of 23 Apr 2019

%\cite{Cabo-Bizet:2019osg}
\bibitem{Cabo-Bizet:2019osg}
  A.~Cabo-Bizet, D.~Cassani, D.~Martelli and S.~Murthy,
  %``The asymptotic growth of states of the 4d N=1 superconformal index,''
  arXiv:1904.05865 [hep-th].
  %%CITATION = ARXIV:1904.05865;%%

%\cite{Larsen:2019oll}
\bibitem{Larsen:2019oll}
  F.~Larsen, J.~Nian and Y.~Zeng,
  %``AdS$_5$ Black Hole Entropy near the BPS Limit,''
  arXiv:1907.02505 [hep-th].
  %%CITATION = ARXIV:1907.02505;%%
  %2 citations counted in INSPIRE as of 28 Jul 2019

%\cite{Choi:2019miv}
\bibitem{Choi:2019miv}
  S.~Choi and S.~Kim,
  %``Large AdS$_6$ black holes from CFT$_5$,''
  arXiv:1904.01164 [hep-th].
  %%CITATION = ARXIV:1904.01164;%%
  %1 citations counted in INSPIRE as of 23 Apr 2019

%\cite{Aharony:2012ns}
\bibitem{Aharony:2012ns}
  O.~Aharony, S.~Giombi, G.~Gur-Ari, J.~Maldacena and R.~Yacoby,
  %``The Thermal Free Energy in Large N Chern-Simons-Matter Theories,''
  JHEP {\bf 1303}, 121 (2013)
  doi:10.1007/JHEP03(2013)121
  [arXiv:1211.4843 [hep-th]].
  %%CITATION = doi:10.1007/JHEP03(2013)121;%%
  %83 citations counted in INSPIRE as of 08 Jun 2019

%\cite{Jain:2013py}
\bibitem{Jain:2013py}
  S.~Jain, S.~Minwalla, T.~Sharma, T.~Takimi, S.~R.~Wadia and S.~Yokoyama,
  %``Phases of large $N$ vector Chern-Simons theories on $S^2 \times S^1$,''
  JHEP {\bf 1309}, 009 (2013)
  doi:10.1007/JHEP09(2013)009
  [arXiv:1301.6169 [hep-th]].
  %%CITATION = doi:10.1007/JHEP09(2013)009;%%
  %66 citations counted in INSPIRE as of 08 Jun 2019



%\cite{Bagger:2006sk}
\bibitem{Bagger:2006sk}
  J.~Bagger and N.~Lambert,
  %``Modeling Multiple M2's,''
  Phys.\ Rev.\ D {\bf 75}, 045020 (2007)
  doi:10.1103/PhysRevD.75.045020
  [hep-th/0611108];
  %%CITATION = doi:10.1103/PhysRevD.75.045020;%%
  %714 citations counted in INSPIRE as of 23 Apr 2019
%\cite{Bagger:2007jr}
%\bibitem{Bagger:2007jr}
  J.~Bagger and N.~Lambert,
  %``Gauge symmetry and supersymmetry of multiple M2-branes,''
  Phys.\ Rev.\ D {\bf 77}, 065008 (2008)
  doi:10.1103/PhysRevD.77.065008
  [arXiv:0711.0955 [hep-th]].
  %%CITATION = doi:10.1103/PhysRevD.77.065008;%%
  %933 citations counted in INSPIRE as of 23 Apr 2019

%\cite{Gustavsson:2007vu}
\bibitem{Gustavsson:2007vu}
  A.~Gustavsson,
  %``Algebraic structures on parallel M2-branes,''
  Nucl.\ Phys.\ B {\bf 811}, 66 (2009)
  doi:10.1016/j.nuclphysb.2008.11.014
  [arXiv:0709.1260 [hep-th]].
  %%CITATION = doi:10.1016/j.nuclphysb.2008.11.014;%%
  %856 citations counted in INSPIRE as of 23 Apr 2019

%\cite{Aharony:2008ug}
\bibitem{Aharony:2008ug}
  O.~Aharony, O.~Bergman, D.~L.~Jafferis and J.~Maldacena,
  %``N=6 superconformal Chern-Simons-matter theories, M2-branes and their gravity duals,''
  JHEP {\bf 0810}, 091 (2008)
  doi:10.1088/1126-6708/2008/10/091
  [arXiv:0806.1218 [hep-th]].
  %%CITATION = doi:10.1088/1126-6708/2008/10/091;%%
  %1756 citations counted in INSPIRE as of 04 Sep 2018



%\cite{Yoshida:2014ssa}
\bibitem{Yoshida:2014ssa}
  Y.~Yoshida and K.~Sugiyama,
  %``Localization of 3d $\mathcal{N}=2$ Supersymmetric Theories on $S^1 \times D^2$,''
  arXiv:1409.6713 [hep-th].
  %%CITATION = ARXIV:1409.6713;%%
  %36 citations counted in INSPIRE as of 11 Jun 2018


%\cite{Gaiotto:2012xa}
\bibitem{Gaiotto:2012xa}
  D.~Gaiotto, L.~Rastelli and S.~S.~Razamat,
  %``Bootstrapping the superconformal index with surface defects,''
  JHEP {\bf 1301}, 022 (2013)
  doi:10.1007/JHEP01(2013)022
  [arXiv:1207.3577 [hep-th]].
  %%CITATION = doi:10.1007/JHEP01(2013)022;%%
  %145 citations counted in INSPIRE as of 04 Aug 2019


%\cite{Gaiotto:2008ak}
\bibitem{Gaiotto:2008ak}
  D.~Gaiotto and E.~Witten,
  %``S-Duality of Boundary Conditions In N=4 Super Yang-Mills Theory,''
  Adv.\ Theor.\ Math.\ Phys.\  {\bf 13}, no. 3, 721 (2009)
  doi:10.4310/ATMP.2009.v13.n3.a5
  [arXiv:0807.3720 [hep-th]].
  %%CITATION = doi:10.4310/ATMP.2009.v13.n3.a5;%%
  %373 citations counted in INSPIRE as of 03 Aug 2019


%\cite{Kim:2012uz}
\bibitem{Kim:2012uz}
  H.~C.~Kim, J.~Kim, S.~Kim and K.~Lee,
  %``Vortices and 3 dimensional dualities,''
  arXiv:1204.3895 [hep-th].
  %%CITATION = ARXIV:1204.3895;%%
  %13 citations counted in INSPIRE as of 11 Jun 2018

\bibitem{Pasquetti:2011fj}
  S.~Pasquetti,
  %``Factorisation of N = 2 Theories on the Squashed 3-Sphere,''
  JHEP {\bf 1204} (2012) 120
  [arXiv:1111.6905 [hep-th]];
  %%CITATION = doi:10.1007/JHEP04(2012)120;%%
%\bibitem{Beem:2012mb}
  C.~Beem, T.~Dimofte and S.~Pasquetti,
%  ``Holomorphic Blocks in Three Dimensions,''
  JHEP {\bf 1412} (2014) 177
  [arXiv:1211.1986 [hep-th]];
  %%CITATION = doi:10.1007/JHEP12(2014)177;%%
%\bibitem{Hwang:2012jh}
  C.~Hwang, H.~C.~Kim and J.~Park,
  %``Factorization of the 3d superconformal index,''
  JHEP {\bf 1408} (2014) 018
  [arXiv:1211.6023 [hep-th]];
  %%CITATION = doi:10.1007/JHEP08(2014)018;%%
%\bibitem{Taki:2013opa}
  M.~Taki,
  %``Holomorphic Blocks for 3d Non-abelian Partition Functions,''
  arXiv:1303.5915 [hep-th];
  %%CITATION = ARXIV:1303.5915;%%
%\bibitem{Fujitsuka:2013fga}
  M.~Fujitsuka, M.~Honda and Y.~Yoshida,
  %``Higgs branch localization of 3d ?? = 2 theories,''
  PTEP {\bf 2014} (2014) no.12,  123B02
  [arXiv:1312.3627 [hep-th]];
  %%CITATION = doi:10.1093/ptep/ptu158;%%
%\bibitem{Benini:2013yva}
  F.~Benini and W.~Peelaers,
  %``Higgs branch localization in three dimensions,''
  JHEP {\bf 1405} (2014) 030
  [arXiv:1312.6078 [hep-th]];
  %%CITATION = doi:10.1007/JHEP05(2014)030;%%
%\bibitem{Benini:2015noa}
  F.~Benini and A.~Zaffaroni,
  %``A topologically twisted index for three-dimensional supersymmetric theories,''
  JHEP {\bf 1507} (2015) 127
  [arXiv:1504.03698 [hep-th]];
  %%CITATION = doi:10.1007/JHEP07(2015)127;%%
%\bibitem{Hwang:2015wna}
  C.~Hwang and J.~Park,
  %``Factorization of the 3d superconformal index with an adjoint matter,''
  JHEP {\bf 1511} (2015) 028
  [arXiv:1506.03951 [hep-th]].
  %%CITATION = doi:10.1007/JHEP11(2015)028;%%

\bibitem{Hwang:2018uyj}
  C.~Hwang, H.~Kim and J.~Park,
  %``On 3d Seiberg-like Dualities with Two Adjoints,''
  arXiv:1807.06198 [hep-th].
  %%CITATION = ARXIV:1807.06198;%%



\bibitem{Imamura:2011su}
  Y.~Imamura and S.~Yokoyama,
  %``Index for three dimensional superconformal field theories with general R-charge assignments,''
  JHEP {\bf 1104} (2011) 007
  doi:10.1007/JHEP04(2011)007
  [arXiv:1101.0557 [hep-th]].
  %%CITATION = doi:10.1007/JHEP04(2011)007;%%

%\cite{Dimofte:2011py}
\bibitem{Dimofte:2011py}
  T.~Dimofte, D.~Gaiotto and S.~Gukov,
  %``3-Manifolds and 3d Indices,''
  Adv.\ Theor.\ Math.\ Phys.\  {\bf 17}, no. 5, 975 (2013)
  doi:10.4310/ATMP.2013.v17.n5.a3
  [arXiv:1112.5179 [hep-th]].
  %%CITATION = doi:10.4310/ATMP.2013.v17.n5.a3;%%
  %169 citations counted in INSPIRE as of 07 Aug 2019


\bibitem{Pasquetti:2019uop}
  S.~Pasquetti and M.~Sacchi,
  %``From 3$d$ dualities to 2$d$ free field correlators and back,''
  arXiv:1903.10817 [hep-th].
  %%CITATION = ARXIV:1903.10817;%%


%\cite{Aharony:2003sx}
\bibitem{Aharony:2003sx}
  O.~Aharony, J.~Marsano, S.~Minwalla, K.~Papadodimas and M.~Van Raamsdonk,
  %``The Hagedorn - deconfinement phase transition in weakly coupled large N gauge theories,''
  Adv.\ Theor.\ Math.\ Phys.\  {\bf 8}, 603 (2004)
  doi:10.4310/ATMP.2004.v8.n4.a1
  [hep-th/0310285].
  %%CITATION = doi:10.4310/ATMP.2004.v8.n4.a1;%%
  %434 citations counted in INSPIRE as of 04 Aug 2019


%\cite{Fredenhagen:2004cj}
\bibitem{Fredenhagen:2004cj}
  S.~Fredenhagen and V.~Schomerus,
%  ``Boundary Liouville theory at c = 1,''
  {JHEP {\bf 0505}, 025 (2005)}
  [{arXiv:hep-th/0409256}].
  %%CITATION = doi:10.1088/1126-6708/2005/05/025;%%
  %27 citations counted in INSPIRE as of 12 Apr 2018




\bibitem{2001math.....12035R} Rains, E.~M.\ 2001, arXiv Mathematics e-prints, math/0112035



%\cite{Nekrasov:2002qd}
\bibitem{Nekrasov:2002qd}
  N.~A.~Nekrasov,
  %``Seiberg-Witten prepotential from instanton counting,''
  Adv.\ Theor.\ Math.\ Phys.\  {\bf 7}, no. 5, 831 (2003)
  doi:10.4310/ATMP.2003.v7.n5.a4
  [hep-th/0206161].
  %%CITATION = doi:10.4310/ATMP.2003.v7.n5.a4;%%
  %1037 citations counted in INSPIRE as of 22 Jun 2018

%\cite{Iqbal:2008ra}
\bibitem{Iqbal:2008ra}
  A.~Iqbal, C.~Kozcaz and K.~Shabbir,
  %``Refined Topological Vertex, Cylindric Partitions and the U(1) Adjoint Theory,''
  Nucl.\ Phys.\ B {\bf 838}, 422 (2010)
  doi:10.1016/j.nuclphysb.2010.06.010
  [arXiv:0803.2260 [hep-th]].
  %%CITATION = doi:10.1016/j.nuclphysb.2010.06.010;%%
  %36 citations counted in INSPIRE as of 22 Jun 2018

%\cite{Kim:2011mv}
\bibitem{Kim:2011mv}
  H.~C.~Kim, S.~Kim, E.~Koh, K.~Lee and S.~Lee,
  %``On instantons as Kaluza-Klein modes of M5-branes,''
  JHEP {\bf 1112}, 031 (2011)
  doi:10.1007/JHEP12(2011)031
  [arXiv:1110.2175 [hep-th]].
  %%CITATION = doi:10.1007/JHEP12(2011)031;%%
  %86 citations counted in INSPIRE as of 22 Jun 2018



\end{thebibliography}
\end{document}